\documentclass[letterpaper,12pt]{article}

\usepackage[left=1in,right=1in,top=1in,bottom=1in,headsep=.25in]{geometry}
\usepackage[bottom]{footmisc}

\usepackage[T1]{fontenc}
\usepackage{lmodern}

\usepackage{amsmath,amsthm,amssymb,amsfonts,mathtools,bm}
\usepackage{dsfont,bbm}
\usepackage[mathscr]{euscript}

\usepackage{graphicx}
\usepackage{float}
\usepackage{wrapfig}
\usepackage{subcaption}
\usepackage{booktabs}
\usepackage{lscape}
\usepackage[flushleft]{threeparttable}
\usepackage{tabularx}
\usepackage{colortbl}
\usepackage{adjustbox}

\newcommand{\algorithmmaxheight}{0.9\textheight}

\usepackage{enumitem}

\usepackage{algorithm}
\usepackage{algpseudocode}

\usepackage{listings}
\usepackage{inconsolata}
\usepackage{pythonhighlight}

\usepackage{tikz}
\usetikzlibrary{arrows,backgrounds,snakes}

\usepackage{xcolor}
\definecolor{greenText}{rgb}{0.5,0.7,0.5}
\definecolor{greyText}{rgb}{0.5,0.5,0.5}
\definecolor{codeFrame}{rgb}{0.5,0.7,0.5}
\definecolor{Gray}{gray}{0.9}

\usepackage[round]{natbib}
\usepackage{url}

\usepackage{fancyhdr}
\newcommand{\headtitle}{Regression Adjustment for Ratios of Treatment Effects in Randomized Experiments}

\usepackage{titlesec}

\titleformat{\section}
  {\centering\normalfont\fontsize{12}{12}\selectfont\scshape}
  {\thesection.}
  {0.6em}
  {\MakeUppercase}

\titlespacing*{\section}
  {0pt}
  {2.0ex plus .4ex minus .2ex}
  {1.0ex plus .2ex}

\titleformat{\subsection}
  {\normalfont\fontsize{12}{12}\selectfont\bfseries}
  {\thesubsection.}
  {0.6em}
  {}

\titlespacing*{\subsection}
  {0pt}
  {1.6ex plus .3ex minus .2ex}
  {0.8ex plus .2ex}
\newcommand{\papertitleblock}{%
  \begin{center}
    {\fontsize{12}{14}\selectfont\bfseries
    REGRESSION ADJUSTMENT FOR RATIOS OF TREATMENT EFFECTS IN RANDOMIZED EXPERIMENTS\\[24pt]}
    {\fontsize{11}{13}\selectfont
    HARUO KAKEHI\footnote{haruo.kakehi@wisc.edu, Department of Economics, University of Wisconsin-Madison. I am grateful to Jack Porter for his invaluable advice on this project. I also thank Yong Cai, Harold Chiang, Bruce Hansen, Hyunseung Kang, Taisuke Otsu, Xiaoxia Shi, and Kohei Yata, as well as participants at the New York Econometrics Camp XX, SETA 2026, and the lunch seminar at the University of Wisconsin--Madison, for their valuable comments.}\par\par}
  \end{center}
  \vspace{24pt}
}

\renewenvironment{abstract}
{\small\begin{quote}\noindent{\scshape Abstract.}\ }
{\end{quote}}

\usepackage{hyperref}
\hypersetup{colorlinks=true,linkcolor=blue,citecolor=blue}

\lstdefinestyle{code}{
  frame=single, rulecolor=\color{codeFrame},
  numbers=left,
  numbersep=8pt,
  numberstyle=\tiny\color{greyText},
  commentstyle=\color{greenText},
  basicstyle=\linespread{1.1}\ttfamily\footnotesize,
  keywordstyle=\ttfamily\footnotesize,
  showstringspaces=false,
  xleftmargin=1.95em,
  framexleftmargin=1.6em,
  breaklines=true,
  postbreak=\mbox{\textcolor{greenText}{$\hookrightarrow$}\space}
}
\newtheoremstyle{assumpstyle}
  {3pt}{3pt}{\itshape}{}{\bfseries}{.}{0.5em}{}
\theoremstyle{assumpstyle}

\newtheorem{theorem}{Theorem}
\newtheorem{assumption}{Assumption}
\newtheorem{proposition}{Proposition}
\newtheorem{lemma}{Lemma}
\newtheorem{corollary}{Corollary}

\newtheoremstyle{remarkstyle}
  {3pt}{3pt}{\normalfont}{}{\bfseries}{.}{0.5em}{}
\theoremstyle{remarkstyle}
\newtheorem{remark}{Remark}
\newtheorem{example}{Example}

\newlist{assumplist}{enumerate}{1}
\setlist[assumplist]{label=(\roman*),leftmargin=*,topsep=0pt,itemsep=0pt,parsep=0pt}

\newenvironment{minilinespace}{\baselineskip = 5mm}{}

\makeatletter
\renewenvironment{abstract}{%
  \begin{quotation}
  \small
  \begin{center}
    \normalfont\scshape\abstractname
  \end{center}
  \noindent
}{%
  \end{quotation}
}
\makeatother

\begin{document}

\papertitleblock
\thispagestyle{empty}

\begin{abstract}
We study regression adjustment for ratios of treatment effects in completely randomized experiments. The existing literature adjusts the numerator and denominator separately, but this approach misses the efficiency target, loss of which matters more when richer or high-dimensional covariate information and flexible or regularized prediction methods are used to pursue further efficiency gains. We propose a two-step estimator that directly predicts the relevant transformed outcome. Using cross-fitting, our method accommodates high-dimensional first-stage prediction while keeping the final adjustment fixed-dimensional, thereby avoiding instability from a growing number of adjustment covariates. We further develop persistence-based LASSO theory that does not require sparsity, permits more covariates than observations, and allows cross-validated penalty selection under prediction-level conditions. For additional efficiency gains, our augmented two-step estimator nests the benchmark adjustment spaces and is therefore asymptotically no less efficient than the benchmark estimators, while its feasible conservative variance estimator satisfies the same efficiency ordering. Simulation studies and a real-data analysis demonstrate that our proposed estimator performs better than existing methods.
\end{abstract}


\clearpage

\section{Introduction}
\label{sec:intro}

Randomized controlled trials (RCTs) are the gold standard for causal inference in many applied fields. In such settings, regression adjustment is widely used to improve precision by leveraging baseline covariates while preserving a no-harm property: appropriately constructed regression-adjusted estimators are asymptotically no less precise than the corresponding unadjusted estimator, regardless of misspecification of the conditional mean function \citep{freedman2008regression_a,freedman2008regression_b,lin2013,Guo02012023,cohenfogarty2023,ye2023toward}. This efficiency gain depends on how well baseline covariates predict the relevant potential outcomes. This perspective has recently motivated adjustment with richer, possibly high-dimensional covariate spaces \citep{LeiandDing2020,chiang2025regressionadjustmentcompletelyrandomized,zhu2025design} and with more flexible prediction methods, including nonlinear regression, regularization, and machine learning \citep{Bloniarz2016,wager2016high,guo2021machine,bannick2025general,lu2025}.

Important estimands in RCTs often include ratios of treatment effects, such as the local average treatment effect (LATE) under noncompliance \citep{imbens2015causal} and the incremental cost-effectiveness ratio (ICER), which compares incremental costs with incremental health effects \citep{drummond2015methods}. Existing regression adjustment methods for the ratio of treatment effects construct adjustments for the numerator and denominator separately \citep{Ren02042024,jiang2024improving,liu2026regression}. However, for ratios of treatment effects, regression adjustment is not simply two separate adjustment problems, holding the misspecification of the conditional mean function. For example, in LATE, the first-order variance of the Wald ratio is governed by the transformed potential outcome $Y_i(z)-\tau D_i(z)$, where $\tau$ is the LATE, rather than by the reduced form $Y_i(z)$ and the first stage $D_i(z)$ separately. The problem is that predicting $Y_i(z)$ and $D_i(z)$ separately does not necessarily imply good prediction of the particular combination $Y_i(z)-\tau D_i(z)$ that determines LATE precision. With ordinary least squares using the same projection space, directly predicting this combination is equivalent to combining the separate predictions. This equivalence need not hold for flexible or regularized prediction methods, which are increasingly used with many covariates to improve prediction and thereby reduce asymptotic variance \citep{Bloniarz2016,wager2016high,farrell2021deep,zhu2025design}.

Directly adjusting the transformed outcome $Y_i(z)-\tau D_i(z)$ with rich covariate information raises two challenges. First, using a growing number of covariates directly in the adjustment estimator can introduce nonnegligible bias and complicate variance estimation \citep{LeiandDing2020,chiang2025regressionadjustmentcompletelyrandomized}. Second, verifying the empirical process condition that controls dependence between prediction and adjustment through entropy bounds can be difficult for flexible learners, especially with many covariates and data-dependent tuning \citep{cohenfogarty2023, Ren02042024}.

In this paper, we propose a two-step estimator for ratios of treatment effects that directly predicts the transformed outcome. By using the predicted values of $Y_i(z)-\tau D_i(z)$ as adjustment covariates, we compress rich covariate information into a fixed-dimensional set of covariates, thereby avoiding the many-regressor estimation problem. In addition, we use conditional cross-fitting \citep{lu2025} that accommodates flexible learners and cross-validation under prediction-level conditions. Our LASSO theory uses a persistence framework \citep{greenshteinritov2004} to establish prediction stability without imposing sparsity on the regression coefficients. This accommodates settings in which predictive information is spread across many covariates, complementing the sparsity-based approach of \citet{Bloniarz2016}. We further establish prediction stability when the penalty is selected by cross-validation, extending the guarantee from a prespecified penalty to data-driven tuning within the design-based framework. To retain the efficiency gains of existing estimators, we propose an augmented two-step estimator that adds baseline covariates and separate predictions to the adjustment covariates. Because the augmented estimator nests the benchmark adjustment spaces, it is asymptotically no less efficient than the benchmarks. Finally, we construct a feasible conservative variance estimator whose probability limit satisfies the same efficiency ordering.

This paper contributes to the literature on regression adjustment in randomized experiments, including adjustment under misspecification \citep{freedman2008regression_a,freedman2008regression_b,lin2013,ye2023toward,lu2025tyranny}. Within the literature on experiments with noncompliance \citep{ansel2018ols,BUGNI2023105497,hartman2024improving,bai2025inference,pashley2025improving}, our work is most closely related to \citet{Ren02042024}, who proposes an Indirect Least Squares Estimator (ILS) and Calibrated Oaxaca-Blinder Estimator (COB) estimators under complete randomization, and \citet{jiang2024improving}, who develop parametric, nonparametric, and regularized adjustments under covariate-adaptive randomization, including semiparametrically efficient procedures under additional conditions. Both approaches separately adjust the reduced form and the first stage. Our two-step estimator directly predicts the transformed outcome governing ratio precision, and our augmented estimator provides efficiency gains when separate adjustment misses a relevant prediction direction, with the same efficiency ordering for the conservative variance estimator.

We also contribute to the literature on regression adjustment with many or high-dimensional covariates \citep{Bloniarz2016,wager2016high,LeiandDing2020,chiang2025regressionadjustmentcompletelyrandomized,lu2025debiased}. We keep the final adjustment fixed-dimensional to avoid bias problems in many covariates, and establish inference under prediction-level conditions. Our LASSO theory adopts the persistence framework \citep{greenshteinritov2004}, allowing $p>N$ under suitable prediction conditions without imposing sparsity conditions, which have been noted to be stringent in the literature \citep{wager2016high,LeiandDing2020,lu2025debiased}. This complements the sparsity-free theory of \citet{lu2025debiased} for linear regression adjustment in the moderately high-dimensional regime.

The remainder of the paper is organized as follows. Section~\ref{sec:model} introduces the setup and the unadjusted ratio estimator. Section~\ref{sec:min_var} characterizes the variance-minimization problem and the role of adjustment covariates. Section~\ref{sec:two_step} develops the two-step estimator and conservative variance estimation. Section~\ref{sec:lasso} develops the LASSO extension. Sections~\ref{sec:simulation} and~\ref{sec:application} report the simulation and empirical application, respectively.

\section{Setup}
\label{sec:model}

\subsection{Notation}

We adopt a finite-population, design-based framework. The population consists of $N$ units indexed by $i=1,\ldots,N$, potential responses $Y_i(z)\in\mathbb R$ and $D_i(z)\in\mathbb R$ for $z\in\{0,1\}$, and fixed $p$-dimensional covariates $X_i\in\mathbb R^p$. The assignment is $Z_i\in\{0,1\}$. The observed outcomes are $D_i=D_i(Z_i)$ and $Y_i=Y_i(Z_i)$. We treat $\{Y_i(z),D_i(z),X_i\}_{i=1}^N$ as fixed, so randomness arises only from the assignment $Z=(Z_1,\ldots,Z_N)$. Hence, the observed variables satisfy
\[
Y_i=Y_i(Z_i)=Z_iY_i(1)+(1-Z_i)Y_i(0),\qquad
D_i=D_i(Z_i)=Z_iD_i(1)+(1-Z_i)D_i(0),
\]
and the researcher observes $\{Y_i,D_i,Z_i,X_i\}_{i=1}^N$.

We consider a completely randomized experiment with fixed numbers $N_1>0$ and $N_0>0$ assigned to $Z_i=1$ and $Z_i=0$, respectively, where $N=N_1+N_0$. Let
\[
\mathcal{Z}_{N_1}
=
\left\{
z\in\{0,1\}^N:\ \sum_{i=1}^N z_i=N_1
\right\}.
\]
Under complete randomization, $Z=(Z_1,\ldots,Z_N)$ is drawn uniformly from $\mathcal{Z}_{N_1}$.

For covariates $X_i\in\mathbb{R}^{p}$, define the finite-population mean and variance-covariance matrix $\bar X=\frac{1}{N}\sum_{i=1}^N X_i$ and $S_X^2=\frac{1}{N-1}\sum_{i=1}^{N}(X_i-\bar X)(X_i-\bar X)'$, respectively. For $R\in\{Y,D\}$ and $z\in\{0,1\}$, define $\bar R(z)=\frac{1}{N}\sum_{i=1}^N R_i(z)$ and $S^2_{R(z)}=\frac{1}{N-1}\sum_{i=1}^N \left\{R_i(z)-\bar R(z)\right\}^2$. For $R\in\{Y,D\}$ and $z\in\{0,1\}$, the observed mean is $\bar R_z=\frac{1}{N_z}\sum_{i=1}^{N}\mathbf{1}\{Z_i=z\}R_i$, where $\mathbf{1}\{\cdot\}$ is an indicator function. For scalar sequences $R_i(z)$ and $Q_i(z)$, define
\[
S_{R(z)Q(z)} = \frac{1}{N-1}\sum_{i=1}^N \bigl(R_i(z)-\bar R(z)\bigr) \bigl(Q_i(z)-\bar Q(z)\bigr),
\]
and the finite-population variance of the unit-level treatment effect
\[
S^2_{R(1)-R(0)}=\frac{1}{N-1}\sum_{i=1}^N \left\{\left(R_i(1)-R_i(0)\right)-\left(\bar R(1)-\bar R(0)\right)\right\}^2.
\]
Similarly, for a scalar sequence $R_i(z)$ and covariates $X_i$, define $S_{X,R(z)} = \frac{1}{N-1}\sum_{i=1}^N \bigl(X_i-\bar X\bigr) \bigl(R_i(z)-\bar R(z)\bigr).$ For a sequence $f=(f_1,\ldots,f_N)$ with $f_i\in\mathbb R^m$, define the empirical $L_2$ norm by $\|f\|_N = \left\{ \frac{1}{N} \sum_{i=1}^N \|f_i\|_2^2 \right\}^{1/2},$ where $\|\cdot\|_2$ denotes the Euclidean norm.

\subsection{Estimand}
\label{sec:estimand_estimator}

We focus on the ratio of the average treatment effects of assignment on $Y$ and $D$. Define these effects by
\[
\tau_{D}=\frac{1}{N}\sum_{i=1}^{N}\{D_i(1)-D_i(0)\},\qquad
\tau_{Y}=\frac{1}{N}\sum_{i=1}^{N}\{Y_i(1)-Y_i(0)\}.
\]
We impose a nonvanishing-denominator condition.

\begin{assumption}
\label{ass:id}
$|\tau_D|\ge C>0$, where $C$ is a constant independent of $N$.
\end{assumption}

Under Assumption~\ref{ass:id}, the ratio of average treatment effects is
\[
\tau=\frac{\tau_{Y}}{\tau_{D}}.
\]
Before introducing the estimator, we give two motivating examples of this ratio estimand.

\begin{example}[Cost-effectiveness]
\leavevmode \par
In a randomized experiment of health economics and public health, let $Y_i(z)$ denote the total cost and $D_i(z)$ the health outcome under assignment $z$. Then $\tau_Y$ is the average incremental cost and $\tau_D$ is the average incremental health effect. Under Assumption~\ref{ass:id}, $\tau$ is the incremental cost-effectiveness ratio (ICER), which compares the change in average cost with the change in average health \citep{drummond2015methods}. For example, when health is measured in quality-adjusted life-years, $\tau$ gives the incremental cost per quality-adjusted life-year gained.
\end{example}

\begin{example}[Local average treatment effect]
\leavevmode \par
In experiments with noncompliance, $Z_i$ is an instrument and $D_i(z)\in\{0,1\}$ denotes potential treatment. We introduce potential outcomes $Y_i(z,d)$, with $Y_i(z)=Y_i\bigl(z,D_i(z)\bigr)$. Suppose that the exclusion restriction $Y_i(1,d)=Y_i(0,d)$ holds for every $i$ and $d\in\{0,1\}$, and that monotonicity $D_i(1)\ge D_i(0)$ holds for every $i$. Under exclusion, define the unit-level treatment-received effect
\[
\tau_i^D
=
Y_i(0,1)-Y_i(0,0)
=
Y_i(1,1)-Y_i(1,0).
\]
Under these restrictions and Assumption~\ref{ass:id}, $\tau_D$ is positive and $\tau$ is the average of $\tau_i^D$ among units with $D_i(1)>D_i(0)$, namely the local average treatment effect \citep{imbens2015causal}.
\end{example}

\subsection{Ratio Estimator}
\label{sec:wald_estimator}

For $R\in\{Y,D\}$, define the difference-in-means estimator $\hat{\tau}_R=\bar R_1-\bar R_0$. The ratio estimator, defined when $\hat{\tau}_D\ne0$, is
\[
\hat{\tau} = \frac{\hat{\tau}_{Y}}{\hat{\tau}_{D}}.
\]
Because $\tau$ is a ratio, the first-order behavior of $\hat{\tau}$ is governed by $(\hat\tau_Y-\tau\hat\tau_D)/\tau_D=\hat\tau_A/\tau_D$, where $\hat\tau_A=\hat\tau_Y-\tau\hat\tau_D$. Under the finite-population central limit theorem conditions below \citep{Li02102017}, this statistic has the same first-order asymptotic variance as $\hat\tau$. Define the transformed potential outcome $A_i(z)=Y_i(z)-\tau D_i(z)$ for $z\in\{0,1\}$ and its realized value $A_i=A_i(Z_i)=Y_i-\tau D_i$. Write $\bar A(z)=N^{-1}\sum_{i=1}^N A_i(z)$ and $\bar A_z=N_z^{-1}\sum_{i:Z_i=z}A_i$, and let $\hat\tau_A=\bar A_1-\bar A_0$. Define the finite-population variance 
\[
\sigma^2
=
\frac{N}{\tau_D^2}
\left\{
\frac{S^2_{A(1)}}{N_1}
+
\frac{S^2_{A(0)}}{N_0}
-
\frac{S^2_{A(1)-A(0)}}{N}
\right\}.
\]
We impose the following standard regularity conditions for the finite-population central limit theorem.

\begin{assumption}
\label{ass:asymp}
\leavevmode \par
As $N \rightarrow \infty$, \par
(i) $\frac{N_1}{N} \rightarrow p_1 \in (0,1)$ and $\frac{N_0}{N} \rightarrow p_0 \in (0,1)$.\par
(ii) For $z\in\{0,1\}$, the finite-population means $\bar Y(z)$ and $\bar D(z)$ and the finite-population covariances $S^2_{Y(z)}, S^2_{D(z)}, S_{Y(z)D(z)}, S^2_{Y(1)-Y(0)}, S^2_{D(1)-D(0)}, S_{\{Y(1)-Y(0)\}\{D(1)-D(0)\}}$ converge to finite limits, and the limit of $\sigma^2$ is positive.\par
(iii) For $z \in \{0,1\}$, $\max_{1\le i\le N}\frac{\{A_i(z)-\bar A(z)\}^2}{N}\to 0.$
\end{assumption}

Under Assumptions~\ref{ass:id} and \ref{ass:asymp}, the probability that $\hat\tau_D=0$ tends to zero, and the finite-population central limit theorem implies that $\sqrt N(\hat\tau-\tau)$ is asymptotically normal with variance
\[
V = \lim_{N\rightarrow\infty}\sigma^2 = \lim_{N\rightarrow\infty}\frac{N}{\tau^2_D}\left\{\frac{S^2_{A(1)}}{N_1} + \frac{S^2_{A(0)}}{N_0} - \frac{S^2_{A(1)-A(0)}}{N}\right\}.
\]
\section{Variance Structure of Covariate Adjustment}
\label{sec:min_var}

This section introduces the structure governing the precision of regression adjustment for the ratio estimand. Fix an adjustment vector $G_i=g(X_i)\in\mathbb R^k$. Write $\tilde G_i=G_i-\bar G$, $\bar G_z=N_z^{-1}\sum_{i:Z_i=z}G_i$, and
\[
S_G^2=\frac{1}{N-1}\sum_{i=1}^N\tilde G_i\tilde G_i',
\qquad
S_{G,R}=\frac{1}{N-1}\sum_{i=1}^N\tilde G_i(R_i-\bar R)
\]
for any scalar finite-population sequence $R_i$. We assume that $S_G^2$ is positive definite. For $R\in\{Y,D\}$, the regression-adjusted ratio estimator $\hat\tau_G=\hat\tau_{G,Y}/\hat\tau_{G,D}$ is defined as
\begin{align}
\label{eq:adjustment_estimator}
\hat\tau_{G,R}
&=
\left\{\bar R_1-(\bar G_1-\bar G)'\hat\gamma_R(1)\right\}
-
\left\{\bar R_0-(\bar G_0-\bar G)'\hat\gamma_R(0)\right\},\\
\hat\gamma_R(z)
&=
\arg\min_{\gamma\in\mathbb R^k}
\sum_{i:Z_i=z}
\left\{R_i-\bar R_z-(G_i-\bar G_z)'\gamma\right\}^2, \notag
\end{align} 
Define the population projection coefficients
\[
\gamma_R(z)=(S_G^2)^{-1}S_{G,R(z)},
\qquad R\in\{Y,D\},\quad z\in\{0,1\}.
\]
The oracle transformed-outcome coefficients are
\[
\delta_z=(S_G^2)^{-1}S_{G,A(z)}
=\gamma_Y(z)-\tau\gamma_D(z),
\qquad z\in\{0,1\}.
\]
For $b_1,b_0\in\mathbb R^k$, set $A_{G,i}(z;b_z)=A_i(z)-\tilde G_i'b_z$ and define variances as
\begin{align*}
V_N(G;b_1,b_0)
&=
\frac{N}{\tau_D^2}
\left\{
\frac{S^2_{A_G(1;b_1)}}{N_1}
+
\frac{S^2_{A_G(0;b_0)}}{N_0}
-
\frac{S^2_{A_G(1;b_1)-A_G(0;b_0)}}{N}
\right\},\\
V_{\mathrm{gap},N}(G;b_1,b_0)
&=
\frac{1}{\tau_D^2}S^2_{A_G(1;b_1)-A_G(0;b_0)},\\
V_{\mathrm{cons},N}(G;b_1,b_0)
&=
V_N(G;b_1,b_0)+V_{\mathrm{gap},N}(G;b_1,b_0).
\end{align*}

The oracle first-order variance $V_N$ and the conservative variance $V_{\mathrm{cons},N}$ differ by $V_{\mathrm{gap},N}$, which is generally unidentified because $A_i(1)$ and $A_i(0)$ are never jointly observed. The following theorem shows that the same arm-specific projections minimize both variances and the gap.
\begin{theorem}
\label{thm:late_oracle_variance_min}
Suppose $S_G^2$ is positive definite and $\tau_D\ne0$.
\begin{itemize}
\item[(i)] $(\delta_1,\delta_0)$ uniquely minimizes $V_{\mathrm{cons},N}(G;b_1,b_0)$.
\item[(ii)] For every $(b_1,b_0)$,
\begin{align*}
V_N(G;b_1,b_0)
&=
\frac{N_1}{N_0\tau_D^2}
S^2_{\frac{N_0}{N_1}A_G(1;b_1)+A_G(0;b_0)},\\
V_{\mathrm{gap},N}(G;b_1,b_0)
&=
\frac{1}{\tau_D^2}
S^2_{A_G(1;b_1)-A_G(0;b_0)}.
\end{align*}
The pair $(\delta_1,\delta_0)$ minimizes both criteria.
\item[(iii)] Replacing $G_i$ by $(\tilde G_i'\delta_1,\tilde G_i'\delta_0)'$ leaves the minimum values of $V_N$, $V_{\mathrm{gap},N}$, and $V_{\mathrm{cons},N}$ unchanged.
\end{itemize}
\end{theorem}

\begin{proof}
See Appendix~\ref{app:proof_var_min}.
\end{proof}

Theorem~\ref{thm:late_oracle_variance_min} identifies $A_i(z)$ as the efficiency-relevant target: its arm-specific least-squares projections jointly minimize the first-order variance $V_N$ and the conservativeness gap $V_{\mathrm{gap},N}$. Part~(iii) further shows that the two fitted values $(\tilde G_i'\delta_1,\tilde G_i'\delta_0)'$ preserve all gains available from $\operatorname{span}(G)$, so richer adjustment helps only by improving prediction of $A_i(1)$ or $A_i(0)$. This compression motivates the two-step estimator in Section~\ref{sec:two_step}, which uses rich covariate information for prediction while keeping the final adjustment fixed-dimensional.

\begin{remark}
\label{rem:late_improvement_condition}
\leavevmode \par
The literature uses predicted potential outcomes based on $X_i$ as adjustment covariates \citep{Guo02012023,cohenfogarty2023,ren2024}. Let $\mu_{R,i}(z)$ denote the deterministic fitted-value target for $R\in\{Y,D,A\}$ in arm $z$, and define
\[
W_i=
\left(
\mu_{Y,i}(1),\mu_{Y,i}(0),
\mu_{D,i}(1),\mu_{D,i}(0)
\right)'.
\]
For example, $\mu_{Y,i}(1)$ is the prediction by the oracle projection of $Y_i(1)$ on covariates $G_i$. In equation \eqref{eq:adjustment_estimator}, Indirect Least Squares Estimator (ILS) uses $G_i=X_i$. In the binary-treatment case, Calibrated Oaxaca-Blinder Estimator (COB) uses $G_i=W_i$.  Componentwise prediction induces
\[
\widetilde\mu_{A,i}(z)
=
\mu_{Y,i}(z)-\tau\mu_{D,i}(z)
\in \operatorname{span}(W).
\]
When the componentwise targets are the correct conditional means under the working model, their combination recovers the corresponding conditional mean of $A_i(z)$ by linearity, therefore attains the efficiency bound \cite{jiang2024improving}. With misspecification, it need not equal the target obtained by predicting $A_i(z)$ directly. At the oracle projection level, equality is preserved by least-squares projection onto a fixed common linear space, but generally fails for nonlinear, regularized, or machine-learning rules. Examples~\ref{ex:nonlinear_adjustment}--\ref{ex:machine_learning} illustrate this distinction, and Section~\ref{sec:two_step} exploits it through direct prediction of $A_i(z)$.
\end{remark}

\begin{example}[Nonlinear adjustment]
\label{ex:nonlinear_adjustment}
\leavevmode \par
The COB estimator \citep{Ren02042024} in the binary-treatment setting and the $F$-type nonlinear adjustment \citep{jiang2024improving} model $Y_i(z)$ and $D_i(z)$ separately, using logistic regression for binary $D_i(z)$. Direct prediction of $A_i(z)$ can add a direction outside $\operatorname{span}(X,W)$.
\end{example}

\begin{example}[High-dimensional covariates]
\label{ex:many_regressor}
\leavevmode \par
Experiments with many baseline covariates motivate LASSO adjustment \citep{Bloniarz2016,wager2016high,zhu2025design,jiang2024improving}. Because its $\ell_1$-constrained prediction class is not a linear space, $\|b_1\|_1,\|b_2\|_1\le B_N$ only imply
\[
\|b_1-\tau b_2\|_1\le (1+|\tau|)B_N,
\]
not the original constraint $\|b_1-\tau b_2\|_1\le B_N$.  Direct prediction can therefore add a direction outside the span of the separately fitted values. In an empirical application, \citet{jones2019workplace} use randomized eligibility as an instrument for wellness-program participation and a post-LASSO union of controls selected separately for participation and each outcome.
\end{example}

\begin{example}[Machine learning methods]
\label{ex:machine_learning}
\leavevmode \par
Flexible learners such as random forests and neural networks can reduce asymptotic variance by improving prediction of the efficiency-relevant potential outcome \citep{guo2021machine,su2023decorrelation,bannick2025general,lu2025}. Because these fitting rules are generally nonlinear in the response, predictions trained on $Y_i(z)-\tau D_i(z)$ need not lie in the span of predictions trained separately on $Y_i(z)$ and $D_i(z)$. For application, \citet{farrell2021deep} use neural networks in a randomized direct-mail experiment, while \citet{list2024using} use random forests to adjust the reduced form and first stage separately in the Oregon Health Insurance Experiment.
\end{example}

\begin{remark}
\leavevmode \par
The conservativeness gap has the familiar treatment-effect-heterogeneity interpretation from regression adjustment for the ATE \citep{ding2019decomposing,lu2025tyranny}. For the ratio estimand, the unit-level transformed effect is
\[
A_i(1)-A_i(0)
=
\{Y_i(1)-Y_i(0)\}-\tau\{D_i(1)-D_i(0)\},
\]
so reducing the gap corresponds to explaining variation in the transformed effect across units. Thus, interacted adjustment can reduce the gap when the adjustment covariates predict this variation, whereas a common-slope adjustment leaves it unchanged.
\end{remark}
    
\section{Estimator}
\label{sec:two_step}

\subsection{Two-step estimator}
\label{sec:two_step_estimator}

Theorem~\ref{thm:late_oracle_variance_min} identifies the arm-specific predictions of $A_i(z)=Y_i(z)-\tau D_i(z)$ as the efficiency-relevant target. This section uses that characterization to construct a feasible estimator. Because $\tau$ is unknown, we form fold-specific pseudo-outcomes using preliminary ratio estimators and use conditional cross-fitting \citep{lu2025} to separate first-stage prediction from second-stage adjustment, allowing flexible prediction methods.

Let $Q\ge2$ be fixed. After observing $Z$, randomly partition the treated and control units separately into folds $\mathcal S=\{S^{[1]},\ldots,S^{[Q]}\}$ as described in Algorithm~\ref{alg:cf_late}, and write $S^{[-q]}=\{1,\ldots,N\}\setminus S^{[q]}$. For each $q$, compute $\hat\tau^{[-q]}$ using $S^{[-q]}$, set $\hat A_i^{[-q]}=Y_i-\hat\tau^{[-q]}D_i$ for $i\in S^{[-q]}$, and obtain the predicted potential outcome $\hat\mu_A^{[-q]}(z)$ using the observations in $S^{[-q]}$ assigned to arm $z$. Define
\begin{align*}
\hat M_i^{[-q]}&=\left(\hat\mu_{A,i}^{[-q]}(1),\hat\mu_{A,i}^{[-q]}(0)\right)',
&M_i&=\left(\mu_{A,i}(1),\mu_{A,i}(0)\right)',
\end{align*}
where $M_i$ is a deterministic oracle prediction target of $A_i(z)$. The prediction rules need not estimate correctly specified conditional means.

Write $N^{[q]}=|S^{[q]}|$ and $N_z^{[q]}=\sum_{i\in S^{[q]}}\mathbf 1\{Z_i=z\}$. We impose the following conditions.

\begin{assumption}
\label{ass:conditional_crossfit}
\leavevmode\par
\begin{itemize}
\item[(i)] Conditional on $\mathcal S$, the assignment vectors $Z^{[1]},\ldots,Z^{[Q]}$ are mutually independent, and each $Z^{[q]}$ follows a completely randomized assignment with $N_1^{[q]}$ treated units and $N_0^{[q]}$ control units, where $0<N_z^{[q]}<N^{[q]}$ for $z\in\{0,1\}$.
\item[(ii)] The fold sizes satisfy
\[
\max_{1\le q\le Q}\left|\frac{N_1^{[q]}}{N^{[q]}}-\frac{N_1}{N}\right|=O(N^{-1/2}),
\]
and $N^{[q]}/N$ is bounded away from zero.
\item[(iii)] The first-stage predictions are stable on the fold where they are evaluated:
\[
\max_{1\le q\le Q}
\left\{
\frac{1}{N^{[q]}}
\sum_{i\in S^{[q]}}
\left\|
\hat M_i^{[-q]}-M_i
\right\|_2^2
\right\}^{1/2}
\xrightarrow{p}0.
\]
\end{itemize}
\end{assumption}

Conditional cross-fitting ensures Assumption~\ref{ass:conditional_crossfit}(i)--(ii). Assumption~\ref{ass:conditional_crossfit}(iii) is the central learner-specific requirement in the full-rank case. It asks the fitted predictions to approach deterministic targets on the evaluation fold, without requiring those targets to be correctly specified conditional means or imposing a particular convergence rate beyond vanishing prediction error. Unlike the full-population stability condition in \citet{lu2025}, Assumption~\ref{ass:conditional_crossfit}(iii) requires stability of each prediction rule only over $S^{[q]}$, where $\hat M^{[-q]}$ is evaluated, rather than over the entire finite population. The first-stage prediction model $\hat\mu_{A,i}^{[-q]}(z)$ is constructed from the training sample $\{(\hat A_i^{[-q]},X_i):i\in S^{[-q]},Z_i=z\}$. For example, given a class $\mathcal F_{N,z}$, one may define
\[
\hat\mu_A^{[-q]}(z)
\in
\arg\min_{m\in\mathcal F_{N,z}}
\sum_{i\in S^{[-q]}:Z_i=z}
\{\hat A_i^{[-q]}-m(X_i)\}^2.
\]
The class $\mathcal F_{N,z}$ may be linear, nonparametric, or regularized and may depend on $N$. More generally, $\hat\mu_A^{[-q]}(z)$ can be any prediction rule trained only on $S^{[-q]}$ and evaluated on $S^{[q]}$, with any auxiliary randomization independent of treatment assignment and sample splitting. We do not impose entropy conditions or prediction-unbiasedness as in \citet{cohenfogarty2023}.

For \(i\in S^{[q]}\), set \(\hat M_{i,\mathrm{cf}}=\hat M_i^{[-q]}\). For \(R\in\{Y,D\}\), define \(\hat\tau_{R,\mathrm{cf}}\) as \(\hat\tau_{G,R}\) from the preceding section with \(G_i=\hat M_{i,\mathrm{cf}}\). With $\hat A_i=Y_i-\hat\tau D_i$, our two-step regression adjustment estimator $\hat\tau_{2s}$ is
\[
\hat\tau_{2s}=\hat\tau+\frac{\hat\tau_{\hat A,\mathrm{cf}}}{\hat\tau_{D,\mathrm{cf}}}.
\]
Thus, the first step predicts the transformed potential outcomes, while the second step applies the preceding regression adjustment using the resulting two-dimensional generated covariate $\hat{M}_{i,\mathrm{cf}}$.

For $z\in\{0,1\}$, define $\delta_{M,z}=(S_M^2)^{-1}S_{M,A(z)}$, $A_{2s,i}(z)=A_i(z)-\tilde M_i'\delta_{M,z}$, and $\bar A_{2s}(z)=N^{-1}\sum_{i=1}^N A_{2s,i}(z)$. For the asymptotic analysis, we introduce the following assumptions.

\begin{assumption}
\label{ass:two_step_asymptotics}
\leavevmode\par
\begin{itemize}
\item[(i)] For $z\in\{0,1\}$,
\[
\max_{1\le i\le N}\frac{\{A_{2s,i}(z)-\bar A_{2s}(z)\}^2}{N}\to0.
\]
\item[(ii)]
\[
\max_{1\le i\le N}\frac{\|M_i-\bar M\|^2}{N}\to0.
\]
For each $z\in\{0,1\}$, the finite-population covariances $S_M^2$, $S_{M,A(z)}$, and $S_{M,D(z)}$ have finite limits, and $S_M^2$ and its limit are positive definite.
\item[(iii)] The limit of
\[
\sigma_{2s}^2=\frac{N}{\tau_D^2}\left\{\frac{S^2_{A_{2s}(1)}}{N_1}+\frac{S^2_{A_{2s}(0)}}{N_0}-\frac{S^2_{A_{2s}(1)-A_{2s}(0)}}{N}\right\}
\]
exists and is positive.
\end{itemize}
\end{assumption}

Assumption~\ref{ass:two_step_asymptotics} imposes standard moment, maximum, and rank conditions for the final adjustment, together with a positive limiting variance.

\begin{theorem}
\label{thm:two_step_cf}
\leavevmode \par
Under Assumptions~\ref{ass:id}, \ref{ass:asymp}, \ref{ass:conditional_crossfit}, and \ref{ass:two_step_asymptotics},
\begin{itemize}
\item[(i)] $\hat\tau_{2s}-\tau\xrightarrow{p}0$.
\item[(ii)] $\sqrt N(\hat\tau_{2s}-\tau)\xrightarrow{d}\mathcal N(0,V_{2s})$, where
\[
V_{2s}=\lim_{N\to\infty}\frac{N}{\tau_D^2}\left\{\frac{S^2_{A_{2s}(1)}}{N_1}+\frac{S^2_{A_{2s}(0)}}{N_0}-\frac{S^2_{A_{2s}(1)-A_{2s}(0)}}{N}\right\}.
\]
\item[(iii)] $V_{2s}\le V$.
\end{itemize}
\end{theorem}

\begin{proof}
See Appendix~\ref{app:proof_two_step_cf}.
\end{proof}

Theorem~\ref{thm:two_step_cf} is the main inference result: conditional cross-fitting and prediction stability yield first-order equivalence to adjustment by $M_i$. Theorem~\ref{thm:late_oracle_variance_min} then ensures that the two-step estimator is asymptotically no less efficient than the unadjusted ratio, even under misspecification.

\begin{remark}
\label{rem:sieve_ils}
\leavevmode\par
Let $K_N$ denote the dimension of the covariate vector used for first-stage prediction. When the same covariate space is used, ILS and direct prediction of $A_i(z)$ attain the same oracle first-order and conservative variances. However, by using the $K_N$ covariates only for prediction, the proposed estimator retains a fixed-dimensional final adjustment and avoids the growing-dimensional problems of ILS as in \cite{LeiandDing2020}. Under favorable leverage, \citet{chiang2025regressionadjustmentcompletelyrandomized} establish asymptotic normality for their cross-fitted regression adjustment estimator. Our estimator can use larger first-stage covariate sets whenever the required prediction-stability condition holds, while only their low-dimensional predictions enter the final adjustment. In addition, our estimator can use LASSO, which accommodates more first-stage covariates, including $K_N\geq N$. The $F$-type estimator in \citet{jiang2024improving} also includes $X_i$ in the final adjustment, so its final-stage dimension grows with the number of covariates.
\end{remark}

In the noncompliance case with $D_i(z)\in\{0,1\}$, we can augment $M_i$ with the covariates used by ILS and a conditionally cross-fitted analogue of the COB estimator of \citet{Ren02042024}. ILS uses $X_i$, while the COB comparison uses predictions of the potential outcomes. For $i\in S^{[q]}$, define $\hat W_{i,\mathrm{cf}}=\left(\hat\mu_{Y,i}^{[-q]}(1),\hat\mu_{Y,i}^{[-q]}(0),\hat\mu_{D,i}^{[-q]}(1),\hat\mu_{D,i}^{[-q]}(0)\right)'$ using working models fitted on $S^{[-q]}$. Let $\hat\tau_{\mathrm{ils}}=\hat\tau_X$ and $\hat\tau_{\mathrm{cob,cf}}=\hat\tau_{\hat W_{\mathrm{cf}}}$. Under the regularity conditions below, they are consistent and have asymptotic variances $V_{\mathrm{ils}}$ and $V_{\mathrm{cob,cf}}$ no larger than $V$. Define the augmented covariates
\[
G_i^+=(X_i',W_i',M_i')',
\]
and define $\hat\tau_{2s,\mathrm{extra}}$ and its asymptotic variance $V_{2s,\mathrm{extra}}$ by replacing $\hat M_{i,\mathrm{cf}}$ with the cross-fitted counterpart of $G_i^+$.

\begin{corollary}
\label{cor:ils_cob_two_step}
\leavevmode\par
Suppose $D_i(z)\in\{0,1\}$ for every $i$ and $z$,
Assumptions~\ref{ass:id}, \ref{ass:asymp}, and
\ref{ass:conditional_crossfit}(i)--(ii) hold, and
Assumption~\ref{ass:conditional_crossfit}(iii) holds for $G_i^+$
and its cross-fitted counterpart. Suppose
Assumption~\ref{ass:two_step_asymptotics} holds with $M_i$ replaced
by $X_i$, $W_i$, and $G_i^+$. Then
\[
\hat\tau_{2s,\mathrm{extra}}-\tau\xrightarrow{p}0,
\qquad
\sqrt N(\hat\tau_{2s,\mathrm{extra}}-\tau)
\xrightarrow{d}
\mathcal N(0,V_{2s,\mathrm{extra}}).
\]
Moreover,
\[
V_{2s,\mathrm{extra}}\le V_{\mathrm{ils}}\le V,
\qquad
V_{2s,\mathrm{extra}}\le V_{\mathrm{cob,cf}}\le V.
\]
\end{corollary}

\begin{proof}
See Appendix~\ref{app:proof_cor_ils_cob_two_step}.
\end{proof}

Corollary~\ref{cor:ils_cob_two_step} provides a way around the lack of a general efficiency ranking between ILS and COB noted by \citet{Ren02042024}: stacking their adjustment covariates yields an estimator no less efficient than either, while adding $M_i$ can provide further gains by directly targeting the transformed outcome. This ordering applies to the augmented estimator; the two-step estimator based only on $M_i$ need not dominate ILS or conditionally cross-fitted COB. The corollary requires $G_i^+$ to be fixed-dimensional with a positive-definite limiting covariance matrix.

\subsection{Variance Estimator}
\label{subsec:variance_estimator}

The previous results establish asymptotic normality and efficiency orderings for the point estimators. For statistical inference, we construct conservative variance estimators by dropping the finite-population variance of adjusted unit-level effects. 

\medskip
\noindent\textbf{Conventional variance estimator}
\leavevmode\par
Let $\hat A_i=Y_i-\hat\tau D_i$. For $z\in\{0,1\}$, define
\begin{align*}
\hat S^2(z)=\frac{1}{N_z-1}\sum_{i:Z_i=z}(\hat A_i-\bar{\hat A}_z)^2,
\end{align*}
where $\bar{\hat A}_z=\frac{1}{N_z}\sum_{i:Z_i=z}\hat A_i.$

The conservative ratio variance estimator is
\[
\hat V=\frac{N}{\hat\tau_D^2}\left\{\frac{\hat S^2(1)}{N_1}+\frac{\hat S^2(0)}{N_0}\right\}.
\]

\medskip
\noindent\textbf{Two-step variance estimator}
\leavevmode\par
For $z\in\{0,1\}$ and $i$ with $Z_i=z$, define
\[
\hat A_{2s,i}=\hat A_i-\frac{\hat\tau_{\hat A,\mathrm{cf}}}{\hat\tau_{D,\mathrm{cf}}}D_i-(\hat M_{i,\mathrm{cf}}-\bar{\hat M}_{z,\mathrm{cf}})'\left\{\hat\gamma_{\hat A,\mathrm{cf}}(z)-\frac{\hat\tau_{\hat A,\mathrm{cf}}}{\hat\tau_{D,\mathrm{cf}}}\hat\gamma_{D,\mathrm{cf}}(z)\right\},
\]
and
\begin{align*}
\hat S^2_{2s}(z)=\frac{1}{N_z-1}\sum_{i:Z_i=z}(\hat A_{2s,i}-\bar{\hat A}_{2s,z})^2,
\end{align*}
where $\bar{\hat A}_{2s,z}=\frac{1}{N_z}\sum_{i:Z_i=z}\hat A_{2s,i}.$

The conservative two-step variance estimator is
\[
\hat V_{2s}=\frac{N}{\hat\tau_{D,\mathrm{cf}}^2}\left\{\frac{\hat S^2_{2s}(1)}{N_1}+\frac{\hat S^2_{2s}(0)}{N_0}\right\}.
\]

The estimators $\hat V_{\mathrm{ils}}$, $\hat V_{\mathrm{cob,cf}}$, and $\hat V_{2s,\mathrm{extra}}$ are defined analogously using the adjustment vectors $X_i$, $\hat W_{i,\mathrm{cf}}$, and $(X_i',\hat W_{i,\mathrm{cf}}',\hat M_{i,\mathrm{cf}}')'$, respectively, together with the corresponding adjusted contrasts and coefficients. Let $V_{\mathrm{cons}}$, $V_{\mathrm{cons,ils}}$, $V_{\mathrm{cons,cob,cf}}$, $V_{\mathrm{cons,2s}}$, and $V_{\mathrm{cons,2s,extra}}$ denote the corresponding conservative asymptotic variance limits. The following theorem establishes consistency and the same ordering for the conservative variance limits.

\begin{theorem}
\label{thm:variance_estimator}
\leavevmode\par
Suppose Assumptions~\ref{ass:id} and \ref{ass:asymp} hold. For each claim, additionally impose Assumption~\ref{ass:two_step_asymptotics} with $M_i$ replaced by $X_i$ for ILS, the assumptions of Theorem~\ref{thm:two_step_cf} for two-step, and those of Corollary~\ref{cor:ils_cob_two_step} for COB and the augmented estimator, as applicable.
\begin{itemize}
\item[(i)] $\hat V$, $\hat V_{\mathrm{ils}}$, $\hat V_{\mathrm{cob,cf}}$, $\hat V_{2s}$, and $\hat V_{2s,\mathrm{extra}}$ are consistent for $V_{\mathrm{cons}}$, $V_{\mathrm{cons,ils}}$, $V_{\mathrm{cons,cob,cf}}$, $V_{\mathrm{cons,2s}}$, and $V_{\mathrm{cons,2s,extra}}$, respectively.

\item[(ii)]
\[
V_{\mathrm{cons,2s,extra}}\le V_{\mathrm{cons,2s}}\le V_{\mathrm{cons}}.
\]

\item[(iii)]
\[
V_{\mathrm{cons,2s,extra}}\le V_{\mathrm{cons,ils}}\le V_{\mathrm{cons}},
\qquad
V_{\mathrm{cons,2s,extra}}\le V_{\mathrm{cons,cob,cf}}\le V_{\mathrm{cons}}.
\]
\end{itemize}
\end{theorem}

\begin{proof}
See Appendix~\ref{app:proof_variance_estimator}.
\end{proof}

Theorem~\ref{thm:variance_estimator} establishes consistency for the conservative variance limits, which may exceed the actual asymptotic variances. The rankings follow from the same oracle projection argument as in Theorem~\ref{thm:late_oracle_variance_min}. In particular, Theorem~\ref{thm:late_oracle_variance_min}(ii) implies that adding the $A$-based generated covariates weakly reduces both the oracle first-order variance and the residual variation in adjusted transformed effects. For the binary-treatment setting, these results support $\hat\tau_{2s,\mathrm{extra}}$ when the stacked adjustment satisfies the conditions of Corollary~\ref{cor:ils_cob_two_step}.

\section{High-dimensional Covariates}
\label{sec:lasso}

LASSO is often used for adjustment estimators with many or high-dimensional covariates \citep{Bloniarz2016,wager2016high,zhu2025design,jiang2024improving}. We provide sufficient conditions for prediction stability using LASSO in the finite-population framework. We initially treat $\lambda$ as fixed before randomization and discuss its selection by cross-validation below.

We allow the first-stage covariates $X_i$ to have dimension $p=p_N$ that may increase with $N$. Define $\tilde X_i=X_i-\bar X$ and write
$
\Sigma_{X,N}=\frac{1}{N}\sum_{i=1}^N\tilde X_i\tilde X_i'.
$
For $z\in\{0,1\}$ and $\lambda>0$, we define the LASSO prediction as
\begin{align*}
\beta_A(z;\lambda)&\in\arg\min_{b\in\mathbb R^{p_{N}}}\left[\frac{1}{N}\sum_{i=1}^N\{A_i(z)-\bar A(z)-\tilde X_i'b\}^2+\lambda\|b\|_1\right],\\
\mu_{A,i}(z;\lambda)&=\bar A(z)+\tilde X_i'\beta_A(z;\lambda).
\end{align*}
The coefficient vector need not be unique when $p_N\ge N$, but the fitted-value vector is unique because the squared loss is strictly convex in fitted values and the penalty is convex.

\begin{assumption}
\label{ass:lasso_persistence}
\leavevmode\par
For deterministic sequences $K_{A,N},K_{X,N},\lambda_N>0$:
\begin{itemize}
\item[(i)]
\begin{align*}
\max_{z\in\{0,1\}}\max_{1\le i\le N}|A_i(z)|&\le K_{A,N},\\
\max_{1\le i\le N}\max_{1\le j\le p_N}|X_{ij}-\bar X_j|&\le K_{X,N}.
\end{align*}
\item[(ii)]
\[
\frac{K_{X,N}}{\lambda_N}\left\{K_{A,N}\sqrt{\frac{\log(p_N+1)}{N}}+\frac{1}{\sqrt N}\right\}\to0.
\]
\end{itemize}
\end{assumption}

Assumption \ref{ass:lasso_persistence} (i) is a regularity condition and (ii) is the key stability condition: the penalty must dominate both cross-moment sampling error and preliminary-ratio error. This stabilizes predictions without requiring coefficient consistency or support recovery.

\begin{remark}
\leavevmode\par
Assumption~\ref{ass:lasso_persistence}(i) is only a sufficient condition for controlling the outcome--covariate cross moments. The proof requires, up to the preliminary-ratio error,
\[
\max_{q,z}
\left\|
\frac{1}{N_z^{[-q]}}
\sum_{i\in S^{[-q]}:Z_i=z}
\tilde X_i A_i(z)
-
\frac{1}{N}\sum_{i=1}^N
\tilde X_i A_i(z)
\right\|_\infty
=
o_p(\lambda_N).
\]
Thus, the outcome bound in part (i) can be replaced by suitable uniform sub-exponential conditions on $\tilde X_{ij}A_i(z)$, allowing unbounded potential outcomes with a corresponding choice of $\lambda_N$.
\end{remark}

Let $N_z^{[-q]}=\sum_{i\in S^{[-q]}}\mathbf 1\{Z_i=z\}$, and let $\bar X_z^{[-q]}$ and $\bar{\hat A}_z^{[-q]}$ denote the corresponding sample means. The conventional armwise LASSO minimizes, over $b\in\mathbb R^{p_N}$,
\[
\frac{1}{N_z^{[-q]}}\sum_{i\in S^{[-q]}:Z_i=z}\left\{\hat A_i^{[-q]}-\bar{\hat A}_z^{[-q]}-(X_i-\bar X_z^{[-q]})'b\right\}^2+\lambda\|b\|_1.
\]
This criterion uses an arm-specific sample Gram matrix. Because the full covariate matrix is fixed and observed under the finite population, we instead center $X_i$ at $\bar X$, use $\Sigma_{X,N}$ for the quadratic term, and estimate the outcome--covariate cross moment from the assignment arm. Hence, the resulting LASSO is
\begin{align*}
\hat\beta_A^{[-q]}(z;\lambda)&\in\arg\min_{b\in\mathbb R^{p_N}}\left[b'\Sigma_{X,N}b-\frac{2}{N_z^{[-q]}}\sum_{i\in S^{[-q]}:Z_i=z}\hat A_i^{[-q]}\tilde X_i'b+\lambda\|b\|_1\right],\\
\hat\mu_{A,i}^{[-q]}(z;\lambda)&=\bar{\hat A}_z^{[-q]}+\tilde X_i'\hat\beta_A^{[-q]}(z;\lambda),\qquad i=1,\ldots,N.
\end{align*}
Write $\hat\mu_A^{[-q]}(z;\lambda)$ and $\mu_A(z;\lambda)$ for the corresponding vectors of $N$ predictions. The fit uses outcomes and assignments only from $S^{[-q]}$; the covariates used in $\Sigma_{X,N}$ are fixed and observed for all units. The common Gram matrix leaves only the cross moment to be estimated, which is key to the prediction bound.

\begin{theorem}
\label{thm:lasso_persistence}
\leavevmode\par
Suppose Assumptions~\ref{ass:id}, \ref{ass:asymp}(i)--(ii), and \ref{ass:conditional_crossfit}(i)--(ii) hold, the conditional split is generated by Algorithm~\ref{alg:cf_late}, and Assumption~\ref{ass:lasso_persistence} holds. Then, for each $z\in\{0,1\}$,
\[
\max_{1\le q\le Q}\|\hat\mu_A^{[-q]}(z;\lambda_N)-\mu_A(z;\lambda_N)\|_N^2\xrightarrow{p}0.
\]
Assumption~\ref{ass:conditional_crossfit}(iii) holds with $\hat\mu_A^{[-q]}(z)=\hat\mu_A^{[-q]}(z;\lambda_N)$ and $M_i=\bigl(\mu_{A,i}(1;\lambda_N),\mu_{A,i}(0;\lambda_N)\bigr)'$.
\end{theorem}

\begin{proof}
See Appendix~\ref{app:proof_lasso_persistence}.
\end{proof}

\begin{remark}
\label{rem:lasso_adjustment}
\leavevmode\par
Theorem~\ref{thm:lasso_persistence} establishes prediction stability without sparsity, restricted eigenvalue conditions, or support recovery in the canonical LASSO. Existing design-based LASSO adjustments allow $p_N>N$ under sparsity conditions \citep{Bloniarz2016,zhu2025design}, whose strength has been noted in subsequent work \citep{wager2016high,LeiandDing2020,lu2025debiased}. Sparsity-free design-based results instead restrict dimensionality: under favorable leverage conditions, \citet{LeiandDing2020} allow $p_N=o\{N^{2/3}/(\log N)^{1/3}\}$, \citet{chiang2025regressionadjustmentcompletelyrandomized} allow $p_N=o\{N^{3/4}/(\log N)^{1/2}\}$, and \citet{lu2025debiased} allow $p_N\asymp N$. Our method keeps the final adjustment fixed-dimensional and regularizes only the prediction stage. If $K_{A,N}$ and $K_{X,N}$ are bounded, Assumption~\ref{ass:lasso_persistence}(ii) is satisfied by
\[
\lambda_N
=
c_N\sqrt{\frac{\log(p_N+1)}{N}},
\qquad
c_N\to\infty,
\]
whenever $\log p_N=o(N).$ Hence $p_N=\exp\left(o(N)\right)$, including $p_N\gg N$, is permitted without sparsity.
\end{remark}

Next, we consider whether choosing the penalty by ordinary cross-validation preserves prediction stability. 
Cross-validation is widely used to tune LASSO in empirical applications, including regression adjustment in randomized experiments, while \citet{LeiandDing2020} note that theoretical guarantees for the resulting data-dependent penalty have remained unclear; we therefore establish prediction stability for the ordinary cross-validation procedure. \footnote{\citet{Bloniarz2016,zhu2025design} heuristically use cross-validation in their implementations and show efficiency gains. Here we verify prediction stability for the penalty selected by the procedure specified below.} Our focus on prediction risk follows the persistence framework for cross-validated LASSO studied by \citet{homrighausen2017risk}. Training and refitting continue to use $\Sigma_{X,N}$.

Fix an integer $K_{\mathrm{cv}}\ge2$ that does not depend on $N$. For each $q\in\{1,\ldots,Q\}$ and $z\in\{0,1\}$, conditional on $\{i\in S^{[-q]}:Z_i=z\}$, draw a partition into disjoint validation folds $V_{z,1}^{[-q]},\ldots,V_{z,K_{\mathrm{cv}}}^{[-q]}$ uniformly over all partitions with fixed fold sizes $N_{z,1}^{[-q]},\ldots,N_{z,K_{\mathrm{cv}}}^{[-q]}$. We use auxiliary randomness independent of the treatment assignment and the outer split. Then we have $\{i\in S^{[-q]}:Z_i=z\}=\bigcup_{k=1}^{K_{\mathrm{cv}}}V_{z,k}^{[-q]}$ and $\sum_{k=1}^{K_{\mathrm{cv}}}N_{z,k}^{[-q]}=N_z^{[-q]}.$

Let $U_{z,k}^{[-q]}
=
\{i\in S^{[-q]}:Z_i=z\}\setminus V_{z,k}^{[-q]}.$ For $\lambda>0$, define the inner training fit by
\begin{align*}
\hat\beta_A^{[-q,-k]}(z;\lambda)
&\in
\arg\min_{b\in\mathbb R^{p_N}}
\left[
b'\Sigma_{X,N}b
-\frac{2}{|U_{z,k}^{[-q]}|}
\sum_{i\in U_{z,k}^{[-q]}}
\hat A_i^{[-q]}\tilde X_i'b
+\lambda\|b\|_1
\right],
\\
\hat\mu_{A,i}^{[-q,-k]}(z;\lambda)
&=
\frac{1}{|U_{z,k}^{[-q]}|}
\sum_{j\in U_{z,k}^{[-q]}}
\hat A_j^{[-q]}
+
\tilde X_i'\hat\beta_A^{[-q,-k]}(z;\lambda),
\qquad i=1,\ldots,N.
\end{align*}
The same outer-training estimate $\hat\tau^{[-q]}$ is used to construct $\hat A_i^{[-q]}$ for all inner folds. Define the cross-validation criterion by evaluating, on each validation fold, the predictions fitted using the remaining observations in the outer training sample:
\[
\widehat{\operatorname{CV}}_{z}^{[-q]}(\lambda)
=
\frac{1}{N_z^{[-q]}}
\sum_{k=1}^{K_{\mathrm{cv}}}
\sum_{i\in V_{z,k}^{[-q]}}
\left\{
\hat A_i^{[-q]}
-\hat\mu_{A,i}^{[-q,-k]}(z;\lambda)
\right\}^2.
\]

\begin{assumption}
\label{ass:lasso_cv_grid}
\leavevmode\par
\begin{itemize}
\item[(i)]
For some constant $c_{\mathrm{cv}}>0$ and all sufficiently large $N$,
\[
\min_{\substack{1\le q\le Q\\z\in\{0,1\}\\1\le k\le K_{\mathrm{cv}}}}
\min\left\{
N_{z,k}^{[-q]},
N_z^{[-q]}-N_{z,k}^{[-q]}
\right\}
\ge c_{\mathrm{cv}}N.
\]
\item[(ii)]
Let $\Lambda_N\subset[\lambda_N,\infty)$ be a deterministic nonempty
finite penalty grid such that $\lambda_N\in\Lambda_N$.
\end{itemize}
\end{assumption}

Assumption~\ref{ass:lasso_cv_grid}(i) is a nondegeneracy condition requiring both inner validation and training samples to contain a positive fraction of the finite population uniformly over folds and arms. With fixed $K_{\mathrm{cv}}$, approximately equal inner folds satisfy this condition whenever the outer training samples are nondegenerate. Assumption~\ref{ass:lasso_cv_grid}(ii) links the cross-validation grid to Assumption~\ref{ass:lasso_persistence}(ii) by requiring all candidate penalties to satisfy $\lambda\ge\lambda_N$ and $\lambda_N$ itself to be included in the grid. No restriction is imposed on the number of grid points.

\begin{theorem}
\label{thm:lasso_cv}
\leavevmode\par
Suppose the conditions of Theorem~\ref{thm:lasso_persistence} and Assumption~\ref{ass:lasso_cv_grid} hold, and construct the inner folds as specified above. For each $q\in\{1,\ldots,Q\}$ and $z\in\{0,1\}$, let $\hat\lambda_{z,N}^{[-q]}\in\Lambda_N$ satisfy
\[
\widehat{\operatorname{CV}}_{z}^{[-q]}(\hat\lambda_{z,N}^{[-q]})
\le
\min_{\lambda\in\Lambda_N}
\widehat{\operatorname{CV}}_{z}^{[-q]}(\lambda)
+\xi_{z,N}^{[-q]},
\]
where $\xi_{z,N}^{[-q]}\ge0$ and
\[
\max_{\substack{1\le q\le Q\\z\in\{0,1\}}}
\xi_{z,N}^{[-q]}
=
o_p(1).
\]
Then, for each $z\in\{0,1\}$,
\[
\max_{1\le q\le Q}
\|\hat\mu_A^{[-q]}(z;\hat\lambda_{z,N}^{[-q]})
-\mu_A(z;\lambda_N)\|_N^2
\xrightarrow{p}0.
\]
Consequently, Assumption~\ref{ass:conditional_crossfit}(iii) holds with
\[
\hat\mu_A^{[-q]}(z)
=
\hat\mu_A^{[-q]}(z;\hat\lambda_{z,N}^{[-q]}),
\qquad
M_i
=
\bigl(
\mu_{A,i}(1;\lambda_N),
\mu_{A,i}(0;\lambda_N)
\bigr)'.
\]
\end{theorem}

\begin{proof}
See Appendix~\ref{app:proof_lasso_cv}.
\end{proof}

Theorem~\ref{thm:lasso_cv} gives convergence to the deterministic predictions at $\lambda_N$, without requiring convergence of the selected penalties or coefficients. The grid's lower bound keeps all candidate penalties within the stability range. These results establish prediction stability; two-step inference additionally requires Assumption~\ref{ass:two_step_asymptotics}, with the growing-dimensional covariates entering the final adjustment through their two predictions.

\section{Simulation}
\label{sec:simulation}

We evaluate the finite-sample performance of the two-step estimator in a randomized experiment with noncompliance in which baseline covariates predict compliance, potential outcomes, and treatment-effect heterogeneity. We set $N=500$ and $p=8$. Draw four mutually independent vectors $\beta_D,\beta_Y,\gamma_D,\gamma_Y\in\mathbb R^p$ by independently drawing their coordinates from the standard normal distribution and normalizing each vector to have unit Euclidean norm. These parameters are drawn once and held fixed. 

Independently across units and coordinates, draw $X_{i,j}$ from a $t_3$ distribution truncated at $[-4,4]$, and center and scale each coordinate within the finite population. 
For $R\in\{D,Y\}$, define linear and nonlinear terms as 
\begin{align*}
L_{R,i}&=X_i'\beta_R,\\ Q_{R,i}&=\sum_{j=1}^p\gamma_{R,j}g(X_{i,j}),\qquad \text{where }g(x)=\sin(1.4x)+\cos(1.4x).
\end{align*}
Independently draw $\nu_i\sim N(0,10^2)$ and set \[ s_i=6L_{D,i}+2Q_{D,i}+\nu_i. \] Let $c_1$ and $c_0$ be the empirical $0.225$ and $0.775$ quantiles of $\{s_i\}_{i=1}^N$, respectively, and define 
\[ D_i(z)=\mathbf 1\{s_i>c_z\},\qquad z\in\{0,1\}. \]
Thus, the compliance share is approximately 55\%.
Similarly, we independently draw $\varepsilon_i\sim N(0,0.35^2)$, and under exclusion, define \begin{align*} Y_i(z)&=L_{Y,i}+Q_{Y,i}+D_i(z)\left\{10+1.7Q_{D,i}+\varepsilon_i\right\},\qquad z\in\{0,1\}. \end{align*}

We generate the finite population once and hold it fixed across $5,000$ complete randomizations with $N_1=N_0=250$. Prediction-based estimators use five-fold conditional cross-fitting, with each assignment arm split separately, except for the F and R constructions described below. For the LASSO learners tuned by cross-validation, the penalty parameter is selected by an additional five-fold cross-validation within each arm-specific outer training sample to minimize held-out mean squared prediction error. Transformed-outcome prediction uses $\hat A_i^{[-q]}=Y_i-\hat\tau^{[-q]}D_i$, with the Wald pilot $\hat\tau^{[-q]}$ computed only from the corresponding outer training sample. We compare Wald, ILS + COB, polynomial ILS, COB, two-step, polynomial ILS + COB, polynomial ILS + two-step, two-step + COB, and two-step extra. Their adjustment spaces contain, respectively, no covariates; $X_i$ together with the four COB predictions; the full polynomial covariates; the four cross-fitted predictions of $Y_i(z)$ and $D_i(z)$; the two cross-fitted predictions of $A_i(z)$; the full polynomial covariates together with the four COB predictions; the full polynomial covariates together with the two transformed-outcome predictions; all six predictions; and $X_i$ together with all six prediction scores. Thus, two-step extra nests the final-stage adjustment spaces of ILS + COB, COB, two-step, and two-step + COB.

For each $d\in\{2,3\}$, prediction uses the polynomials with total degree at most $d$, excluding the constant. Its dimension equals $44$ and $164$ for $d=2$ and $3$, respectively. For $d=2$, we use least squares for outcome and transformed-outcome prediction and ordinary logistic regression for potential treatment prediction, as in \cite{Ren02042024}. For $d=3$, we use LASSO for outcome and transformed-outcome prediction and logistic LASSO for $D_i(z)$ prediction. Polynomial ILS and the other polynomial specifications are reported only for $d=2$. We additionally report F estimator at $d=2$, using the polynomial basis together with two full-arm logistic predictions of $D_i(z)$, and the R augmented-ratio estimator at $d=3$, using separate full-arm LASSO fits for $Y_i(z)$ and $D_i(z)$ following \cite{jiang2024improving}.

\begin{table}[p]
  \caption{Simulation results of estimators}
  \centering
  \label{tab:simulation}
  \scalebox{0.83}{
  \begin{tabular}{lrrrrrrr}
\toprule
Method & Median Bias & Oracle SD & SD & SE & True Cons & Coverage & CI Length \\
\midrule
Wald & 0.007 & 0.345 & 0.344 & 0.358 & 0.091 & 0.962 & 1.405 \\
\midrule
\multicolumn{8}{l}{Panel A: $d=2$, Polynomial} \\
COB & 0.001 & 0.189 & 0.248 & 0.263 & 0.080 & 0.963 & 1.031 \\
Polynomial ILS & 0.006 & 0.168 & 0.201 & 0.187 & 0.071 & 0.928 & 0.732 \\
Polynomial ILS + COB & 0.007 & 0.162 & 0.207 & 0.172 & 0.071 & 0.893 & 0.673 \\
Polynomial ILS + Two-step & 0.017 & 0.168 & 0.219 & 0.144 & 0.071 & 0.807 & 0.564 \\
F & -0.026 & 0.162 & 0.203 & 0.187 & 0.071 & 0.926 & 0.732 \\
ILS + COB & 0.001 & 0.184 & 0.216 & 0.229 & 0.077 & 0.959 & 0.898 \\
Two-step & -0.002 & 0.168 & 0.200 & 0.217 & 0.071 & 0.969 & 0.851 \\
Two-step + COB & -0.002 & 0.162 & 0.198 & 0.214 & 0.071 & 0.965 & 0.840 \\
Two-step extra & 0.002 & 0.162 & 0.196 & 0.210 & 0.071 & 0.961 & 0.821 \\
\midrule
\multicolumn{8}{l}{Panel B: $d=3$, LASSO} \\
COB & 0.004 & 0.132 & 0.235 & 0.250 & 0.073 & 0.964 & 0.980 \\
R  & -0.018 & 0.230 & 0.255 & 0.261 & 0.098 & 0.955 & 1.022 \\
ILS + COB & 0.003 & 0.130 & 0.214 & 0.226 & 0.071 & 0.963 & 0.888 \\
Two-step & -0.002 & 0.111 & 0.151 & 0.172 & 0.063 & 0.976 & 0.675 \\
Two-step + COB & -0.002 & 0.109 & 0.150 & 0.170 & 0.063 & 0.975 & 0.668 \\
Two-step extra & -0.004 & 0.107 & 0.150 & 0.168 & 0.062 & 0.974 & 0.658 \\
\bottomrule
\end{tabular}
  }
  \caption*{\footnotesize Notes: Results are based on 5,000 complete randomizations of one fixed finite population with $N=500$ under balanced assignment. Panel A uses degree-two polynomials, whereas Panel B uses LASSO; logistic models are used for potential treatment $D_i(z)$. Wald does not depend on $d$ and is reported in the first row. Oracle SD is the finite-population oracle standard deviation, SD is the Monte Carlo standard deviation, and SE is the average estimated conservative standard error. True Cons is the true conservativeness term in our DGP. Coverage and CI Length refer to nominal 95\% confidence intervals. F and R use the same fixed-finite-population variance convention as the other methods.}
\end{table}

Table~\ref{tab:simulation} reports results from 5,000 complete randomizations of one fixed finite population with $N=500$. Panel A considers the degree-two polynomial design. All estimators have small median bias relative to SD. Our result shows the limitation of Polynomial ILS, which uses many covariates into $G_i$. Its mean SE is $0.187$, but coverage is only $0.928$. This undercoverage is consistent with the many-regressor problems studied by \citet{LeiandDing2020,chiang2025regressionadjustmentcompletelyrandomized}. Adding COB or two-step predictions to Polynomial ILS exacerbates the problem. The mean SE falls to $0.172$ and $0.144$, respectively, while coverage deteriorates to $0.893$ and $0.807$. F also undercovers, with a mean SE of $0.187$ and coverage of $0.926$. These smaller reported SEs therefore do not represent reliable efficiency gains. The results highlight the importance of separating prediction from adjustment: a rich polynomial basis can be used to construct a small number of predictions without including the full basis in the final adjustment regression.

The methods that retain a low-dimensional final adjustment such as COB, ILS + COB, two-step, two-step + COB, and two-step extra, maintain nominal coverage. Two-step reduces the mean SE by $39.4\%$ and $17.5\%$ relative to Wald and COB, respectively. Similarly, two-step extra reduces the mean SE by $41.3\%$ and $20.2\%$ relative to the same benchmarks. Thus, two-step extra provides the greatest estimated precision among the estimators that maintain nominal coverage.

Panel B considers $d=3$. Because adjustment using the full polynomial basis becomes unstable in this design, LASSO is instead used for first-stage prediction while the final adjustment remains fixed-dimensional. Two-step extra has the smallest Oracle SD and Monte Carlo SD among the LASSO-based estimators, with coverage of $0.974$. Two-step reduces the mean SE by $52.0\%$, $31.2\%$, and $34.1\%$ relative to Wald, COB, and R, respectively. Similarly, two-step extra reduces the mean SE by $53.1\%$, $32.8\%$, and $35.6\%$ relative to the same benchmarks. Thus, two-step extra provides the greatest estimated precision among the estimators considered in this high-dimensional design while maintaining nominal coverage. Along the nested sequence from COB to two-step + COB and then to two-step extra, both Oracle SD and True Cons decrease. The simultaneous reductions in Oracle SD and True Cons illustrate the two implications of Theorem~\ref{thm:late_oracle_variance_min}: enlarging the adjustment space improves first-order efficiency and reduces the conservativeness of design-based inference.

\section{Application}
\label{sec:application}

Following \cite{jiang2024improving}, we revisit the result of \cite{dupas2018banking}, which examines whether savings-account access improves financial outcomes in Uganda, Malawi, and Chile. In Uganda, 2,160 households were randomized within 41 strata defined by gender, occupation, and bank branch, with half assigned vouchers covering account-opening costs in each stratum. Among assigned households, 41.87\% opened an account and made at least one deposit within two years. Let $Z_i$ denote voucher assignment and $D_i$ the measure of active use. We estimate the LATE for voucher-induced users on the eight first-follow-up savings-stock outcomes examined by \cite{jiang2024improving}, measured in 2010 US dollars.

Conditional on households, potential outcomes, covariates, and strata, the design is a CRE within each stratum with fixed assignment counts. We therefore apply the stratified extension in Appendix~\ref{app:stratification}, allowing adjustment coefficients to vary across strata. The LATE interpretation requires exclusion, no interference, monotonicity, and a nonzero aggregate first stage; reduced account-opening costs make monotonicity plausible.

We use the data and benchmark code from \texttt{drcarlate}, version 1.2.0 \cite{jiang2023drcarlate}, reproducing its unadjusted estimator (NA) and F results and applying our estimator to the same observations and primitive covariates. Excluding missing outcomes and strata with fewer than ten observations leaves 1,952--1,975 households in 32 strata. The three covariates are baseline savings, baseline total income, and a missing-value indicator. Baseline savings corresponds to each outcome, except that baseline total savings is used for formal-institution savings, mobile money, and total formal savings. First-stage dictionaries contain all monomials of total degree at most three, excluding redundant terms, interacted with stratum indicators. We fit LASSO for $Y_i,A_i$ and $D_i$, with stratum intercepts and cross-validated penalties over the simulation's fixed grid. The final vector $G^+=(X,\widehat W,\widehat M)$ contains three primitive covariates and six generated predictions. Within each split, final adjustment pools the out-of-fold predictions across both folds, with separate slopes by assignment arm and stratum. We average estimates over ten two-fold splits constructed within assignment-arm-by-stratum cells and report the square root of the average within-split variance.

\begin{table}[p]
  \caption{Application results in \cite{dupas2018banking}}
  \centering
  \label{tab:application}
  \scalebox{1}{
  \begin{tabular}{lcccc}
\toprule
$\text{Outcome}$ & $n$ & NA & F & Two-step Extra \\
\midrule
Formal fin. inst.
    & 1,968 & 20.558 & 22.737 & 22.024 \\[-0.35em]
    &       & (3.067) & (2.951) & (2.878) \\[0.55em]
Mobile money
    & 1,972 & -0.208 & -0.310 & -0.267 \\[-0.35em]
    &       & (0.223) & (0.209) & (0.208) \\[0.55em]
Total formal
    & 1,966 & 20.399 & 22.325 & 21.657 \\[-0.35em]
    &       & (3.089) & (2.964) & (2.870) \\[0.55em]
Cash at home
    & 1,971 & -10.826 & -8.691 & -9.212 \\[-0.35em]
    &       & (5.003) & (4.403) & (4.262) \\[0.55em]
ROSCA/VSLA
    & 1,975 & -1.933 & -1.551 & -1.771 \\[-0.35em]
    &       & (1.971) & (1.808) & (1.725) \\[0.55em]
Friends/family
    & 1,974 & -3.621 & -1.427 & -2.536 \\[-0.35em]
    &       & (2.040) & (1.869) & (1.866) \\[0.55em]
Total informal
    & 1,960 & -17.643 & -15.512 & -15.158 \\[-0.35em]
    &       & (6.200) & (5.113) & (5.073) \\[0.55em]
Total savings
    & 1,952 & 2.787 & 8.481 & 6.798 \\[-0.35em]
    &       & (7.290) & (6.139) & (6.083) \\
\bottomrule
\end{tabular}
  }
  \caption*{\footnotesize Notes: The table reports LATE estimates of opening a bank account on saving stocks. Standard errors are in parentheses. NA and F are obtained by executing the empirical-application code and data distributed in the CRAN package \texttt{drcarlate}, version 1.2.0 by \cite{jiang2023drcarlate}. Two-step Extra uses degree-three first-stage polynomial dictionaries and separate final slopes by assignment arm and stratum. It averages estimates and within-split variances over ten repeated two-fold sample splits.}
\end{table}

Table~\ref{tab:application} compares the unadjusted estimator (NA), the further-efficiency-improvement estimator (F) of \cite{jiang2024improving}, and Two-step Extra. First, Two-step Extra has smaller reported standard errors than NA for all eight outcomes. Second, the precision gains extend to the F benchmark, which \cite{jiang2024improving} recommend for parametric regression adjustment. Relative to F, the standard errors decrease by 0.14--4.59\% and the largest reduction is for ROSCA/VSLA savings, from 1.808 to 1.725. Finally, the point estimate remains similar: account use increases formal savings and reduces informal savings, including cash held at home, while the effect on total savings remains statistically insignificant. For cash at home, the estimate changes from $-8.691$ to $-9.212$ and the standard error falls from 4.403 to 4.262.

\clearpage

\begin{minilinespace}
\bibliography{design}

@article{ye2023toward,
  title={Toward better practice of covariate adjustment in analyzing randomized clinical trials},
  author={Ye, Ting and Shao, Jun and Yi, Yanyao and Zhao, Qingyuan},
  journal={Journal of the American Statistical Association},
  volume={118},
  number={544},
  pages={2370--2382},
  year={2023},
  publisher={Taylor \& Francis}
}

@book{drummond2015methods,
  title={Methods for the economic evaluation of health care programmes},
  author={Drummond, Michael F and Sculpher, Mark J and Claxton, Karl and Stoddart, Greg L and Torrance, George W},
  year={2015},
  publisher={Oxford university press}
}

@article{liu2026regression,
  title={Regression methods for cost-effectiveness analysis with different censoring times or terminating events for survival time and costs},
  author={Liu, Dingning and Chen, Shuai},
  journal={Biometrics},
  volume={82},
  number={2},
  pages={ujag073},
  year={2026},
  publisher={Oxford University Press}
}

@Manual{jiang2023drcarlate,
  title  = {drcarlate: Improving Estimation Efficiency in {CAR}
            with Imperfect Compliance},
  author = {Liang Jiang and Oliver B. Linton and Haihan Tang
            and Yichong Zhang},
  year   = {2023},
  note   = {R package version 1.2.0},
  url    = {https://CRAN.R-project.org/package=drcarlate}
}

@article{dupas2018banking,
  title={Banking the unbanked? Evidence from three countries},
  author={Dupas, Pascaline and Karlan, Dean and Robinson, Jonathan and Ubfal, Diego},
  journal={American Economic Journal: Applied Economics},
  volume={10},
  number={2},
  pages={257--297},
  year={2018},
  publisher={American Economic Association 2014 Broadway, Suite 305, Nashville, TN 37203-2425}
}

@article{guo2021machine,
  title={Machine learning for variance reduction in online experiments},
  author={Guo, Yongyi and Coey, Dominic and Konutgan, Mikael and Li, Wenting and Schoener, Chris and Goldman, Matt},
  journal={Advances in Neural Information Processing Systems},
  volume={34},
  pages={8637--8648},
  year={2021}
}

@article{jones2019workplace,

  author  = {Jones, Damon and Molitor, David and Reif, Julian},

  title   = {What Do Workplace Wellness Programs Do? Evidence from the Illinois Workplace Wellness Study},

  journal = {The Quarterly Journal of Economics},

  year    = {2019},

  volume  = {134},

  number  = {4},

  pages   = {1747--1791},

  doi     = {10.1093/qje/qjz023}

}

@article{lu2025tyranny,
  title={Tyranny-of-the-minority regression adjustment in randomized experiments},
  author={Lu, Xin and Liu, Hanzhong},
  journal={Journal of the American Statistical Association},
  volume={120},
  number={550},
  pages={846--858},
  year={2025},
  publisher={Taylor \& Francis}
}

@article{bardenet2015concentration_hoeffding,
  title={Concentration inequalities for sampling without replacement},
  author={Bardenet, R{\'e}mi and Maillard, Odalric-Ambrym},
  year={2015}
}

@article{homrighausen2017risk,
  author  = {Homrighausen, Darren and McDonald, Daniel J.},
  title   = {Risk Consistency of Cross-Validation with {Lasso}-Type Procedures},
  journal = {Statistica Sinica},
  year    = {2017},
  volume  = {27},
  number  = {3},
  pages   = {1017--1036},
  doi     = {10.5705/ss.202015.0355}
}

@article{wager2016high,
  title={High-dimensional regression adjustments in randomized experiments},
  author={Wager, Stefan and Du, Wenfei and Taylor, Jonathan and Tibshirani, Robert J},
  journal={Proceedings of the National Academy of Sciences},
  volume={113},
  number={45},
  pages={12673--12678},
  year={2016},
  publisher={National Academy of Sciences}
}

@book{imbens2015causal,
  title={Causal inference in statistics, social, and biomedical sciences},
  author={Imbens, Guido W and Rubin, Donald B},
  year={2015},
  publisher={Cambridge university press}
}

@article{freedman2008regression_a,
  title={ON REGRESSION ADJUSTMENTS IN EXPERIMENTS WITH SEVERAL TREATMENTS},
  author={Freedman, David A},
  journal={The Annals of applied statistics},
  volume={2},
  number={1},
  pages={176--196},
  year={2008}
}

@article{freedman2008regression_b,
  title={On regression adjustments to experimental data},
  author={Freedman, David A},
  journal={Advances in Applied Mathematics},
  volume={40},
  number={2},
  pages={180--193},
  year={2008},
  publisher={Elsevier}
}

@article{LeiandDing2020,
    author = {Lei, Lihua and Ding, Peng},
    title = {Regression adjustment in completely randomized experiments with a diverging number of covariates},
    journal = {Biometrika},
    volume = {108},
    number = {4},
    pages = {815-828},
    year = {2020},
    month = {12},
    issn = {0006-3444},
    doi = {10.1093/biomet/asaa103},
    url = {https://doi.org/10.1093/biomet/asaa103},
    eprint = {https://academic.oup.com/biomet/article-pdf/108/4/815/41153623/asaa103.pdf},
}

@misc{chiang2025regressionadjustmentcompletelyrandomized,
      title={Regression adjustment in completely randomized experiments with many covariates}, 
      author={Harold D Chiang and Yukitoshi Matsushita and Taisuke Otsu},
      year={2025},
      eprint={2302.00469},
      archivePrefix={arXiv},
      primaryClass={econ.EM},
      url={https://arxiv.org/abs/2302.00469}, 
}

@article{lin2013,
 ISSN = {19326157, 19417330},
 URL = {http://www.jstor.org/stable/23566512},
 author = {Winston Lin},
 journal = {The Annals of Applied Statistics},
 number = {1},
 pages = {295--318},
 publisher = {Institute of Mathematical Statistics},
 title = {AGNOSTIC NOTES ON REGRESSION ADJUSTMENTS TO EXPERIMENTAL DATA: REEXAMINING FREEDMAN'S CRITIQUE},
 urldate = {2026-02-16},
 volume = {7},
 year = {2013}
}

@misc{lu2025,
      title={Conditional cross-fitting for unbiased machine-learning-assisted covariate adjustment in randomized experiments}, 
      author={Xin Lu and Lei Shi and Hanzhong Liu and Peng Ding},
      year={2025},
      eprint={2508.15664},
      archivePrefix={arXiv},
      primaryClass={stat.ME},
      url={https://arxiv.org/abs/2508.15664}, 
}

@article{
Bloniarz2016,
author = {Adam Bloniarz  and Hanzhong Liu  and Cun-Hui Zhang  and Jasjeet S. Sekhon  and Bin Yu },
title = {Lasso adjustments of treatment effect estimates in randomized experiments},
journal = {Proceedings of the National Academy of Sciences},
volume = {113},
number = {27},
pages = {7383-7390},
year = {2016},
doi = {10.1073/pnas.1510506113},
URL = {https://www.pnas.org/doi/abs/10.1073/pnas.1510506113},
eprint = {https://www.pnas.org/doi/pdf/10.1073/pnas.1510506113}}

@article{Guo02012023,
author = {Kevin Guo and Guillaume Basse},
title = {The Generalized Oaxaca-Blinder Estimator},
journal = {Journal of the American Statistical Association},
volume = {118},
number = {541},
pages = {524--536},
year = {2023},
publisher = {Taylor \& Francis},
doi = {10.1080/01621459.2021.1941053},


URL = { 
    
        https://doi.org/10.1080/01621459.2021.1941053
    
    

},
eprint = { 
    
        https://doi.org/10.1080/01621459.2021.1941053
    
    

}

}

@article{cohenfogarty2023,
    author = {Cohen, P L and Fogarty, C B},
    title = {No-harm calibration for generalized Oaxaca–Blinder estimators},
    journal = {Biometrika},
    volume = {111},
    number = {1},
    pages = {331-338},
    year = {2023},
    month = {06},
    issn = {1464-3510},
    doi = {10.1093/biomet/asad036},
    url = {https://doi.org/10.1093/biomet/asad036},
    eprint = {https://academic.oup.com/biomet/article-pdf/111/1/331/56665720/asad036.pdf},
}

@article{ren2024,
  title={Design-Based Covariate Adjustment for Causal Inference with Interference and Noncompliance},
  author={Ren, Jiyang and Liu, Hanzhong and Ma, Yingying},
  journal={Available at SSRN:4783803},
  year={2024}
}

@article{bai2025inference,
  title={Inference in experiments with matched pairs and imperfect compliance},
  author={Bai, Yuehao and Guo, Hongchang and Shaikh, Azeem M and Tabord-Meehan, Max},
  journal={Journal of Business \& Economic Statistics},
  volume={43},
  number={3},
  pages={627--642},
  year={2025},
  publisher={Taylor \& Francis}
}

@article{pashley2025improving,
  title={Improving instrumental variable estimators with poststratification},
  author={Pashley, Nicole E and Keele, Luke and Miratrix, Luke W},
  journal={Journal of the Royal Statistical Society Series A: Statistics in Society},
  volume={188},
  number={3},
  pages={765--790},
  year={2025},
  publisher={Oxford University Press UK}
}

@article{hartman2024improving,
  title={Improving precision through design and analysis in experiments with noncompliance},
  author={Hartman, Erin and Huang, Melody},
  journal={Political Science Research and Methods},
  volume={12},
  number={3},
  pages={557--572},
  year={2024},
  publisher={Cambridge University Press}
}

@article{zhu2025design,
  title={Design-based theory for Lasso adjustment in randomized block experiments and rerandomized experiments},
  author={Zhu, Ke and Liu, Hanzhong and Yang, Yuehan},
  journal={Journal of Business \& Economic Statistics},
  volume={43},
  number={3},
  pages={544--555},
  year={2025},
  publisher={Taylor \& Francis}
}

@article{lu2025debiased,
  title={Debiased regression adjustment in completely randomized experiments with moderately high-dimensional covariates},
  author={Lu, Xin and Yang, Fan and Wang, Yuhao},
  journal={The Annals of Statistics},
  volume={53},
  number={4},
  pages={1535--1558},
  year={2025},
  publisher={Institute of Mathematical Statistics}
}

@article{ansel2018ols,
  title={OLS and 2SLS in randomized and conditionally randomized experiments},
  author={Ansel, Jason and Hong, Han and Jessie Li, and},
  journal={Jahrb{\"u}cher f{\"u}r National{\"o}konomie und Statistik},
  volume={238},
  number={3-4},
  pages={243--293},
  year={2018},
  publisher={De Gruyter}
}

@article{Li02102017,
author = {Xinran Li and Peng Ding},
title = {General Forms of Finite Population Central Limit Theorems with Applications to Causal Inference},
journal = {Journal of the American Statistical Association},
volume = {112},
number = {520},
pages = {1759--1769},
year = {2017},
publisher = {ASA Website},
doi = {10.1080/01621459.2017.1295865},


URL = { 
    
        https://doi.org/10.1080/01621459.2017.1295865
    
    

},
eprint = { 
    
        https://doi.org/10.1080/01621459.2017.1295865
    
    

}

}

@article{ding2019decomposing,
  author = {Ding, Peng and Feller, Avi and Miratrix, Luke},
  title = {Decomposing Treatment Effect Variation},
  journal = {Journal of the American Statistical Association},
  year = {2019},
  volume = {114},
  number = {525},
  pages = {304--317},
  doi = {10.1080/01621459.2017.1407322}
}

@book{hansen2022econometrics,
  author    = {Hansen, Bruce E.},
  title     = {Econometrics},
  publisher = {Princeton University Press},
  year      = {2022}
}

@article{farrell2021deep,
  title={Deep neural networks for estimation and inference},
  author={Farrell, Max H and Liang, Tengyuan and Misra, Sanjog},
  journal={Econometrica},
  volume={89},
  number={1},
  pages={181--213},
  year={2021},
  publisher={Wiley Online Library}
}

@article{list2024using,
  title={Using machine learning for efficient flexible regression adjustment in economic experiments},
  author={List, John A and Muir, Ian and Sun, Gregory},
  journal={Econometric reviews},
  volume={44},
  number={1},
  pages={2--40},
  year={2024},
  publisher={Taylor \& Francis}
}

@article{su2023decorrelation,
  title={A decorrelation method for general regression adjustment in randomized experiments},
  author={Su, Fangzhou and Mou, Wenlong and Ding, Peng and Wainwright, Martin J},
  journal={arXiv preprint arXiv:2311.10076},
  year={2023}
}

@article{greenshteinritov2004,
  author  = {Greenshtein, Eitan and Ritov, Ya'Acov},
  title   = {Persistence in High-Dimensional Linear Predictor Selection and the Virtue of Overparametrization},
  journal = {Bernoulli},
  year    = {2004},
  volume  = {10},
  number  = {6},
  pages   = {971--988},
  doi     = {10.3150/bj/1106314846}
}

@article{BUGNI2023105497,
title = {Inference under covariate-adaptive randomization with imperfect compliance},
journal = {Journal of Econometrics},
volume = {237},
number = {1},
pages = {105497},
year = {2023},
issn = {0304-4076},
doi = {https://doi.org/10.1016/j.jeconom.2023.105497},
url = {https://www.sciencedirect.com/science/article/pii/S0304407623002130},
author = {Federico A. Bugni and Mengsi Gao}
}

@article{jiang2024improving,
  author = {Jiang, Liang and Linton, Oliver B. and Tang, Haihan and Zhang, Yichong},
  title = {Improving Estimation Efficiency via Regression-Adjustment in Covariate-Adaptive Randomizations with Imperfect Compliance},
  journal = {The Review of Economics and Statistics},
  volume = {108},
  number = {3},
  pages = {774--791},
  year = {2026},
  doi = {10.1162/rest_a_01417}
}

@article{Ren02042024,
author = {Jiyang Ren},
title = {Model-Assisted Complier Average Treatment Effect Estimates in Randomized Experiments with Noncompliance},
journal = {Journal of Business \& Economic Statistics},
volume = {42},
number = {2},
pages = {707--718},
year = {2024},
publisher = {ASA Website},
doi = {10.1080/07350015.2023.2224851},


URL = { 
    
        https://doi.org/10.1080/07350015.2023.2224851
    
    

},
eprint = { 
    
        https://doi.org/10.1080/07350015.2023.2224851
    
    

}

}

@article{bannick2025general,
  author = {Bannick, Marlena S. and Shao, Jun and Liu, Jingyi and Du, Yu and Yi, Yanyao and Ye, Ting},
  title = {A General Form of Covariate Adjustment in Clinical Trials under Covariate-Adaptive Randomization},
  journal = {Biometrika},
  volume = {112},
  number = {3},
  pages = {asaf029},
  year = {2025},
  doi = {10.1093/biomet/asaf029}
}
\end{minilinespace}

\appendix
\clearpage
\setcounter{table}{0}
\renewcommand{\thetable}{A\arabic{table}}
\setcounter{figure}{0}
\renewcommand{\thefigure}{A\arabic{figure}}

\section*{Appendix}

\section{Estimation Algorithm}

Algorithm~\ref{alg:cf_late} implements the estimator in Section~\ref{sec:two_step_estimator}. The training-fold pilot is used to construct the first-stage pseudo-outcomes, while the full-sample pilot is used only in the final adjustment. By Lemma~\ref{lem:adjusted_ratio_reduction}, the returned estimator is exactly the ratio of the adjusted $Y$ and $D$ contrasts when its denominator is nonzero.

\begin{algorithm}[H]
\caption{Cross-fitted two-step ratio estimator}
\label{alg:cf_late}

\begin{adjustbox}{
    max width=\linewidth,
    max totalheight=\algorithmmaxheight,
    center
}
\begin{minipage}{\linewidth}
\footnotesize

\setlength{\abovedisplayskip}{3pt}
\setlength{\belowdisplayskip}{3pt}
\setlength{\abovedisplayshortskip}{2pt}
\setlength{\belowdisplayshortskip}{2pt}
\setlength{\jot}{2pt}

\begin{algorithmic}[1]
\Require Data \(\{(Y_i,D_i,Z_i,X_i)\}_{i=1}^N\).

\State Compute \(\hat\tau=\hat\tau_Y/\hat\tau_D\), where \(\hat\tau_R=\bar R_1-\bar R_0\), \(R\in\{Y,D\}\), and set \(\hat A_i=Y_i-\hat\tau D_i\).

\State Split the sample by assignment arm: for each \(z\in\{0,1\}\), randomly partition \(\{i:Z_i=z\}\) into \(S_z^{[q]}\), \(q\in\{1,\ldots,Q\}\), with fold sizes chosen to satisfy Assumption~\ref{ass:conditional_crossfit}(ii). Let \(S^{[q]}=S_1^{[q]}\cup S_0^{[q]}\) and \(S^{[-q]}=\{1,\ldots,N\}\setminus S^{[q]}\).

\For{\(q\in\{1,\ldots,Q\}\)}
    \State Compute the training-fold ratio estimator
    \[
    \hat\tau^{[-q]}
    =
    \frac{\hat\tau_Y^{[-q]}}{\hat\tau_D^{[-q]}},
    \qquad
    \hat\tau_R^{[-q]}
    =
    \bar R_1^{[-q]}-\bar R_0^{[-q]},
    \qquad
    R\in\{Y,D\},
    \]
    where
    \[
    \bar R_z^{[-q]}
    =
    \frac{1}{N_z^{[-q]}}
    \sum_{i\in S^{[-q]}:Z_i=z}R_i,
    \qquad
    N_z^{[-q]}
    =
    \sum_{i\in S^{[-q]}}\mathbf 1\{Z_i=z\}.
    \]

    \State Set \(\hat A_i^{[-q]}=Y_i-\hat\tau^{[-q]}D_i\) for \(i\in S^{[-q]}\).

    \State For each \(z\in\{0,1\}\), estimate \(\hat\mu_{A,i}^{[-q]}(z)\) using only observations in \(S^{[-q]}\) with \(Z_i=z\).

    \State Construct
    \[
    \hat M_i^{[-q]}
    =
    \left(
    \hat\mu_{A,i}^{[-q]}(1),
    \hat\mu_{A,i}^{[-q]}(0)
    \right)'
    \]
    for \(i=1,\ldots,N\).
\EndFor

\State Construct the cross-fitted generated covariate and its means:
\[
\hat M_{i,\mathrm{cf}}
=
\hat M_i^{[-q]}
\quad
\text{for }i\in S^{[q]},
\qquad
\bar{\hat M}_{\mathrm{cf}}
=
\frac{1}{N}\sum_{i=1}^N\hat M_{i,\mathrm{cf}},
\qquad
\bar{\hat M}_{z,\mathrm{cf}}
=
\frac{1}{N_z}\sum_{i:Z_i=z}\hat M_{i,\mathrm{cf}}.
\]

\For{\(R\in\{\hat A,D\}\)}
    \State Set \(\bar R_z=N_z^{-1}\sum_{i:Z_i=z}R_i\) for \(z\in\{0,1\}\).
    \For{\(z\in\{0,1\}\)}
        \State Estimate the pooled adjustment coefficient
        \[
        \hat\gamma_{R,\mathrm{cf}}(z)
        =
        \arg\min_\gamma
        \sum_{i:Z_i=z}
        \left\{
        R_i-\bar R_z
        -
        \left(
        \hat M_{i,\mathrm{cf}}
        -
        \bar{\hat M}_{z,\mathrm{cf}}
        \right)'\gamma
        \right\}^2.
        \]
    \EndFor

    \State Compute the adjusted difference
    \[
    \hat\tau_{R,\mathrm{cf}}
    =
    \left\{
    \bar R_1
    -
    \left(
    \bar{\hat M}_{1,\mathrm{cf}}
    -
    \bar{\hat M}_{\mathrm{cf}}
    \right)'
    \hat\gamma_{R,\mathrm{cf}}(1)
    \right\}
    -
    \left\{
    \bar R_0
    -
    \left(
    \bar{\hat M}_{0,\mathrm{cf}}
    -
    \bar{\hat M}_{\mathrm{cf}}
    \right)'
    \hat\gamma_{R,\mathrm{cf}}(0)
    \right\}.
    \]
\EndFor

\State \Return
\[
\hat\tau_{2s}
=
\hat\tau
+
\frac{\hat\tau_{\hat A,\mathrm{cf}}}
{\hat\tau_{D,\mathrm{cf}}}.
\]
\end{algorithmic}

\end{minipage}
\end{adjustbox}
\end{algorithm}

\clearpage

\section{Stratified randomized experiments}
\label{app:stratification}
Stratification is widely used in empirical experiments \citep{imbens2015causal}. This section extends the LATE special case of our results to stratified randomized experiments with a fixed number of large strata, including the experiment of \citet{dupas2018banking} analyzed in Section~\ref{sec:application}. Here $D_{h,i}(z)\in\{0,1\}$, and the exclusion and monotonicity restrictions of the LATE example in Section~\ref{sec:estimand_estimator} hold. Suppose that the finite population is partitioned into strata $h\in\{1,\ldots,H\}$, where stratum $h$ contains $N_h$ units, $N_{h,z}$ units are assigned to arm $z$, and $\pi_h=N_h/N$. Treatment assignment is completely randomized within each stratum and independent across strata. For $R\in\{Y,D\}$, define $\tau_{R,h}=N_h^{-1}\sum_{i=1}^{N_h}\{R_{h,i}(1)-R_{h,i}(0)\}$ and $\tau_R=\sum_{h=1}^H\pi_h\tau_{R,h}$.

Under exclusion and monotonicity, let $\tau_h$ denote the stratum-specific LATE when $\tau_{D,h}>0$, and set $\tau_h=0$ by convention when $\tau_{D,h}=0$. The overall LATE can be written as
\[
\tau
=\frac{\sum_{h=1}^H\pi_h\tau_{Y,h}}{\sum_{h=1}^H\pi_h\tau_{D,h}}
=\sum_{h=1}^H\frac{\pi_h\tau_{D,h}}{\tau_D}\tau_h.
\]
This is the complier-share-weighted average of the stratum-specific LATEs, with weights proportional to $\pi_h\tau_{D,h}$. Strata without compliers have $\tau_{Y,h}=\tau_{D,h}=0$ and receive zero weight, so the convention does not affect the overall estimand, which is invariant to the partition of a given finite population. Under monotonicity, Assumption~\ref{ass:id} requires only the aggregate relevance condition $\tau_D\ge C>0$, without requiring each $\tau_{D,h}$ to be bounded away from zero. Other notation follows the main text with a stratum subscript $h$; transformed outcomes retain the overall $\tau$, and centering and projection are performed within strata. In particular, $A_{h,i}(z)=Y_{h,i}(z)-\tau D_{h,i}(z)$ and $A_{2s,h,i}(z)=A_{h,i}(z)-(M_{h,i}-\bar M_h)'\delta_{M,h,z}$, where $\delta_{M,h,z}=(S_{M,h}^2)^{-1}S_{M,A_h(z)}$.

Let $Q\ge2$ be the number of cross-fitting folds in Algorithm~\ref{alg:cf_late}, formed approximately equally within each assignment-arm-by-stratum cell. Predictions may pool training observations across strata, excluding the held-out fold in every stratum. Use the ratio of the training-sample contrasts aggregated with the full-population weights $\pi_h$ as the preliminary estimator, and fit final adjustment coefficients separately by arm and stratum, pooling out-of-fold predictions across folds. For $R\in\{Y,D\}$, the within-stratum adjustment is
\begin{align*}
\hat\tau_{R,h,\mathrm{cf}}
&=\left\{\bar R_{h,1}-(\bar{\hat M}_{h,1,\mathrm{cf}}-\bar{\hat M}_{h,\mathrm{cf}})'\hat\gamma_{R,h,\mathrm{cf}}(1)\right\}-\left\{\bar R_{h,0}-(\bar{\hat M}_{h,0,\mathrm{cf}}-\bar{\hat M}_{h,\mathrm{cf}})'\hat\gamma_{R,h,\mathrm{cf}}(0)\right\},\\
\hat\gamma_{R,h,\mathrm{cf}}(z)
&=\arg\min_{\gamma\in\mathbb R^2}\sum_{i:Z_{h,i}=z}
\left\{R_{h,i}-\bar R_{h,z}-(\hat M_{h,i,\mathrm{cf}}-\bar{\hat M}_{h,z,\mathrm{cf}})'\gamma\right\}^2.
\end{align*}
The stratified two-step estimator is
\[
\hat\tau_{R,\mathrm{cf}}^{\mathrm{str}}
=\sum_{h=1}^H\pi_h\hat\tau_{R,h,\mathrm{cf}},
\qquad R\in\{Y,D\},
\qquad
\hat\tau_{2s}^{\mathrm{str}}
=\frac{\hat\tau_{Y,\mathrm{cf}}^{\mathrm{str}}}
       {\hat\tau_{D,\mathrm{cf}}^{\mathrm{str}}}.
\]

\begin{assumption}
\label{ass:stratified_two_step}
\leavevmode\par
\begin{itemize}
\item[(i)] Assumption~\ref{ass:id} holds for the overall population. $H$ and $Q$ are fixed, and, for every $h$ and $z$,
\[
\lim_{N\to\infty}\pi_h>0,
\qquad
\frac{N_{h,z}}{N_h}\to p_{h,z}\in(0,1).
\]
\item[(ii)] The mean and covariance convergence conditions in Assumption~\ref{ass:asymp}(ii), excluding its positive variance-limit requirement, and the maximum condition in Assumption~\ref{ass:asymp}(iii) hold within each stratum, with $N$ replaced by $N_h$ and the overall $\tau$ retained. Assumption~\ref{ass:two_step_asymptotics}(i)--(ii) likewise holds within each stratum, with projections formed within strata.
\item[(iii)] Assumption~\ref{ass:conditional_crossfit} holds within each stratum conditional on the full partition $\mathcal S$, with mutually independent assignments across all stratum-fold cells and $N_h^{[q]}/N_h\to1/Q$ for every $h,q$.
\item[(iv)] The variance limit $V_{2s}^{\mathrm{str}}$ below exists and lies in $(0,\infty)$.
\end{itemize}
\end{assumption}

\begin{proposition}
\label{prop:stratified_two_step}
Under Assumption~\ref{ass:stratified_two_step},
\begin{itemize}
\item[(i)] $\hat\tau_{2s}^{\mathrm{str}}-\tau\xrightarrow{p}0$.
\item[(ii)] $\sqrt N(\hat\tau_{2s}^{\mathrm{str}}-\tau) \xrightarrow{d}\mathcal N(0,V_{2s}^{\mathrm{str}})$, where
\[
V_{2s}^{\mathrm{str}}
=\lim_{N\to\infty}\frac{N}{\tau_D^2}\sum_{h=1}^H\pi_h^2
\left\{
\frac{S^2_{A_{2s,h}(1)}}{N_{h,1}}
+\frac{S^2_{A_{2s,h}(0)}}{N_{h,0}}
-\frac{S^2_{A_{2s,h}(1)-A_{2s,h}(0)}}{N_h}
\right\}.
\]
\item[(iii)] $V_{2s}^{\mathrm{str}}\le V^{\mathrm{str}}$, where $V^{\mathrm{str}}$ is the same expression with $A_{2s,h}(z)$ replaced by $A_h(z)$.
\end{itemize}
\end{proposition}

\begin{proof}
Because $\hat\tau_{2s}^{\mathrm{str}}$ is a ratio of stratum-weighted contrasts, its denominator converges in probability to the aggregate $\tau_D$, which is bounded away from zero by Assumption~\ref{ass:id}. Apply the oracle expansion in the proof of Theorem~\ref{thm:two_step_cf} to the component contrasts within each stratum, without forming stratum-specific ratios. The $h$th transformed-outcome contrast is centered at $\tau_{Y,h}-\tau\tau_{D,h}$, and the weighted sum of these centering terms is zero. Because $H$ is fixed, the centered oracle contributions are independent across strata; those with positive limiting variance satisfy the finite-population central limit theorem under the within-stratum maximum conditions, whereas contributions with zero limiting variance are $o_p(1)$ after $\sqrt N$ scaling by Chebyshev's inequality. Their aggregate variance has a positive limit by Assumption~\ref{ass:stratified_two_step}(iv), giving (i)--(ii). Applying the projection inequality underlying Theorem~\ref{thm:late_oracle_variance_min} within each stratum and scaling by the common $\tau_D^2$ gives (iii).
\end{proof}

Under the corresponding within-stratum moment, maximum, and rank conditions, the required out-of-fold prediction stability, and positive aggregate variance limits for the adjustment vectors in Corollary~\ref{cor:ils_cob_two_step}, the same weighted-aggregation argument extends its consistency, asymptotic normality, and efficiency comparisons to the stratified estimators. ILS uses the deterministic covariates $X_i$ and does not require cross-fitting; COB and the augmented estimator use conditionally cross-fitted predictions. No stratum-specific first-stage bound is required.

\section{Proofs}
\label{app:proof}

\subsection{Proof of Theorem~\ref{thm:late_oracle_variance_min}}
\label{app:proof_var_min}

\begin{proof}
The positive definiteness of $S_G^2$ assumed in Theorem~\ref{thm:late_oracle_variance_min} makes the projection coefficients well defined, and the condition $\tau_D\ne0$ makes the variance scaling factors positive. For $z\in\{0,1\}$, let
\[
e_{G,i}(z)
=
A_i(z)-\tilde G_i'\delta_z.
\]
By Lemma~\ref{lem:projection_geometry}, the normal equations imply
\[
S_{G,e_G(z)}=0
\]
and, for every $b_z\in\mathbb R^k$,
\begin{equation}
\label{eq:oracle-pythagorean}
S^2_{A_G(z;b_z)}
=
S^2_{e_G(z)}
+
(b_z-\delta_z)'S_G^2(b_z-\delta_z).
\end{equation}
\textbf{Proof of part~(i).}
\leavevmode\par
First note that
\begin{align*}
V_{\mathrm{cons},N}(G;b_1,b_0)
&=
\frac{N}{\tau_D^2}
\left\{
\frac{S^2_{A_G(1;b_1)}}{N_1}
+
\frac{S^2_{A_G(0;b_0)}}{N_0}
\right\}.
\end{align*}
Applying \eqref{eq:oracle-pythagorean} in the two arms gives
\begin{align*}
V_{\mathrm{cons},N}(G;b_1,b_0)
&=
\frac{N}{\tau_D^2}
\left[
\frac{
S^2_{e_G(1)}
+
(b_1-\delta_1)'S_G^2(b_1-\delta_1)
}{N_1}
+
\frac{
S^2_{e_G(0)}
+
(b_0-\delta_0)'S_G^2(b_0-\delta_0)
}{N_0}
\right]
\\
&=
\frac{N}{\tau_D^2}
\left\{
\frac{S^2_{e_G(1)}}{N_1}
+
\frac{S^2_{e_G(0)}}{N_0}
\right\}
+
\frac{N}{\tau_D^2}
\left\{
\frac{(b_1-\delta_1)'S_G^2(b_1-\delta_1)}{N_1}
+
\frac{(b_0-\delta_0)'S_G^2(b_0-\delta_0)}{N_0}
\right\}
\\
&=
\frac{N}{\tau_D^2}
\left\{
\frac{S^2_{A_G(1;\delta_1)}}{N_1}
+
\frac{S^2_{A_G(0;\delta_0)}}{N_0}
\right\}
+
\frac{N}{\tau_D^2}
\left\{
\frac{(b_1-\delta_1)'S_G^2(b_1-\delta_1)}{N_1}
+
\frac{(b_0-\delta_0)'S_G^2(b_0-\delta_0)}{N_0}
\right\}
\\
&=
V_{\mathrm{cons},N}(G;\delta_1,\delta_0)
+
\frac{N}{\tau_D^2}
\left\{
\frac{(b_1-\delta_1)'S_G^2(b_1-\delta_1)}{N_1}
+
\frac{(b_0-\delta_0)'S_G^2(b_0-\delta_0)}{N_0}
\right\}.
\end{align*}
Therefore,
\begin{align*}
&V_{\mathrm{cons},N}(G;b_1,b_0)
-
V_{\mathrm{cons},N}(G;\delta_1,\delta_0)=
\frac{N}{\tau_D^2}
\left\{
\frac{(b_1-\delta_1)'S_G^2(b_1-\delta_1)}{N_1}
+
\frac{(b_0-\delta_0)'S_G^2(b_0-\delta_0)}{N_0}
\right\}.
\end{align*}
Because $S_G^2$ is positive definite, the right-hand side is nonnegative and equals zero only when $b_1=\delta_1$ and $b_0=\delta_0$. This proves part~(i).

\textbf{Proof of part~(ii).}
\leavevmode\par
Using the relationship between the variance and covariance,
\[
S^2_{A_G(1;b_1)-A_G(0;b_0)}
=
S^2_{A_G(1;b_1)}
+
S^2_{A_G(0;b_0)}
-
2S_{A_G(1;b_1),A_G(0;b_0)},
\]
we obtain
\begin{align*}
V_N(G;b_1,b_0)
&=
\frac{N}{\tau_D^2}
\left\{
\frac{S^2_{A_G(1;b_1)}}{N_1}
+
\frac{S^2_{A_G(0;b_0)}}{N_0}
-
\frac{S^2_{A_G(1;b_1)-A_G(0;b_0)}}{N}
\right\}
\\
&=
\frac{N}{\tau_D^2}
\left[
\frac{S^2_{A_G(1;b_1)}}{N_1}
+
\frac{S^2_{A_G(0;b_0)}}{N_0}
-
\frac{
S^2_{A_G(1;b_1)}
+
S^2_{A_G(0;b_0)}
-
2S_{A_G(1;b_1),A_G(0;b_0)}
}{N}
\right]
\\
&=
\frac{1}{\tau_D^2}
\left\{
\left(\frac{N}{N_1}-1\right)S^2_{A_G(1;b_1)}
+
\left(\frac{N}{N_0}-1\right)S^2_{A_G(0;b_0)}
+
2S_{A_G(1;b_1),A_G(0;b_0)}
\right\}
\\
&=
\frac{1}{\tau_D^2}
\left\{
\frac{N_0}{N_1}S^2_{A_G(1;b_1)}
+
\frac{N_1}{N_0}S^2_{A_G(0;b_0)}
+
2S_{A_G(1;b_1),A_G(0;b_0)}
\right\}
\\
&=
\frac{N_1}{N_0\tau_D^2}
\left\{
\left(\frac{N_0}{N_1}\right)^2
S^2_{A_G(1;b_1)}
+
S^2_{A_G(0;b_0)}
+
2\frac{N_0}{N_1}
S_{A_G(1;b_1),A_G(0;b_0)}
\right\}
\\
&=
\frac{N_1}{N_0\tau_D^2}
S^2_{
\frac{N_0}{N_1}A_G(1;b_1)+A_G(0;b_0)
}
\\
&=
\frac{N_1}{N_0\tau_D^2}
S^2_{
\frac{N_0}{N_1}
\{A(1)-\tilde G'b_1\}
+
\{A(0)-\tilde G'b_0\}
}
\\
&=
\frac{N_1}{N_0\tau_D^2}
S^2_{
\frac{N_0}{N_1}A(1)+A(0)
-
\tilde G'
\left(
\frac{N_0}{N_1}b_1+b_0
\right)
}.
\end{align*}
Here, the fourth equality uses \(N=N_1+N_0\), and the sixth equality follows from the variance formula for a linear combination of two scalar sequences. Similarly, starting from the definition of \(V_{\mathrm{gap},N}\),
\begin{align*}
V_{\mathrm{gap},N}(G;b_1,b_0)
&=
\frac{1}{\tau_D^2}
S^2_{A_G(1;b_1)-A_G(0;b_0)}
\\
&=
\frac{1}{\tau_D^2}
S^2_{
\{A(1)-\tilde G'b_1\}
-
\{A(0)-\tilde G'b_0\}
}
\\
&=
\frac{1}{\tau_D^2}
S^2_{
A(1)-A(0)-\tilde G'(b_1-b_0)
}.
\end{align*}
The projection coefficients for the two scalar targets are
\begin{align*}
(S_G^2)^{-1}S_{G,\frac{N_0}{N_1}A(1)+A(0)}
&=
\frac{N_0}{N_1}\delta_1+\delta_0,\\
(S_G^2)^{-1}S_{G,A(1)-A(0)}
&=
\delta_1-\delta_0.
\end{align*}
The projection coefficient for the scalar target determining \(V_N\) is
\begin{align*}
(S_G^2)^{-1}
S_{G,\frac{N_0}{N_1}A(1)+A(0)}
&=
(S_G^2)^{-1}
\left\{
\frac{N_0}{N_1}S_{G,A(1)}
+
S_{G,A(0)}
\right\}
\\
&=
\frac{N_0}{N_1}
(S_G^2)^{-1}S_{G,A(1)}
+
(S_G^2)^{-1}S_{G,A(0)}
\\
&=
\frac{N_0}{N_1}\delta_1+\delta_0.
\end{align*}
The projection coefficient for the scalar target determining
\(V_{\mathrm{gap},N}\) is
\begin{align*}
(S_G^2)^{-1}S_{G,A(1)-A(0)}
&=
(S_G^2)^{-1}
\left\{
S_{G,A(1)}-S_{G,A(0)}
\right\}
\\
&=
(S_G^2)^{-1}S_{G,A(1)}
-
(S_G^2)^{-1}S_{G,A(0)}
\\
&=
\delta_1-\delta_0.
\end{align*}

Applying Lemma~\ref{lem:projection_geometry} to the two preceding scalar targets yields
\begin{align*}
&V_N(G;b_1,b_0)
-
V_N(G;\delta_1,\delta_0)
=
\frac{N_1}{N_0\tau_D^2}
\left\{
\frac{N_0}{N_1}(b_1-\delta_1)
+
(b_0-\delta_0)
\right\}'
S_G^2
\left\{
\frac{N_0}{N_1}(b_1-\delta_1)
+
(b_0-\delta_0)
\right\}
\ge 0,\\
&V_{\mathrm{gap},N}(G;b_1,b_0)
-
V_{\mathrm{gap},N}(G;\delta_1,\delta_0)
=
\frac{1}{\tau_D^2}
\left\{
(b_1-\delta_1)-(b_0-\delta_0)
\right\}'
S_G^2
\left\{
(b_1-\delta_1)-(b_0-\delta_0)
\right\}
\ge 0.
\end{align*}
Hence \((\delta_1,\delta_0)\) minimizes both \(V_N\) and \(V_{\mathrm{gap},N}\), which proves part~(ii).

\textbf{Proof of part~(iii).}
\leavevmode\par
Every linear adjustment based on
$
\left(
\tilde G_i'\delta_1,
\tilde G_i'\delta_0
\right)'
$
is also available from \(G_i\), because both coordinates belong to the centered linear span generated by \(G_i\). Thus, for each of the three criteria, the minimum based on \(G_i\) is no larger than the minimum based on
$
\left(
\tilde G_i'\delta_1,
\tilde G_i'\delta_0
\right)'.$

Conversely, use the coefficient vector \((1,0)'\) in arm one and the coefficient vector \((0,1)'\) in arm zero. The resulting arm-specific fitted values are exactly
\[
\tilde G_i'\delta_1
\qquad\text{and}\qquad
\tilde G_i'\delta_0.
\]
They attain the minimum of \(V_{\mathrm{cons},N}\) by part~(i). For the point-variance target, their weighted sum is
\[
\frac{N_0}{N_1}\tilde G_i'\delta_1
+
\tilde G_i'\delta_0
=
\tilde G_i'
\left(
\frac{N_0}{N_1}\delta_1+\delta_0
\right),
\]
and, for the gap target, their difference is
\[
\tilde G_i'\delta_1-\tilde G_i'\delta_0
=
\tilde G_i'(\delta_1-\delta_0).
\]
Part~(ii) therefore implies that the same two arm-specific fitted values attain the minima of \(V_N\) and \(V_{\mathrm{gap},N}\). Hence the reverse inequalities hold, proving equality of the three minimum values.
\end{proof}

\subsection{Proof of Theorem~\ref{thm:two_step_cf}}
\label{app:proof_two_step_cf}

\begin{proof}

We first express the estimation error as a ratio, approximate its numerator and denominator by differences in means of fixed residual potential outcomes, and replace the denominator by $\tau_D$.

Lemma~\ref{lem:adjusted_ratio_reduction}, applied with \(G_i=\hat M_{i,\mathrm{cf}}\) and \(s=\hat\tau\), gives
\[
\hat\tau_{\hat A,\mathrm{cf}}
=
\hat\tau_{A,\mathrm{cf}}
-
(\hat\tau-\tau)\hat\tau_{D,\mathrm{cf}}.
\]
Together with the definition of \(\hat\tau_{2s}\), this yields the exact identity
\begin{equation}
\label{eq:two-step-exact-ratio}
\hat\tau_{2s}-\tau
=
\frac{\hat\tau_{A,\mathrm{cf}}}{\hat\tau_{D,\mathrm{cf}}},
\end{equation}
whenever \(\hat\tau_{D,\mathrm{cf}}\ne0\).

Define
\[
\hat\tau_{A_{2s}}
=
\frac{1}{N_1}\sum_{i:Z_i=1}A_{2s,i}(1)
-
\frac{1}{N_0}\sum_{i:Z_i=0}A_{2s,i}(0).
\]
The conditions of Lemma~\ref{lem:cf_oracle_equivalence} follow from Assumptions~\ref{ass:conditional_crossfit} and \ref{ass:two_step_asymptotics}(ii), together with Assumptions~\ref{ass:id} and \ref{ass:asymp}(i)--(ii), which ensure nonvanishing treatment fractions, $\tau=O(1)$, and bounded finite-population second moments of $A_i(z)$. Applying that lemma with \(\hat G_i^{[-q]}=\hat M_i^{[-q]}\), \(G_i=M_i\), and \(R=A\) gives
\begin{equation}
\label{eq:two-step-A-oracle}
\hat\tau_{A,\mathrm{cf}}
=
\hat\tau_{A_{2s}}
+
o_p(N^{-1/2}).
\end{equation}

Similarly, define
\[
D_{2s,i}(z)
=
D_i(z)-\tilde M_i'\gamma_D(z),
\qquad
\gamma_D(z)
=
(S_M^2)^{-1}S_{M,D(z)},
\]
and
\[
\hat\tau_{D_{2s}}
=
\frac{1}{N_1}\sum_{i:Z_i=1}D_{2s,i}(1)
-
\frac{1}{N_0}\sum_{i:Z_i=0}D_{2s,i}(0).
\]
Applying Lemma~\ref{lem:cf_oracle_equivalence} with \(R=D\) gives
\[
\hat\tau_{D,\mathrm{cf}}
=
\hat\tau_{D_{2s}}
+
o_p(N^{-1/2}).
\]
The full-population means of the adjustment terms are zero. The finite-population variance formula, Assumption~\ref{ass:asymp}(i), and Assumption~\ref{ass:two_step_asymptotics}(ii) therefore imply
\[
\hat\tau_{D_{2s}}-\tau_D
=
O_p(N^{-1/2}).
\]
Consequently,
\begin{equation}
\label{eq:two-step-D-limit}
\hat\tau_{D,\mathrm{cf}}
=
\tau_D+o_p(1).
\end{equation}

Because \(N^{-1}\sum_i\tilde M_i=0\),
\[
\bar A_{2s}(1)-\bar A_{2s}(0)
=
\bar A(1)-\bar A(0)
=
\tau_Y-\tau\tau_D
=
0.
\]
Under Assumptions~\ref{ass:id}, \ref{ass:asymp}(i)--(ii), and \ref{ass:two_step_asymptotics}(ii), the finite-population variance formula also gives
\[
\sqrt N\hat\tau_{A_{2s}}=O_p(1).
\]
Combining \eqref{eq:two-step-exact-ratio}--\eqref{eq:two-step-D-limit} and using \(|\tau_D|\ge C>0\) from Assumption~\ref{ass:id} gives
\begin{equation}
\label{eq:two-step-linearization}
\sqrt N(\hat\tau_{2s}-\tau)
=
\frac{1}{\tau_D}\sqrt N\hat\tau_{A_{2s}}
+
o_p(1).
\end{equation}

\textbf{Proof of parts~(i) and~(ii).}
\leavevmode\par
Assumption~\ref{ass:two_step_asymptotics}(i) is the finite-population maximal condition for the two potential residual sequences, and Assumption~\ref{ass:two_step_asymptotics}(iii) gives a positive limiting variance. Assumption~\ref{ass:asymp}(i) gives limiting treatment fractions in $(0,1)$, while Assumptions~\ref{ass:id}, \ref{ass:asymp}(ii), and \ref{ass:two_step_asymptotics}(ii) ensure convergence of the residual covariances. Applying the finite-population central limit theorem \citep{Li02102017} to the residual sequences scaled by $\tau_D^{-1}$ therefore yields
\[
\frac{\sqrt N\hat\tau_{A_{2s}}}{\tau_D}
\xrightarrow{d}
\mathcal N(0,V_{2s}).
\]
Together with \eqref{eq:two-step-linearization}, this proves
\[
\sqrt N(\hat\tau_{2s}-\tau)
\xrightarrow{d}
\mathcal N(0,V_{2s}).
\]
The same expansion implies $\hat\tau_{2s}-\tau=O_p(N^{-1/2})$, and hence consistency.

\textbf{Proof of part~(iii).}
\leavevmode\par
For each $N$, the variance scale of the leading term in \eqref{eq:two-step-linearization} is $V_N(M;\delta_{M,1},\delta_{M,0})$. By Theorem~\ref{thm:late_oracle_variance_min}(ii),
\[
V_N(M;\delta_{M,1},\delta_{M,0})
\le
V_N(M;0,0)
=
\sigma^2.
\]
The limits exist by Assumptions~\ref{ass:two_step_asymptotics}(iii) and \ref{ass:asymp}(ii). Taking limits gives $V_{2s}\le V$.
\end{proof}

\subsection{Proof of Theorem~\ref{thm:variance_estimator}}
\label{app:proof_variance_estimator}

\begin{proof}

\textbf{Proof of part~(i).}
\leavevmode\par
For the unadjusted ratio estimator, Assumptions~\ref{ass:id} and \ref{ass:asymp} give $\hat\tau-\tau=O_p(N^{-1/2})$ and $\hat\tau_D-\tau_D=O_p(N^{-1/2})$. Assumptions~\ref{ass:asymp}(i)--(ii) imply $N^{-1}\sum_iD_i(z)^2=O(1)$ and $N_z/N$ is bounded away from zero, so $N_z^{-1}\sum_{i:Z_i=z}D_i(z)^2=O(1)$. Hence $\hat A_i-A_i(z)=-(\hat\tau-\tau)D_i(z)$ has arm-specific empirical squared mean $o_p(1)$. Assumption~\ref{ass:asymp}(iii) and Lemma~\ref{lem:sample_variance_stability} give $\hat S^2(z)-S^2_{A(z)}\xrightarrow{p}0$. Together with $\hat\tau_D-\tau_D=o_p(1)$, this proves $\hat V-V_{\mathrm{cons}}\xrightarrow{p}0$.

For ILS, set $G_i=X_i$ in the main-text notation without sample splitting. Assumptions~\ref{ass:asymp}(i)--(ii) and Assumption~\ref{ass:two_step_asymptotics}(ii) with $M_i$ replaced by $X_i$ imply convergence of the arm-specific sample means and OLS covariance matrices for $X_i$, $Y_i(z)$, and $D_i(z)$ to their finite-population counterparts. In particular, the sampling variances of cross-product means vanish because $\max_i\|X_i-\bar X\|_2^2/N=o(1)$ and the finite-population second moments of $Y_i(z)$ and $D_i(z)$ are bounded. Positive definiteness then gives $\hat\gamma_R(z)-\gamma_R(z)=o_p(1)$ for $R\in\{Y,D\}$. Since $\bar X_z-\bar X=O_p(N^{-1/2})$, it follows that $\hat\tau_{X,D}-\tau_D=o_p(1)$, $\hat\tau_{\mathrm{ils}}-\tau=o_p(1)$, and $\hat\gamma_Y(z)-\hat\tau_{\mathrm{ils}}\hat\gamma_D(z)-\delta_z=o_p(1)$. Assumption~\ref{ass:two_step_asymptotics}(i) with $M_i$ replaced by $X_i$ supplies the maximal condition for $A_{G,i}(z;\delta_z)$ with $G_i=X_i$. Lemma~\ref{lem:sample_variance_stability}, with the second-moment bound on $D_i(z)$ established above, yields $\hat V_{\mathrm{ils}}-V_{\mathrm{cons,ils}}\xrightarrow{p}0$.

We consider the remaining estimators using a generic conditionally cross-fitted adjustment vector
\(\hat G_{i,\mathrm{cf}}\) with deterministic target \(G_i\). This notation
covers
\[
\hat W_{i,\mathrm{cf}},
\qquad
\hat M_{i,\mathrm{cf}},
\qquad
(X_i',\hat W_{i,\mathrm{cf}}',\hat M_{i,\mathrm{cf}}')'.
\]
For each \(z\in\{0,1\}\), write
\[
\delta_{G,z}
=
(S_G^2)^{-1}S_{G,A(z)}.
\]
Let
\[
\hat\tau_{\hat G_{\mathrm{cf}}}
=
\frac{
\hat\tau_{\hat G_{\mathrm{cf}},Y}
}{
\hat\tau_{\hat G_{\mathrm{cf}},D}
}
\]
and
\[
\hat b_{\hat G_{\mathrm{cf}},z}
=
\hat\gamma_{\hat G_{\mathrm{cf}},Y}(z)
-
\hat\tau_{\hat G_{\mathrm{cf}}}
\hat\gamma_{\hat G_{\mathrm{cf}},D}(z).
\]
For \(i\) with \(Z_i=z\), define the feasible residual
\begin{equation}
\label{eq:generic-feasible-residual}
\hat e_{\hat G_{\mathrm{cf}},i}
=
Y_i
-
\hat\tau_{\hat G_{\mathrm{cf}}}D_i
-
(\hat G_{i,\mathrm{cf}}-\bar{\hat G}_{z,\mathrm{cf}})'
\hat b_{\hat G_{\mathrm{cf}},z}.
\end{equation}

Write $\hat S_G^2(z)$ for the arm-$z$ sample variance of $\hat e_{\hat G_{\mathrm{cf}},i}$ and $\hat V_G$ for the corresponding conservative plug-in variance with denominator $\hat\tau_{\hat G_{\mathrm{cf}},D}^2$. For the two-step estimator, Lemma~\ref{lem:adjusted_ratio_reduction} gives
\[
\hat\tau_{\hat G_{\mathrm{cf}}}
=
\hat\tau
+
\frac{
\hat\tau_{\hat G_{\mathrm{cf}},\hat A}
}{
\hat\tau_{\hat G_{\mathrm{cf}},D}
}
\]
when \(\hat G_{i,\mathrm{cf}}=\hat M_{i,\mathrm{cf}}\). Hence
\eqref{eq:generic-feasible-residual} coincides with the residual used
to define \(\hat V_{2s}\). The same identity applies to the augmented
two-step estimator.

Under Assumptions~\ref{ass:id}, \ref{ass:asymp}(i)--(ii), and the corresponding versions of Assumptions~\ref{ass:conditional_crossfit} and \ref{ass:two_step_asymptotics}(ii), Lemmas~\ref{lem:adjusted_ratio_reduction} and \ref{lem:cf_oracle_equivalence} imply
\[
\hat\tau_{\hat G_{\mathrm{cf}}}
\xrightarrow{p}
\tau
\]
and, for each \(z\in\{0,1\}\),
\[
\hat\gamma_{\hat G_{\mathrm{cf}},Y}(z)
=
\gamma_{G,Y}(z)+o_p(1),
\qquad
\hat\gamma_{\hat G_{\mathrm{cf}},D}(z)
=
\gamma_{G,D}(z)+o_p(1).
\]
Therefore,
\begin{align*}
\hat b_{\hat G_{\mathrm{cf}},z}
&=
\hat\gamma_{\hat G_{\mathrm{cf}},Y}(z)
-
\hat\tau_{\hat G_{\mathrm{cf}}}
\hat\gamma_{\hat G_{\mathrm{cf}},D}(z)
\\
&=
\gamma_{G,Y}(z)
-
\tau\gamma_{G,D}(z)
+
o_p(1)
\\
&=
\delta_{G,z}
+
o_p(1).
\end{align*}

Under Assumptions~\ref{ass:asymp}(i)--(ii) and the corresponding versions of Assumptions~\ref{ass:conditional_crossfit} and \ref{ass:two_step_asymptotics}(ii), Lemma~\ref{lem:cf_moment_stability} also gives
\[
\frac{1}{N_z}
\sum_{i:Z_i=z}
\|\hat G_{i,\mathrm{cf}}-G_i\|_2^2
=
o_p(1),
\qquad
\bar{\hat G}_{z,\mathrm{cf}}-\bar G
=
o_p(1).
\]
Under the corresponding version of Assumption~\ref{ass:two_step_asymptotics}(ii),
\[
\|\delta_{G,z}\|_2=O(1),
\qquad
\frac{1}{N}
\sum_{i=1}^N
\|G_i-\bar G\|_2^2
=
O(1),
\]
and the oracle residual
\[
A_{G,i}(z;\delta_{G,z})
=
A_i(z)-(G_i-\bar G)'\delta_{G,z}
\]
satisfies, by Assumptions~\ref{ass:id} and \ref{ass:asymp}(ii) and the corresponding version of Assumption~\ref{ass:two_step_asymptotics}(i)--(ii),
\[
S^2_{A_G(z;\delta_{G,z})}
=
O(1),
\qquad
\max_{1\le i\le N}
\frac{
\left\{
A_{G,i}(z;\delta_{G,z})
-
\bar A_G(z;\delta_{G,z})
\right\}^2
}{N}
\to0.
\]
Lemma~\ref{lem:sample_variance_stability}, applied with
\[
\hat b_z=\hat b_{\hat G_{\mathrm{cf}},z},
\qquad
\delta_z=\delta_{G,z},
\qquad
e_i(z)=A_{G,i}(z;\delta_{G,z}),
\]
therefore yields
\[
\frac{1}{N_z}
\sum_{i:Z_i=z}
\left\{
\hat e_{\hat G_{\mathrm{cf}},i}
-
A_{G,i}(z;\delta_{G,z})
\right\}^2
=
o_p(1)
\]
and
\begin{equation}
\label{eq:generic-sample-variance-limit}
\hat S_G^2(z)
-
S^2_{A_G(z;\delta_{G,z})}
\xrightarrow{p}0.
\end{equation}

Lemma~\ref{lem:cf_oracle_equivalence} applied to \(R=D\) also gives
\[
\hat\tau_{\hat G_{\mathrm{cf}},D}
=
\tau_D+o_p(1).
\]
Combining this result with \eqref{eq:generic-sample-variance-limit}, using Assumption~\ref{ass:asymp}(i) and the lower bound on $|\tau_D|$ in Assumption~\ref{ass:id}, gives
\[
\hat V_G
-
\frac{N}{\tau_D^2}
\left\{
\frac{
S^2_{A_G(1;\delta_{G,1})}
}{N_1}
+
\frac{
S^2_{A_G(0;\delta_{G,0})}
}{N_0}
\right\}
\xrightarrow{p}0.
\]
By the definition of \(V_{\mathrm{cons},N}\),
\[
\frac{N}{\tau_D^2}
\left\{
\frac{
S^2_{A_G(1;\delta_{G,1})}
}{N_1}
+
\frac{
S^2_{A_G(0;\delta_{G,0})}
}{N_0}
\right\}
=
V_{\mathrm{cons},N}
(G;\delta_{G,1},\delta_{G,0}).
\]
Taking limits therefore proves consistency of
\(\hat V_{\mathrm{cob,cf}}\), \(\hat V_{2s}\), and
\(\hat V_{2s,\mathrm{extra}}\) for their corresponding conservative
asymptotic variance limits. Together with the unadjusted ratio and ILS results above, this proves part~(i).

\textbf{Proof of parts~(ii) and~(iii).}
\leavevmode\par

The centered adjustment spaces generated by \(M_i\), \(X_i\), and
\(W_i\) are all contained in the centered adjustment space generated by
\[
G_i^+=(X_i',W_i',M_i')'.
\]
Lemma~\ref{lem:projection_geometry} and
Theorem~\ref{thm:late_oracle_variance_min} therefore imply
\[
V_{\mathrm{cons,2s,extra}}
\le
V_{\mathrm{cons,2s}},
\qquad
V_{\mathrm{cons,2s,extra}}
\le
V_{\mathrm{cons,ils}},
\qquad
V_{\mathrm{cons,2s,extra}}
\le
V_{\mathrm{cons,cob,cf}}.
\]
Zero adjustment is feasible in each of the three benchmark adjustment
spaces, so
\[
V_{\mathrm{cons,2s}}
\le
V_{\mathrm{cons}},
\qquad
V_{\mathrm{cons,ils}}
\le
V_{\mathrm{cons}},
\qquad
V_{\mathrm{cons,cob,cf}}
\le
V_{\mathrm{cons}}.
\]
Combining these inequalities proves parts~(ii) and~(iii).
\end{proof}

\subsection{Proof of Theorem~\ref{thm:lasso_persistence}}
\label{app:proof_lasso_persistence}

\begin{proof}
Fix $z\in\{0,1\}$. Let
\[
\beta=\beta_A(z;\lambda_N),
\qquad
\hat\beta_q=\hat\beta_A^{[-q]}(z;\lambda_N).
\]
By the definitions of the sample and population prediction functions,
\[
\hat\mu_{A,i}^{[-q]}(z;\lambda_N)
-
\mu_{A,i}(z;\lambda_N)
=
\{\bar{\hat A}_z^{[-q]}-\bar A(z)\}
+
\tilde X_i'(\hat\beta_q-\beta).
\]
Since $\sum_{i=1}^N\tilde X_i=0$, for every $q$,
\[
\|\hat\mu_A^{[-q]}(z;\lambda_N)-\mu_A(z;\lambda_N)\|_N^2
=
\{\bar{\hat A}_z^{[-q]}-\bar A(z)\}^2
+
\frac{1}{N}\sum_{i=1}^N
\{\tilde X_i'(\hat\beta_q-\beta)\}^2.
\]
Therefore, it suffices to show
\[
\max_{1\le q\le Q}
|\bar{\hat A}_z^{[-q]}-\bar A(z)|
=o_p(1)
\]
and
\[
\max_{1\le q\le Q}
\frac{1}{N}\sum_{i=1}^N
\{\tilde X_i'(\hat\beta_q-\beta)\}^2
=o_p(1).
\]

\textbf{Step 1: Mean component.}
\leavevmode\par
For $i\in S^{[-q]}$ with $Z_i=z$,
\[
\hat A_i^{[-q]}
=
A_i(z)-(\hat\tau^{[-q]}-\tau)D_i(z).
\]
Therefore,
\begin{align*}
\max_{1\le q\le Q}
|\bar{\hat A}_z^{[-q]}-\bar A(z)|
&\le
\max_q
\left|
\frac{1}{N_z^{[-q]}}
\sum_{i\in S^{[-q]}:Z_i=z}
A_i(z)
-
\bar A(z)
\right|
\\
&\quad+
\max_q|\hat\tau^{[-q]}-\tau|
\max_q
\left|
\frac{1}{N_z^{[-q]}}
\sum_{i\in S^{[-q]}:Z_i=z}
D_i(z)
\right|.
\end{align*}
The inequality follows by substituting the preceding identity and applying
the triangle inequality. It therefore suffices to control the two terms on
the right-hand side.

Assumptions~\ref{ass:id} and \ref{ass:asymp}(ii) imply
$\tau=O(1)$ and
\[
\frac{1}{N}\sum_{i=1}^N
\{A_i(z)^2+D_i(z)^2\}
=
O(1).
\]
Moreover, Lemma~\ref{lem:split_sampling},
Assumptions~\ref{ass:asymp}(i) and
\ref{ass:conditional_crossfit}(ii), and fixed $Q$ imply that
\[
\min_{q,z}\frac{N_z^{[-q]}}{N}\ge c>0
\]
for some constant $c>0$ and all sufficiently large $N$. Hence, for
$R\in\{A,D\}$,
\begin{align*}
\max_{1\le q\le Q}
\frac{1}{N_z^{[-q]}}
\sum_{i\in S^{[-q]}:Z_i=z}
R_i(z)^2
&\le
\frac{N}{\min_q N_z^{[-q]}}
\frac{1}{N}\sum_{i=1}^N R_i(z)^2
\\
&=
O(1).
\end{align*}

For the first term, Lemma~\ref{lem:split_sampling} implies that, for each
$q$, the set $\{i\in S^{[-q]}:Z_i=z\}$ is marginally a simple random
sample without replacement of size $N_z^{[-q]}$ from the finite population.
Hence,
\begin{align*}
\mathbb E\left[
\frac{1}{N_z^{[-q]}}
\sum_{i\in S^{[-q]}:Z_i=z}
A_i(z)
\right]
&=
\bar A(z),
\\
\operatorname{Var}\left(
\frac{1}{N_z^{[-q]}}
\sum_{i\in S^{[-q]}:Z_i=z}
A_i(z)
\right)
&=
\left(
\frac{1}{N_z^{[-q]}}
-
\frac{1}{N}
\right)
S_{A(z)}^2.
\end{align*}
Assumptions~\ref{ass:id} and \ref{ass:asymp}(ii) imply
$\tau=O(1)$ and
\begin{align*}
S_{A(z)}^2
&=
S_{Y(z)}^2
+
\tau^2 S_{D(z)}^2
-
2\tau S_{Y(z)D(z)}
\\
&=
O(1).
\end{align*}
Therefore,
\begin{align*}
\max_q
\operatorname{Var}\left(
\frac{1}{N_z^{[-q]}}
\sum_{i\in S^{[-q]}:Z_i=z}
A_i(z)
\right)
&=
\max_q
\left(
\frac{1}{N_z^{[-q]}}
-
\frac{1}{N}
\right)
S_{A(z)}^2
\\
&\le
\frac{1}{\min_q N_z^{[-q]}}
S_{A(z)}^2
\\
&=
O(N^{-1}).
\end{align*}
Thus, for every $M>0$, Chebyshev's inequality gives
\begin{align*}
&\Pr\left(
\max_q
\left|
\frac{1}{N_z^{[-q]}}
\sum_{i\in S^{[-q]}:Z_i=z}
A_i(z)
-
\bar A(z)
\right|
>
\frac{M}{\sqrt N}
\right)
\\
&\quad\le
\sum_{q=1}^Q
\Pr\left(
\left|
\frac{1}{N_z^{[-q]}}
\sum_{i\in S^{[-q]}:Z_i=z}
A_i(z)
-
\bar A(z)
\right|
>
\frac{M}{\sqrt N}
\right)
\\
&\quad\le
\frac{N}{M^2}
\sum_{q=1}^Q
\operatorname{Var}\left(
\frac{1}{N_z^{[-q]}}
\sum_{i\in S^{[-q]}:Z_i=z}
A_i(z)
\right)
\\
&\quad=
O(M^{-2}),
\end{align*}
where the last bound is uniform in $N$ because $Q$ is fixed.
Letting $M\to\infty$ yields
\[
\max_q
\left|
\frac{1}{N_z^{[-q]}}
\sum_{i\in S^{[-q]}:Z_i=z}
A_i(z)
-
\bar A(z)
\right|
=
O_p(N^{-1/2}).
\]

For the second term, Lemma~\ref{lem:training_ratio_rate} gives
\[
\max_{1\le q\le Q}
|\hat\tau^{[-q]}-\tau|
=
O_p(N^{-1/2}).
\]
In addition, Cauchy--Schwarz and the preceding second-moment bound give
\begin{align*}
&\max_q|\hat\tau^{[-q]}-\tau|
\max_q
\left|
\frac{1}{N_z^{[-q]}}
\sum_{i\in S^{[-q]}:Z_i=z}
D_i(z)
\right|
\\
&\qquad\le
\max_q|\hat\tau^{[-q]}-\tau|
\max_q
\left\{
\frac{1}{N_z^{[-q]}}
\sum_{i\in S^{[-q]}:Z_i=z}
D_i(z)^2
\right\}^{1/2}
\\
&\qquad=
O_p(N^{-1/2}).
\end{align*}
Combining the two bounds yields
\[
\max_{1\le q\le Q}
|\bar{\hat A}_z^{[-q]}-\bar A(z)|
=
O_p(N^{-1/2})
=
o_p(1).
\]

\textbf{Step 2: Slope prediction component.}
\leavevmode\par
It remains to show
\[
\max_q
\frac{1}{N}
\sum_{i=1}^N
\{\tilde X_i'(\hat\beta_q-\beta)\}^2
=
o_p(1),
\]
where
\[
\beta=\beta_A(z;\lambda_N),
\qquad
\hat\beta_q=\hat\beta_A^{[-q]}(z;\lambda_N).
\]
Define
\[
g_{N,z}
=
\frac{1}{N}
\sum_{i=1}^N
\tilde X_iA_i(z),
\qquad
\hat g_{q,z}
=
\frac{1}{N_z^{[-q]}}
\sum_{i\in S^{[-q]}:Z_i=z}
\tilde X_i\hat A_i^{[-q]}.
\]
The population and sample LASSO criteria have the same quadratic term
$\Sigma_{X,N}$ and differ only through $g_{N,z}$ and $\hat g_{q,z}$.
Their optimality conditions and the subgradient inequality imply
\begin{align*}
\frac{1}{N}
\sum_{i=1}^N
\{\tilde X_i'(\hat\beta_q-\beta)\}^2
&=
(\hat\beta_q-\beta)'
\Sigma_{X,N}
(\hat\beta_q-\beta)
\\
&\le
2(\hat g_{q,z}-g_{N,z})'
(\hat\beta_q-\beta)
\\
&\le
2
\|\hat g_{q,z}-g_{N,z}\|_\infty
\|\hat\beta_q-\beta\|_1
\\
&\le
2
\|\hat g_{q,z}-g_{N,z}\|_\infty
\left(
\|\hat\beta_q\|_1+\|\beta\|_1
\right).
\end{align*}
The second inequality is H\"older's inequality, and the third is the
triangle inequality. Thus, it suffices to control the cross-moment
discrepancy and the $\ell_1$ norms of the two minimizers.

To control the $\ell_1$ sizes of the two minimizers, we compare their
objective values with the feasible choice $b=0$. Since a minimizer cannot
have a larger objective value than the unadjusted choice $b=0$, while the
penalty contributes $\lambda_N\|b\|_1$, this comparison will yield
\[
\|\beta\|_1+\max_q\|\hat\beta_q\|_1
=
O_p(\lambda_N^{-1}).
\]
Hence the preceding bound is $o_p(1)$ if
\[
\max_q
\|\hat g_{q,z}-g_{N,z}\|_\infty
=
o_p(\lambda_N).
\]
The factor $\lambda_N$ therefore appears because the penalty controls
the $\ell_1$ sizes of the minimizers at order $\lambda_N^{-1}$.
We establish these two bounds in turn.

First, consider the cross-moment discrepancy. Using
\[
\hat A_i^{[-q]}
=
A_i(z)-(\hat\tau^{[-q]}-\tau)D_i(z),
\]
we obtain
\begin{align*}
\|\hat g_{q,z}-g_{N,z}\|_\infty
&=
\left\|
\frac{1}{N_z^{[-q]}}
\sum_{i\in S^{[-q]}:Z_i=z}
\tilde X_i
\left\{
A_i(z)-(\hat\tau^{[-q]}-\tau)D_i(z)
\right\}
-
g_{N,z}
\right\|_\infty
\\
&=
\left\|
\left\{
\frac{1}{N_z^{[-q]}}
\sum_{i\in S^{[-q]}:Z_i=z}
\tilde X_iA_i(z)
-
g_{N,z}
\right\}
-
(\hat\tau^{[-q]}-\tau)
\frac{1}{N_z^{[-q]}}
\sum_{i\in S^{[-q]}:Z_i=z}
\tilde X_iD_i(z)
\right\|_\infty
\\
&\le
\left\|
\frac{1}{N_z^{[-q]}}
\sum_{i\in S^{[-q]}:Z_i=z}
\tilde X_iA_i(z)
-
g_{N,z}
\right\|_\infty
\\
&\quad+
|\hat\tau^{[-q]}-\tau|
\left\|
\frac{1}{N_z^{[-q]}}
\sum_{i\in S^{[-q]}:Z_i=z}
\tilde X_iD_i(z)
\right\|_\infty.
\end{align*}
The first equality substitutes the expression for
$\hat A_i^{[-q]}$, the second separates the terms involving
$A_i(z)$ and $D_i(z)$, and the inequality follows from the
triangle inequality.

For every $j\in\{1,\ldots,p_N\}$,
\begin{align*}
\left|
\frac{1}{N_z^{[-q]}}
\sum_{i\in S^{[-q]}:Z_i=z}
\tilde X_{ij}D_i(z)
\right|
&\le
\left\{
\frac{1}{N_z^{[-q]}}
\sum_{i\in S^{[-q]}:Z_i=z}
\tilde X_{ij}^2
\right\}^{1/2}
\left\{
\frac{1}{N_z^{[-q]}}
\sum_{i\in S^{[-q]}:Z_i=z}
D_i(z)^2
\right\}^{1/2}
\\
&\le
K_{X,N}
\left\{
\frac{1}{N_z^{[-q]}}
\sum_{i\in S^{[-q]}:Z_i=z}
D_i(z)^2
\right\}^{1/2}.
\end{align*}
The first inequality is Cauchy--Schwarz, and the second follows from
Assumption~\ref{ass:lasso_persistence}(i). Hence,
\begin{align*}
\max_q\|\hat g_{q,z}-g_{N,z}\|_\infty
&\le
\max_q
\left\|
\frac{1}{N_z^{[-q]}}
\sum_{i\in S^{[-q]}:Z_i=z}
\tilde X_iA_i(z)
-
g_{N,z}
\right\|_\infty
\\
&\quad+
K_{X,N}
\max_q|\hat\tau^{[-q]}-\tau|
\max_q
\left\{
\frac{1}{N_z^{[-q]}}
\sum_{i\in S^{[-q]}:Z_i=z}
D_i(z)^2
\right\}^{1/2}.
\end{align*}

For the first term on the right-hand side, similar to Step~1,
Lemma~\ref{lem:split_sampling} implies that each
$\{i\in S^{[-q]}:Z_i=z\}$ is marginally a simple random sample
without replacement from the finite population. Moreover,
Assumptions~\ref{ass:asymp}(i) and
\ref{ass:conditional_crossfit}(ii), together with fixed $Q$, imply
\[
\min_{q,z}\frac{N_z^{[-q]}}{N}\ge c>0.
\]
Assumption~\ref{ass:lasso_persistence}(i) gives
\[
|\tilde X_{ij}A_i(z)|
\le
K_{X,N}K_{A,N}.
\]
Therefore, Lemma~\ref{lem:finite_population_concentration}, applied
with
\[
v_{i,j}=\tilde X_{ij}A_i(z),
\qquad
H_N=K_{X,N}K_{A,N},
\qquad
J_N=p_N,
\]
gives
\[
\max_q
\left\|
\frac{1}{N_z^{[-q]}}
\sum_{i\in S^{[-q]}:Z_i=z}
\tilde X_iA_i(z)
-
g_{N,z}
\right\|_\infty
=
O_p\left\{
K_{X,N}K_{A,N}
\sqrt{\frac{\log(p_N+1)}{N}}
\right\}.
\]

For the second term, Lemma~\ref{lem:training_ratio_rate} gives
\[
\max_q
|\hat\tau^{[-q]}-\tau|
=
O_p(N^{-1/2}).
\]
In addition, Assumptions~\ref{ass:id} and
\ref{ass:asymp}(ii), together with
Lemma~\ref{lem:split_sampling},
Assumptions~\ref{ass:asymp}(i) and
\ref{ass:conditional_crossfit}(ii), imply the second-moment bound
established in Step~1:
\[
\max_q
\frac{1}{N_z^{[-q]}}
\sum_{i\in S^{[-q]}:Z_i=z}
D_i(z)^2
=
O(1).
\]
Consequently,
\begin{align*}
&K_{X,N}
\max_q|\hat\tau^{[-q]}-\tau|
\max_q
\left\{
\frac{1}{N_z^{[-q]}}
\sum_{i\in S^{[-q]}:Z_i=z}
D_i(z)^2
\right\}^{1/2}
\\
&\qquad=
O_p\left(\frac{K_{X,N}}{\sqrt N}\right).
\end{align*}
Combining the two bounds gives
\begin{align*}
\max_q
\|\hat g_{q,z}-g_{N,z}\|_\infty
&=
O_p\left[
K_{X,N}
\left\{
K_{A,N}
\sqrt{\frac{\log(p_N+1)}{N}}
+
\frac{1}{\sqrt N}
\right\}
\right]
\\
&=
o_p(\lambda_N),
\end{align*}
where the last equality follows from
Assumption~\ref{ass:lasso_persistence}(ii).

Next, we bound the $\ell_1$ norms of the population and sample
minimizers. For the population minimizer, optimality relative to the
feasible choice $b=0$ gives
\begin{align*}
\lambda_N\|\beta\|_1
&\le
\frac{1}{N}
\sum_{i=1}^N
\{A_i(z)-\bar A(z)-\tilde X_i'\beta\}^2
+
\lambda_N\|\beta\|_1
\\
&\le
\frac{1}{N}
\sum_{i=1}^N
\{A_i(z)-\bar A(z)\}^2
\\
&=
O(1).
\end{align*}
The first inequality uses nonnegativity of the squared loss, the
second follows because $\beta$ minimizes the population criterion and
$b=0$ is feasible, and the last equality follows from
Assumptions~\ref{ass:id} and \ref{ass:asymp}(ii). Hence,
\[
\|\beta\|_1
=
O(\lambda_N^{-1}).
\]

For the sample minimizer, optimality relative to $b=0$ gives
\begin{align*}
\lambda_N\|\hat\beta_q\|_1
&\le
\frac{2}{N_z^{[-q]}}
\sum_{i\in S^{[-q]}:Z_i=z}
\hat A_i^{[-q]}
\tilde X_i'\hat\beta_q
-
\frac{1}{N}
\sum_{i=1}^N
(\tilde X_i'\hat\beta_q)^2
\\
&\le
2
\left\{
\frac{N}{(N_z^{[-q]})^2}
\sum_{i\in S^{[-q]}:Z_i=z}
(\hat A_i^{[-q]})^2
\right\}^{1/2}
\left\{
\frac{1}{N}
\sum_{i=1}^N
(\tilde X_i'\hat\beta_q)^2
\right\}^{1/2}
\\
&\quad-
\frac{1}{N}
\sum_{i=1}^N
(\tilde X_i'\hat\beta_q)^2
\\
&\le
\frac{N}{(N_z^{[-q]})^2}
\sum_{i\in S^{[-q]}:Z_i=z}
(\hat A_i^{[-q]})^2
\\
&\le
\frac{2N}{(N_z^{[-q]})^2}
\sum_{i\in S^{[-q]}:Z_i=z}
\left\{
A_i(z)^2
+
(\hat\tau^{[-q]}-\tau)^2D_i(z)^2
\right\}.
\end{align*}
The first inequality follows from sample optimality relative to
$b=0$. The second follows from Cauchy--Schwarz and from bounding the
training-sample sum of $(\tilde X_i'\hat\beta_q)^2$ by its
full-population sum. The third uses $2uv-v^2\le u^2$, and the last uses
\[
\hat A_i^{[-q]}
=
A_i(z)-(\hat\tau^{[-q]}-\tau)D_i(z)
\]
and $(a-b)^2\le2a^2+2b^2$.

By the second-moment bound established in Step~1,
\[
\max_q
\frac{1}{N_z^{[-q]}}
\sum_{i\in S^{[-q]}:Z_i=z}
\{A_i(z)^2+D_i(z)^2\}
=
O(1),
\]
where this bound follows from Assumptions~\ref{ass:id} and
\ref{ass:asymp}(ii), Lemma~\ref{lem:split_sampling},
Assumptions~\ref{ass:asymp}(i) and
\ref{ass:conditional_crossfit}(ii), and fixed $Q$.
Moreover, Lemma~\ref{lem:training_ratio_rate} gives
\[
\max_q
|\hat\tau^{[-q]}-\tau|
=
O_p(N^{-1/2}),
\]
and the preceding lower bound on $N_z^{[-q]}/N$ gives
\[
\max_q\frac{N}{N_z^{[-q]}}
=
O(1).
\]
Therefore,
\[
\max_q
\lambda_N\|\hat\beta_q\|_1
=
O_p(1),
\]
and hence
\[
\max_q
\|\hat\beta_q\|_1
=
O_p(\lambda_N^{-1}).
\]
Together with the population bound,
\[
\|\beta\|_1
+
\max_q\|\hat\beta_q\|_1
=
O_p(\lambda_N^{-1}).
\]

We have now established both bounds required by the inequality at the
beginning of Step~2:
\[
\max_q
\|\hat g_{q,z}-g_{N,z}\|_\infty
=
o_p(\lambda_N),
\qquad
\|\beta\|_1+\max_q\|\hat\beta_q\|_1
=
O_p(\lambda_N^{-1}).
\]
Therefore,
\begin{align*}
\max_q
\frac{1}{N}
\sum_{i=1}^N
\{\tilde X_i'(\hat\beta_q-\beta)\}^2
&\le
2
\max_q
\|\hat g_{q,z}-g_{N,z}\|_\infty
\left(
\max_q\|\hat\beta_q\|_1+\|\beta\|_1
\right)
\\
&=
O_p\left[
\frac{K_{X,N}}{\lambda_N}
\left\{
K_{A,N}
\sqrt{\frac{\log(p_N+1)}{N}}
+
\frac{1}{\sqrt N}
\right\}
\right]
\\
&=
o_p(1),
\end{align*}
where the final equality follows from
Assumption~\ref{ass:lasso_persistence}(ii).
This proves the slope prediction component.

Combining Steps~1 and~2 with the initial decomposition yields
\[
\max_{1\le q\le Q}
\|\hat\mu_A^{[-q]}(z;\lambda_N)-\mu_A(z;\lambda_N)\|_N^2
=
o_p(1).
\]

Because the argument applies to both $z=0$ and $z=1$, let
\[
M_i
=
\bigl(
\mu_{A,i}(1;\lambda_N),
\mu_{A,i}(0;\lambda_N)
\bigr)',
\]
and, for $i\in S^{[q]}$, define
\[
\hat M_i^{[-q]}
=
\left(
\hat\mu_{A,i}^{[-q]}(1;\lambda_N),
\hat\mu_{A,i}^{[-q]}(0;\lambda_N)
\right)'.
\]
Then
\begin{align*}
\max_q
\frac{1}{N^{[q]}}
\sum_{i\in S^{[q]}}
\|\hat M_i^{[-q]}-M_i\|_2^2
&\le
\max_q\frac{N}{N^{[q]}}
\sum_{z=0}^1
\max_q
\|\hat\mu_A^{[-q]}(z;\lambda_N)
-\mu_A(z;\lambda_N)\|_N^2
\\
&=
o_p(1),
\end{align*}
where $N/N^{[q]}=O(1)$ uniformly over $q$ by
Assumption~\ref{ass:conditional_crossfit}(ii).
This verifies Assumption~\ref{ass:conditional_crossfit}(iii).
\end{proof}

\subsection{Proof of Theorem~\ref{thm:lasso_cv}}
\label{app:proof_lasso_cv}

\begin{proof}
Fix $z\in\{0,1\}$. We first reduce the desired prediction stability
to two components. For every $q$,
\begin{align*}
&\|\hat\mu_A^{[-q]}(z;\hat\lambda_{z,N}^{[-q]})
-\mu_A(z;\lambda_N)\|_N^2
\\
&\quad\le
2
\|\hat\mu_A^{[-q]}(z;\hat\lambda_{z,N}^{[-q]})
-\mu_A(z;\hat\lambda_{z,N}^{[-q]})\|_N^2
+
2
\|\mu_A(z;\hat\lambda_{z,N}^{[-q]})
-\mu_A(z;\lambda_N)\|_N^2.
\end{align*}
Thus, it suffices to show that both terms are $o_p(1)$ uniformly over
$q$.

Because Assumption~\ref{ass:lasso_cv_grid}(ii) implies
$\hat\lambda_{z,N}^{[-q]}\ge\lambda_N$, the first term is controlled
by the uniform stability result in
Lemma~\ref{lem:lasso_cv_uniform_stability}. For the second term,
Lemma~\ref{lem:lasso_path_geometry} gives, for each $q$,
\begin{align*}
&\|\mu_A(z;\hat\lambda_{z,N}^{[-q]})
-\mu_A(z;\lambda_N)\|_N^2
\\
&\quad\le
L_{N,z}(\hat\lambda_{z,N}^{[-q]})
-
L_{N,z}(\lambda_N).
\end{align*}
Since the theorem requires prediction stability uniformly over the
fixed number of outer folds, it is therefore enough to show that
\[
\max_q
\left\{
L_{N,z}(\hat\lambda_{z,N}^{[-q]})
-
L_{N,z}(\lambda_N)
\right\}
=
o_p(1).
\]

To establish this risk bound, we first show that the cross-validation
criterion uniformly approximates $L_{N,z}(\lambda)$ up to terms that
do not depend on $\lambda$. Approximate CV optimality then transfers
this uniform approximation to the selected penalty. Accordingly, the
remainder of the proof has two steps: Step~1 establishes the uniform
approximation of the CV criterion, and Step~2 derives the population
risk and prediction bounds for the selected penalty.

\textbf{Step 1: Uniform approximation of the cross-validation criterion.}
\leavevmode\par
We compare the cross-validation criterion with the population prediction
risk after removing terms that do not depend on $\lambda$. Since the
validation folds partition the arm-specific outer training sample,
\begin{align*}
\widehat{\operatorname{CV}}_{z}^{[-q]}(\lambda)
&=
\frac{1}{N_z^{[-q]}}
\sum_{i\in S^{[-q]}:Z_i=z}
(\hat A_i^{[-q]})^2
\\
&\quad+
\frac{1}{N_z^{[-q]}}
\sum_{k=1}^{K_{\mathrm{cv}}}
\sum_{i\in V_{z,k}^{[-q]}}
\left[
\hat\mu_{A,i}^{[-q,-k]}(z;\lambda)^2
-
2\hat A_i^{[-q]}
\hat\mu_{A,i}^{[-q,-k]}(z;\lambda)
\right].
\end{align*}
Similarly,
\begin{align*}
L_{N,z}(\lambda)
&=
\frac{1}{N}\sum_{i=1}^N A_i(z)^2
\\
&\quad+
\frac{1}{N}\sum_{i=1}^N
\left[
\mu_{A,i}(z;\lambda)^2
-
2A_i(z)\mu_{A,i}(z;\lambda)
\right].
\end{align*}
The first term in each display is independent of $\lambda$. Therefore,
to compare the two criteria uniformly over the penalty grid, it suffices
to show
\begin{align*}
\varepsilon_{N,z}
&=
\max_q\sup_{\lambda\in\Lambda_N}
\Bigg|
\widehat{\operatorname{CV}}_{z}^{[-q]}(\lambda)
-
\frac{1}{N_z^{[-q]}}
\sum_{i\in S^{[-q]}:Z_i=z}
(\hat A_i^{[-q]})^2
-
L_{N,z}(\lambda)
+
\frac{1}{N}\sum_{i=1}^N A_i(z)^2
\Bigg|
=
o_p(1).
\end{align*}

Consider one validation fold $V_{z,k}^{[-q]}$. By the definition of
$L_{N,z}(\lambda)$,
\begin{align*}
&
\frac{1}{N_{z,k}^{[-q]}}
\sum_{i\in V_{z,k}^{[-q]}}
\left[
\left\{
\hat A_i^{[-q]}
-
\hat\mu_{A,i}^{[-q,-k]}(z;\lambda)
\right\}^2
-
(\hat A_i^{[-q]})^2
\right]
-
L_{N,z}(\lambda)
+
\frac{1}{N}\sum_{i=1}^N A_i(z)^2
\\
&=
\frac{1}{N_{z,k}^{[-q]}}
\sum_{i\in V_{z,k}^{[-q]}}
\left[
\hat\mu_{A,i}^{[-q,-k]}(z;\lambda)^2
-
2\hat A_i^{[-q]}
\hat\mu_{A,i}^{[-q,-k]}(z;\lambda)
\right]
\\
&\quad-
\frac{1}{N}\sum_{i=1}^N
\left[
\mu_{A,i}(z;\lambda)^2
-
2A_i(z)\mu_{A,i}(z;\lambda)
\right]
\\
&=
\left\{
\frac{1}{N_{z,k}^{[-q]}}
\sum_{i\in V_{z,k}^{[-q]}}
\mu_{A,i}(z;\lambda)^2
-
\frac{1}{N}\sum_{i=1}^N
\mu_{A,i}(z;\lambda)^2
\right\}
\\
&\quad-
2
\left\{
\frac{1}{N_{z,k}^{[-q]}}
\sum_{i\in V_{z,k}^{[-q]}}
A_i(z)\mu_{A,i}(z;\lambda)
-
\frac{1}{N}\sum_{i=1}^N
A_i(z)\mu_{A,i}(z;\lambda)
\right\}
\\
&\quad+
\frac{1}{N_{z,k}^{[-q]}}
\sum_{i\in V_{z,k}^{[-q]}}
\left\{
\hat\mu_{A,i}^{[-q,-k]}(z;\lambda)^2
-
\mu_{A,i}(z;\lambda)^2
\right\}
\\
&\quad-
\frac{2}{N_{z,k}^{[-q]}}
\sum_{i\in V_{z,k}^{[-q]}}
A_i(z)
\left\{
\hat\mu_{A,i}^{[-q,-k]}(z;\lambda)
-
\mu_{A,i}(z;\lambda)
\right\}
\\
&\quad+
\frac{2(\hat\tau^{[-q]}-\tau)}
{N_{z,k}^{[-q]}}
\sum_{i\in V_{z,k}^{[-q]}}
D_i(z)
\hat\mu_{A,i}^{[-q,-k]}(z;\lambda).
\end{align*}
The first equality expands the squared validation loss and uses the
definition of $L_{N,z}(\lambda)$. The second substitutes
\[
\hat A_i^{[-q]}
=
A_i(z)-(\hat\tau^{[-q]}-\tau)D_i(z)
\]
and adds and subtracts the corresponding population quantities.
It therefore suffices to show that the five terms on the right-hand
side are uniformly $o_p(1)$.

For the first term,
Lemma~\ref{lem:lasso_validation_concentration} directly gives
\[
\max_{q,k}\sup_{\lambda\in\Lambda_N}
\left|
\frac{1}{N_{z,k}^{[-q]}}
\sum_{i\in V_{z,k}^{[-q]}}
\mu_{A,i}(z;\lambda)^2
-
\frac{1}{N}\sum_{i=1}^N
\mu_{A,i}(z;\lambda)^2
\right|
=
o_p(1).
\]

For the second term, using
\[
\mu_{A,i}(z;\lambda)
=
\bar A(z)+\tilde X_i'\beta_A(z;\lambda),
\]
we obtain
\begin{align*}
&
\frac{1}{N_{z,k}^{[-q]}}
\sum_{i\in V_{z,k}^{[-q]}}
A_i(z)\mu_{A,i}(z;\lambda)
-
\frac{1}{N}\sum_{i=1}^N
A_i(z)\mu_{A,i}(z;\lambda)
\\
&=
\bar A(z)
\left\{
\frac{1}{N_{z,k}^{[-q]}}
\sum_{i\in V_{z,k}^{[-q]}}
A_i(z)
-
\bar A(z)
\right\}+
\beta_A(z;\lambda)'
\left\{
\frac{1}{N_{z,k}^{[-q]}}
\sum_{i\in V_{z,k}^{[-q]}}
\tilde X_iA_i(z)
-
\frac{1}{N}
\sum_{i=1}^N
\tilde X_iA_i(z)
\right\}.
\end{align*}

By Lemma~\ref{lem:split_sampling}, each $V_{z,k}^{[-q]}$ is
marginally a simple random sample without replacement of size
$N_{z,k}^{[-q]}$. Hence,
\begin{align*}
\mathbb E\left[
\frac{1}{N_{z,k}^{[-q]}}
\sum_{i\in V_{z,k}^{[-q]}}A_i(z)
\right]
&=
\bar A(z),
\\
\operatorname{Var}\left(
\frac{1}{N_{z,k}^{[-q]}}
\sum_{i\in V_{z,k}^{[-q]}}A_i(z)
\right)
&=
\left(
\frac{1}{N_{z,k}^{[-q]}}
-
\frac{1}{N}
\right)
S_{A(z)}^2
\\
&=
O(N^{-1}),
\end{align*}
where $S_{A(z)}^2=O(1)$ follows from
Assumptions~\ref{ass:id} and \ref{ass:asymp}(ii), and
$N_{z,k}^{[-q]}\ge c_{\mathrm{cv}}N$ follows from
Assumption~\ref{ass:lasso_cv_grid}(i).
Since $Q$ and $K_{\mathrm{cv}}$ are fixed, Chebyshev's inequality
and the union bound imply
\[
\max_{q,k}
\left|
\frac{1}{N_{z,k}^{[-q]}}
\sum_{i\in V_{z,k}^{[-q]}}A_i(z)
-
\bar A(z)
\right|
=
O_p(N^{-1/2}).
\]
Assumptions~\ref{ass:id} and \ref{ass:asymp}(ii) also imply
$\bar A(z)=O(1)$.

Moreover, Lemma~\ref{lem:lasso_cv_uniform_stability} gives
\[
\sup_{\lambda\ge\lambda_N}
\|\beta_A(z;\lambda)\|_1
=
O(\lambda_N^{-1}),
\]
and Lemma~\ref{lem:finite_population_concentration}, applied with
\[
v_{i,j}=\tilde X_{ij}A_i(z),
\qquad
H_N=K_{X,N}K_{A,N},
\qquad
J_N=p_N,
\]
gives
\[
\max_{q,k}
\left\|
\frac{1}{N_{z,k}^{[-q]}}
\sum_{i\in V_{z,k}^{[-q]}}
\tilde X_iA_i(z)
-
\frac{1}{N}
\sum_{i=1}^N
\tilde X_iA_i(z)
\right\|_\infty
=
O_p\left\{
K_{X,N}K_{A,N}
\sqrt{\frac{\log(p_N+1)}{N}}
\right\}.
\]
Therefore,
\begin{align*}
&\max_{q,k}\sup_{\lambda\in\Lambda_N}
\left|
\frac{1}{N_{z,k}^{[-q]}}
\sum_{i\in V_{z,k}^{[-q]}}
A_i(z)\mu_{A,i}(z;\lambda)
-
\frac{1}{N}\sum_{i=1}^N
A_i(z)\mu_{A,i}(z;\lambda)
\right|
\\
&=
O_p\left[
\frac{1}{\sqrt N}
+
\frac{K_{X,N}K_{A,N}}{\lambda_N}
\sqrt{\frac{\log(p_N+1)}{N}}
\right]
\\
&=
o_p(1),
\end{align*}
where the last equality follows from
Assumption~\ref{ass:lasso_persistence}(ii).

For the third term, Assumption~\ref{ass:lasso_cv_grid}(i) implies
$N_{z,k}^{[-q]}\ge c_{\mathrm{cv}}N$. Therefore,
\begin{align*}
&
\left|
\frac{1}{N_{z,k}^{[-q]}}
\sum_{i\in V_{z,k}^{[-q]}}
\left\{
\hat\mu_{A,i}^{[-q,-k]}(z;\lambda)^2
-
\mu_{A,i}(z;\lambda)^2
\right\}
\right|
\\
&\le
\left\{
\frac{1}{N_{z,k}^{[-q]}}
\sum_{i\in V_{z,k}^{[-q]}}
\left\{
\hat\mu_{A,i}^{[-q,-k]}(z;\lambda)
-
\mu_{A,i}(z;\lambda)
\right\}^2
\right\}^{1/2}
\\
&\quad\times
\left[
\left\{
\frac{1}{N_{z,k}^{[-q]}}
\sum_{i\in V_{z,k}^{[-q]}}
\hat\mu_{A,i}^{[-q,-k]}(z;\lambda)^2
\right\}^{1/2}
+
\left\{
\frac{1}{N_{z,k}^{[-q]}}
\sum_{i\in V_{z,k}^{[-q]}}
\mu_{A,i}(z;\lambda)^2
\right\}^{1/2}
\right]
\\
&\le
c_{\mathrm{cv}}^{-1}
\|\hat\mu_A^{[-q,-k]}(z;\lambda)-\mu_A(z;\lambda)\|_N
\\
&\quad\times
\left\{
\|\hat\mu_A^{[-q,-k]}(z;\lambda)\|_N
+
\|\mu_A(z;\lambda)\|_N
\right\}
\\
&=
o_p(1)
\end{align*}
uniformly over $q,k,\lambda\in\Lambda_N$, where the first inequality
is Cauchy--Schwarz and the final equality follows from
Lemma~\ref{lem:lasso_cv_uniform_stability}.

For the fourth term, Cauchy--Schwarz gives
\begin{align*}
&
\left|
\frac{1}{N_{z,k}^{[-q]}}
\sum_{i\in V_{z,k}^{[-q]}}
A_i(z)
\left\{
\hat\mu_{A,i}^{[-q,-k]}(z;\lambda)
-
\mu_{A,i}(z;\lambda)
\right\}
\right|
\\
&\le
\left\{
\frac{1}{N_{z,k}^{[-q]}}
\sum_{i\in V_{z,k}^{[-q]}}
A_i(z)^2
\right\}^{1/2}
\times
\left\{
\frac{1}{N_{z,k}^{[-q]}}
\sum_{i\in V_{z,k}^{[-q]}}
\left\{
\hat\mu_{A,i}^{[-q,-k]}(z;\lambda)
-
\mu_{A,i}(z;\lambda)
\right\}^2
\right\}^{1/2}
\\
&\le
c_{\mathrm{cv}}^{-1/2}
\left\{
\frac{1}{N_{z,k}^{[-q]}}
\sum_{i\in V_{z,k}^{[-q]}}
A_i(z)^2
\right\}^{1/2}
\|\hat\mu_A^{[-q,-k]}(z;\lambda)-\mu_A(z;\lambda)\|_N
\\
&=
o_p(1)
\end{align*}
uniformly over $q,k,\lambda\in\Lambda_N$.
Here,
\begin{align*}
\max_{q,k}
\frac{1}{N_{z,k}^{[-q]}}
\sum_{i\in V_{z,k}^{[-q]}}A_i(z)^2
&\le
\frac{N}{\min_{q,k}N_{z,k}^{[-q]}}
\frac{1}{N}\sum_{i=1}^N A_i(z)^2
\\
&=
O(1)
\end{align*}
by Assumptions~\ref{ass:id}, \ref{ass:asymp}(ii), and
\ref{ass:lasso_cv_grid}(i).

For the fifth term, Cauchy--Schwarz gives
\begin{align*}
&
|\hat\tau^{[-q]}-\tau|
\left|
\frac{1}{N_{z,k}^{[-q]}}
\sum_{i\in V_{z,k}^{[-q]}}
D_i(z)
\hat\mu_{A,i}^{[-q,-k]}(z;\lambda)
\right|
\\
&\le
|\hat\tau^{[-q]}-\tau|
\left\{
\frac{1}{N_{z,k}^{[-q]}}
\sum_{i\in V_{z,k}^{[-q]}}D_i(z)^2
\right\}^{1/2}
\left\{
\frac{1}{N_{z,k}^{[-q]}}
\sum_{i\in V_{z,k}^{[-q]}}
\hat\mu_{A,i}^{[-q,-k]}(z;\lambda)^2
\right\}^{1/2}
\\
&\le
c_{\mathrm{cv}}^{-1/2}
|\hat\tau^{[-q]}-\tau|
\left\{
\frac{1}{N_{z,k}^{[-q]}}
\sum_{i\in V_{z,k}^{[-q]}}D_i(z)^2
\right\}^{1/2}
\|\hat\mu_A^{[-q,-k]}(z;\lambda)\|_N
\\
&=
o_p(1)
\end{align*}
uniformly over $q,k,\lambda\in\Lambda_N$.
The last equality follows from
Lemma~\ref{lem:training_ratio_rate},
Lemma~\ref{lem:lasso_cv_uniform_stability}, and
\begin{align*}
\max_{q,k}
\frac{1}{N_{z,k}^{[-q]}}
\sum_{i\in V_{z,k}^{[-q]}}D_i(z)^2
&\le
\frac{N}{\min_{q,k}N_{z,k}^{[-q]}}
\frac{1}{N}\sum_{i=1}^N D_i(z)^2
\\
&=
O(1)
\end{align*}
by Assumption~\ref{ass:asymp}(ii) and
Assumption~\ref{ass:lasso_cv_grid}(i).

Thus the foldwise approximation error is uniformly $o_p(1)$.
Because the validation folds partition the arm-specific outer
training sample,
\begin{align*}
\widehat{\operatorname{CV}}_{z}^{[-q]}(\lambda)
&=
\sum_{k=1}^{K_{\mathrm{cv}}}
\frac{N_{z,k}^{[-q]}}{N_z^{[-q]}}
\frac{1}{N_{z,k}^{[-q]}}
\sum_{i\in V_{z,k}^{[-q]}}
\left\{
\hat A_i^{[-q]}
-
\hat\mu_{A,i}^{[-q,-k]}(z;\lambda)
\right\}^2,
\\
\frac{1}{N_z^{[-q]}}
\sum_{i\in S^{[-q]}:Z_i=z}
(\hat A_i^{[-q]})^2
&=
\sum_{k=1}^{K_{\mathrm{cv}}}
\frac{N_{z,k}^{[-q]}}{N_z^{[-q]}}
\frac{1}{N_{z,k}^{[-q]}}
\sum_{i\in V_{z,k}^{[-q]}}
(\hat A_i^{[-q]})^2,
\\
\sum_{k=1}^{K_{\mathrm{cv}}}
\frac{N_{z,k}^{[-q]}}{N_z^{[-q]}}
&=
1.
\end{align*}
Since $K_{\mathrm{cv}}$ is fixed, the preceding foldwise bounds imply
\[
\varepsilon_{N,z}
=
o_p(1).
\]

\textbf{Step 2: Population risk of the selected penalty.}
\leavevmode\par
By Assumption~\ref{ass:lasso_cv_grid}(ii),
\[
\hat\lambda_{z,N}^{[-q]}
\in\Lambda_N
\subset[\lambda_N,\infty),
\]
so
\[
\hat\lambda_{z,N}^{[-q]}
\ge
\lambda_N.
\]
Lemma~\ref{lem:lasso_path_geometry} therefore implies
\[
L_{N,z}(\hat\lambda_{z,N}^{[-q]})
\ge
L_{N,z}(\lambda_N).
\]
It remains to establish the reverse inequality up to $o_p(1)$.

By the definition of $\varepsilon_{N,z}$ and the approximate
cross-validation optimality condition in
Theorem~\ref{thm:lasso_cv},
\begin{align*}
L_{N,z}(\hat\lambda_{z,N}^{[-q]})
&\le
\widehat{\operatorname{CV}}_{z}^{[-q]}
(\hat\lambda_{z,N}^{[-q]})
-
\frac{1}{N_z^{[-q]}}
\sum_{i\in S^{[-q]}:Z_i=z}
(\hat A_i^{[-q]})^2
+
\frac{1}{N}\sum_{i=1}^NA_i(z)^2
+
\varepsilon_{N,z}
\\
&\le
\widehat{\operatorname{CV}}_{z}^{[-q]}(\lambda_N)
-
\frac{1}{N_z^{[-q]}}
\sum_{i\in S^{[-q]}:Z_i=z}
(\hat A_i^{[-q]})^2
+
\frac{1}{N}\sum_{i=1}^NA_i(z)^2
+
\varepsilon_{N,z}
+
\xi_{z,N}^{[-q]}
\\
&\le
L_{N,z}(\lambda_N)
+
2\varepsilon_{N,z}
+
\xi_{z,N}^{[-q]}.
\end{align*}
The first and third inequalities use the uniform approximation from
Step~1, and the second uses the approximate CV optimality condition.
Consequently,
\begin{align*}
0
&\le
\max_q
\left\{
L_{N,z}(\hat\lambda_{z,N}^{[-q]})
-
L_{N,z}(\lambda_N)
\right\}
\\
&\le
2\varepsilon_{N,z}
+
\max_q\xi_{z,N}^{[-q]}
\\
&=
o_p(1).
\end{align*}

Because $\hat\lambda_{z,N}^{[-q]}\ge\lambda_N$,
Lemma~\ref{lem:lasso_path_geometry} further gives
\begin{align*}
\max_q
\|\mu_A(z;\hat\lambda_{z,N}^{[-q]})
-
\mu_A(z;\lambda_N)\|_N^2
&\le
\max_q
\left\{
L_{N,z}(\hat\lambda_{z,N}^{[-q]})
-
L_{N,z}(\lambda_N)
\right\}
\\
&=
o_p(1).
\end{align*}

Combining this result with the initial decomposition and
Lemma~\ref{lem:lasso_cv_uniform_stability} yields
\[
\max_q
\|\hat\mu_A^{[-q]}(z;\hat\lambda_{z,N}^{[-q]})
-
\mu_A(z;\lambda_N)\|_N^2
=
o_p(1).
\]

Because the argument applies to both $z=0$ and $z=1$, let
\[
M_i
=
\bigl(
\mu_{A,i}(1;\lambda_N),
\mu_{A,i}(0;\lambda_N)
\bigr)'.
\]
For $i\in S^{[q]}$, define
\[
\hat M_i^{[-q]}
=
\left(
\hat\mu_{A,i}^{[-q]}(1;\hat\lambda_{1,N}^{[-q]}),
\hat\mu_{A,i}^{[-q]}(0;\hat\lambda_{0,N}^{[-q]})
\right)'.
\]
Then
\begin{align*}
\max_q
\frac{1}{N^{[q]}}
\sum_{i\in S^{[q]}}
\|\hat M_i^{[-q]}-M_i\|_2^2
&\le
\max_q\frac{N}{N^{[q]}}
\sum_{z=0}^1
\max_q
\|\hat\mu_A^{[-q]}(z;\hat\lambda_{z,N}^{[-q]})
-
\mu_A(z;\lambda_N)\|_N^2
\\
&=
o_p(1),
\end{align*}
which verifies
Assumption~\ref{ass:conditional_crossfit}(iii).
\end{proof}

\subsection{Proof of Corollary~\ref{cor:ils_cob_two_step}}
\label{app:proof_cor_ils_cob_two_step}

\begin{proof}
For each $G\in\{X,W,G^+\}$, use the definitions of $\delta_z$ and $A_{G,i}(z;\delta_z)$ in Section~\ref{sec:min_var} with that choice of $G$. Write $\hat G_{i,\mathrm{cf}}$ for $X_i$, $\hat W_{i,\mathrm{cf}}$, or $(X_i',\hat W_{i,\mathrm{cf}}',\hat M_{i,\mathrm{cf}}')'$, respectively. The corresponding adjusted ratios are $\hat\tau_{\mathrm{ils}}$, $\hat\tau_{\mathrm{cob,cf}}$, and $\hat\tau_{2s,\mathrm{extra}}$, with the last identity following from Lemma~\ref{lem:adjusted_ratio_reduction}.

For $G=W$ or $G=G^+$, Assumption~\ref{ass:conditional_crossfit}(iii) for $G^+$ implies the required prediction stability for $G$, and Lemma~\ref{lem:cf_oracle_equivalence} applies under the corresponding moment and rank conditions. For $G=X$, the covariates are fixed, so simple-random-sampling moment convergence and the positive-definiteness condition in Assumption~\ref{ass:two_step_asymptotics}(ii) give $\hat\gamma_R(z)=\gamma_R(z)+o_p(1)$ for $R\in\{Y,D\}$; also $\bar X_z-\bar X=O_p(N^{-1/2})$. Thus the same first-order expansion follows directly for ILS, without cross-fitting. In all three cases, Assumptions~\ref{ass:id} and \ref{ass:asymp}(ii) give $\tau=O(1)$ and $|\tau_D|\ge C$, and the adjusted denominator converges in probability to $\tau_D$. Hence
\[
\sqrt N(\hat\tau_{\hat G_{\mathrm{cf}}}-\tau)
=
\frac{\sqrt N}{\tau_D}
\left[
\frac{1}{N_1}\sum_{i:Z_i=1}A_{G,i}(1;\delta_1)
-
\frac{1}{N_0}\sum_{i:Z_i=0}A_{G,i}(0;\delta_0)
\right]
+
o_p(1).
\]

Under the stated substitutions, Assumption~\ref{ass:two_step_asymptotics}(i) gives the maximal condition, Assumption~\ref{ass:two_step_asymptotics}(ii) gives convergence of the residual covariances, and Assumption~\ref{ass:two_step_asymptotics}(iii) gives a positive limiting variance. Together with Assumption~\ref{ass:asymp}(i), these conditions allow the finite-population central limit theorem to apply to each leading term after division by $\tau_D$. This proves consistency and asymptotic normality, including the asserted result for $\hat\tau_{2s,\mathrm{extra}}$.

The centered spans of $X_i$ and $W_i$ are contained in the centered span of $G_i^+$. Lemma~\ref{lem:projection_geometry} and Theorem~\ref{thm:late_oracle_variance_min}(ii) therefore give
\[
V_{2s,\mathrm{extra}}
\le
V_{\mathrm{ils}},
\qquad
V_{2s,\mathrm{extra}}
\le
V_{\mathrm{cob,cf}}.
\]
Zero adjustment belongs to both the $X_i$ and $W_i$ spaces, so
\[
V_{\mathrm{ils}}\le V,
\qquad
V_{\mathrm{cob,cf}}\le V.
\]
\end{proof}

\section{Lemmas}
\label{app:lemma}

\subsection{Lemma~\ref{lem:cf_moment_stability}: Cross-fitted moment stability}

\begin{lemma}[Cross-fitted moment stability]
\label{lem:cf_moment_stability}
Let \(G_i\in\mathbb R^k\) be deterministic and fixed dimensional. Conditional on \(\mathcal S\), require each \(\hat G^{[-q]}\) to depend only on the observed training data in \(S^{[-q]}\), the fixed covariates, and any auxiliary randomization independent of treatment assignment and sample splitting. For \(i\in S^{[q]}\), set
\[
\hat G_{i,\mathrm{cf}}=\hat G_i^{[-q]}.
\]
Suppose Assumption~\ref{ass:conditional_crossfit}(i)--(ii) holds, the treatment fractions are bounded away from zero, and
\[
\max_{1\le q\le Q}
\left\{
\frac{1}{N^{[q]}}
\sum_{i\in S^{[q]}}
\left\|
\hat G_i^{[-q]}-G_i
\right\|_2^2
\right\}^{1/2}
\xrightarrow{p}0.
\]
Suppose further that
\[
\max_{1\le i\le N}
\frac{\|G_i-\bar G\|_2^2}{N}
\to0,
\qquad
\|S_G^2\|_{\mathrm{op}}=O(1).
\]
For a scalar response with potential values \(R_i(z)\), assume, for \(z\in\{0,1\}\),
\[
S^2_{R(z)}=O(1),
\qquad
\|S_{G,R(z)}\|_2=O(1).
\]
Let \(\widehat S_{\hat G,z}^{\,2}\) and \(\widehat S_{\hat G,R,z}\) denote the arm-\(z\) sample covariance matrix of \(\hat G_{i,\mathrm{cf}}\) and its sample covariance with \(R_i\), respectively, both centered within arm and normalized by \(N_z-1\). Then
\begin{align*}
\bar{\hat G}_{z,\mathrm{cf}}-\bar{\hat G}_{\mathrm{cf}}
&=
\bar G_z-\bar G+o_p(N^{-1/2}),
\\
\bar G_z-\bar G
&=
O_p(N^{-1/2}),
\\
\frac{1}{N_z}
\sum_{i:Z_i=z}
\|\hat G_{i,\mathrm{cf}}-G_i\|_2^2
&=
o_p(1),
\\
\widehat S_{\hat G,z}^{\,2}
&=
S_G^2+o_p(1),
\\
\widehat S_{\hat G,R,z}
&=
S_{G,R(z)}+o_p(1).
\end{align*}
Also,
\[
\bar{\hat G}_{\mathrm{cf}}-\bar G=o_p(1),
\qquad
\bar{\hat G}_{z,\mathrm{cf}}-\bar G=o_p(1).
\]
\end{lemma}

\begin{proof}
\leavevmode\par
\textbf{Step 1: Cross-fitted mean imbalance.}
\leavevmode\par
For \(i\in S^{[q]}\), define
\[
\Delta_i
=
\hat G_i^{[-q]}-G_i.
\]
For each fold and arm, let
\[
\bar\Delta^{[q]}
=
\frac{1}{N^{[q]}}
\sum_{i\in S^{[q]}}\Delta_i,
\qquad
\bar\Delta_z^{[q]}
=
\frac{1}{N_z^{[q]}}
\sum_{i\in S^{[q]}:Z_i=z}\Delta_i.
\]
The generated-covariate imbalance satisfies the exact decomposition
\begin{align}
&
\left(
\bar{\hat G}_{z,\mathrm{cf}}
-
\bar{\hat G}_{\mathrm{cf}}
\right)
-
\left(
\bar G_z-\bar G
\right)
\notag\\
&=
\sum_{q=1}^Q
\frac{N_z^{[q]}}{N_z}
\left(
\bar\Delta_z^{[q]}-\bar\Delta^{[q]}
\right)
+
\sum_{q=1}^Q
\left(
\frac{N_z^{[q]}}{N_z}
-
\frac{N^{[q]}}{N}
\right)
\bar\Delta^{[q]}.
\label{eq:cf-moment-mean-decomposition}
\end{align}

Conditional on the split, the training data, and any auxiliary
randomization used to construct \(\hat G^{[-q]}\), the vectors
\(\{\Delta_i:i\in S^{[q]}\}\) are fixed with respect to the
assignment in fold \(q\). By Assumption~\ref{ass:conditional_crossfit}(i),
the units assigned to arm \(z\) within \(S^{[q]}\) form a simple random
sample of fixed size \(N_z^{[q]}\). The finite-population variance formula
therefore gives
\[
\bar\Delta_z^{[q]}-\bar\Delta^{[q]}
=
O_p\left[
\frac{1}{\sqrt{N^{[q]}}}
\left\{
\frac{1}{N^{[q]}}
\sum_{i\in S^{[q]}}
\|\Delta_i\|_2^2
\right\}^{1/2}
\right].
\]
The stability condition in Lemma~\ref{lem:cf_moment_stability} and the lower bound on $N^{[q]}/N$ in Assumption~\ref{ass:conditional_crossfit}(ii) imply, uniformly over the fixed number of folds,
\[
\bar\Delta_z^{[q]}-\bar\Delta^{[q]}
=
o_p(N^{-1/2}).
\]
Hence the first sum in
\eqref{eq:cf-moment-mean-decomposition} is \(o_p(N^{-1/2})\).

For the second sum, the Cauchy--Schwarz inequality gives
\[
\|\bar\Delta^{[q]}\|_2
\le
\left\{
\frac{1}{N^{[q]}}
\sum_{i\in S^{[q]}}
\|\Delta_i\|_2^2
\right\}^{1/2}
=
o_p(1)
\]
uniformly over \(q\). Moreover, Assumption~\ref{ass:conditional_crossfit}(ii)
and the treatment-fraction bound imply
\[
\max_{q,z}
\left|
\frac{N_z^{[q]}}{N_z}
-
\frac{N^{[q]}}{N}
\right|
=
O(N^{-1/2}).
\]
Since \(Q\) is fixed, the second sum in
\eqref{eq:cf-moment-mean-decomposition} is also \(o_p(N^{-1/2})\).
Therefore,
\begin{equation}
\bar{\hat G}_{z,\mathrm{cf}}
-
\bar{\hat G}_{\mathrm{cf}}
=
\bar G_z-\bar G+o_p(N^{-1/2}).
\label{eq:cf-moment-mean-equivalence}
\end{equation}

The finite-population variance formula and
\(\|S_G^2\|_{\mathrm{op}}=O(1)\) imply
\[
\bar G_z-\bar G
=
O_p(N^{-1/2}).
\]
Also,
\begin{align*}
\|\bar{\hat G}_{\mathrm{cf}}-\bar G\|_2
&=
\left\|
\sum_{q=1}^Q
\frac{N^{[q]}}{N}\bar\Delta^{[q]}
\right\|_2
\\
&\le
\sum_{q=1}^Q
\frac{N^{[q]}}{N}
\left\{
\frac{1}{N^{[q]}}
\sum_{i\in S^{[q]}}
\|\Delta_i\|_2^2
\right\}^{1/2}
\\
&=
o_p(1).
\end{align*}
Together with \eqref{eq:cf-moment-mean-equivalence}, this yields
\[
\bar{\hat G}_{z,\mathrm{cf}}-\bar G
=
o_p(1).
\]

\leavevmode\par
\textbf{Step 2: Pooled empirical \(L_2\) stability.}
\leavevmode\par
Using the stability and treatment-fraction bounds in Lemma~\ref{lem:cf_moment_stability}, for each \(z\in\{0,1\}\),
\begin{align*}
\frac{1}{N_z}
\sum_{i:Z_i=z}
\|\hat G_{i,\mathrm{cf}}-G_i\|_2^2
&\le
\frac{1}{N_z}
\sum_{q=1}^Q
\sum_{i\in S^{[q]}}
\|\hat G_i^{[-q]}-G_i\|_2^2
\\
&=
\sum_{q=1}^Q
\frac{N^{[q]}}{N_z}
\left\{
\frac{1}{N^{[q]}}
\sum_{i\in S^{[q]}}
\|\hat G_i^{[-q]}-G_i\|_2^2
\right\}
\\
&\le
\frac{N}{N_z}
\max_{1\le q\le Q}
\left\{
\frac{1}{N^{[q]}}
\sum_{i\in S^{[q]}}
\|\hat G_i^{[-q]}-G_i\|_2^2
\right\}
\\
&=
o_p(1).
\end{align*}

\leavevmode\par
\textbf{Step 3: Oracle moment convergence.}
\leavevmode\par
We first consider the oracle adjustment vector \(G_i\). For each pair of
coordinates \(j,\ell\),
\begin{align*}
&
\frac{1}{N^2}
\sum_{i=1}^N
\left\{
(G_{ij}-\bar G_j)
(G_{i\ell}-\bar G_\ell)
\right\}^2
\\
&\le
\frac{\max_{1\le i\le N}\|G_i-\bar G\|_2^2}{N}
\left\{
\frac{1}{N}
\sum_{i=1}^N
\|G_i-\bar G\|_2^2
\right\}
\to0.
\end{align*}
The convergence uses the maximal condition in Lemma~\ref{lem:cf_moment_stability}; the second factor is bounded by the fixed-dimension and covariance conditions in that lemma. Likewise, for each coordinate \(j\),
\begin{align*}
&
\frac{1}{N^2}
\sum_{i=1}^N
\left[
(G_{ij}-\bar G_j)
\{R_i(z)-\bar R(z)\}
\right]^2
\\
&\le
\frac{\max_{1\le i\le N}\|G_i-\bar G\|_2^2}{N}
\left[
\frac{1}{N}
\sum_{i=1}^N
\{R_i(z)-\bar R(z)\}^2
\right]
\to0.
\end{align*}
The response-variance bound in Lemma~\ref{lem:cf_moment_stability} gives the last convergence. Under the stated treatment-fraction bound, the finite-population variance formula therefore implies that the
within-arm oracle Gram matrix and cross moment converge in probability
to \(S_G^2\) and \(S_{G,R(z)}\), respectively. The same argument gives
\[
\bar R_z-\bar R(z)
=
O_p(N^{-1/2}),
\qquad
\bar G_z-\bar G
=
O_p(N^{-1/2}).
\]

\leavevmode\par
\textbf{Step 4: Feasible Gram and cross-moment convergence.}
\leavevmode\par

By Step 2 and the Cauchy--Schwarz inequality,
\begin{align*}
&
\left\|
\frac{1}{N_z}
\sum_{i:Z_i=z}
(\hat G_{i,\mathrm{cf}}-G_i)
\{R_i(z)-\bar R_z\}
\right\|_2
\\
&\le
\left\{
\frac{1}{N_z}
\sum_{i:Z_i=z}
\|\hat G_{i,\mathrm{cf}}-G_i\|_2^2
\right\}^{1/2}
\left\{
\frac{1}{N_z}
\sum_{i:Z_i=z}
\{R_i(z)-\bar R_z\}^2
\right\}^{1/2}
\\
&=
o_p(1),
\end{align*}
because the second factor is \(O_p(1)\).

For the Gram matrix, sample centering cannot increase the empirical
\(L_2\) distance, so
\[
\frac{1}{N_z}
\sum_{i:Z_i=z}
\left\|
(\hat G_{i,\mathrm{cf}}-\bar{\hat G}_{z,\mathrm{cf}})
-
(G_i-\bar G_z)
\right\|_2^2
\le
\frac{1}{N_z}
\sum_{i:Z_i=z}
\|\hat G_{i,\mathrm{cf}}-G_i\|_2^2
=
o_p(1).
\]
The oracle within-arm second moment of \(G_i-\bar G_z\) is \(O_p(1)\)
by Step 3. The Cauchy--Schwarz inequality therefore implies that the
difference between the feasible and oracle within-arm Gram matrices is
\(o_p(1)\). Combining this result with Step 3 gives
\[
\widehat S_{\hat G,z}^{\,2}
=
S_G^2+o_p(1).
\]
The cross-moment bound above, together with Step 3 and the corresponding
sample-centering identities, similarly gives
\[
\widehat S_{\hat G,R,z}
=
S_{G,R(z)}+o_p(1).
\]
This completes the proof.
\end{proof}

\subsection{Lemma~\ref{lem:projection_geometry}: Finite-population projection geometry}
\begin{lemma}[Finite-population least-squares projection]
\label{lem:projection_geometry}
Let \(R_1,\ldots,R_N\) be a scalar finite-population sequence, and let
\(G_i\in\mathbb R^k\). Let
\[
\gamma_G
\in
\arg\min_{\gamma\in\mathbb R^k}\sum_{i=1}^{N}
(R_i-\tilde G_i'\gamma)^2,
\qquad
e_{G,i}
=
R_i-\tilde G_i'\gamma_G.
\]
Then the following statements hold.
\begin{itemize}
\item[(i)] The normal equations satisfy $S_{G,e_G}=0$
and, for every \(b\in\mathbb R^k\),
\[
S^2_{R-\tilde G'b}
=
S^2_{e_G}
+
(b-\gamma_G)'S_G^2(b-\gamma_G).
\]

\item[(ii)] If \(S_G^2\) is singular, every least-squares solution yields
the same fitted-value vector and the same residual vector.

\item[(iii)] Let \(H_i\in\mathbb R^m\), and suppose the centered span
generated by \(G_i\) is contained in that generated by \(H_i\). Then
\[
\min_{c\in\mathbb R^m}
S^2_{R-\tilde H'c}
\le
\min_{b\in\mathbb R^k}
S^2_{R-\tilde G'b}.
\]
Equality holds if the least-squares fitted-value vector based on \(H_i\)
belongs to the centered span generated by \(G_i\).
\end{itemize}
\end{lemma}

\begin{proof}
\textbf{Proof of part~(i).}
\leavevmode\par
The normal equations and the stated variance decomposition are the standard least-squares property; see \citet[Sections~3.8 and 3.14--3.15]{hansen2022econometrics}. Applying these identities to the regression of \(R_i-\bar R\) on \(\tilde G_i\) gives
\[
S_{G,e_G}=0
\]
and
\[
S^2_{R-\tilde G'b}=S^2_{e_G}+(b-\gamma_G)'S_G^2(b-\gamma_G),\qquad b\in\mathbb R^k.
\]

\textbf{Proof of part~(ii).}
\leavevmode\par
For the nonuniqueness of least-squares coefficients under collinearity, see \citet[Section~3.24]{hansen2022econometrics}. Let \(\gamma_1\) and \(\gamma_2\) be two least-squares solutions. Because they attain the same minimum, part~(i) gives
\begin{align*}
0
&=S^2_{R-\tilde G'\gamma_2}-S^2_{R-\tilde G'\gamma_1}\\
&=\frac{1}{N-1}\sum_{i=1}^N\left\{\tilde G_i'(\gamma_2-\gamma_1)\right\}^2.
\end{align*}
Hence
\[
\tilde G_i'\gamma_1=\tilde G_i'\gamma_2
\]
for every \(i\), so all least-squares solutions yield the same fitted values and residuals. This proves part~(ii).

\textbf{Proof of part~(iii).}
\leavevmode\par
Every fitted-value vector available from \(G_i\) is also available from \(H_i\) under the stated span containment, which proves the inequality. If an \(H_i\)-based minimizing fitted value belongs to the centered span generated by \(G_i\), it is feasible in both problems, so the reverse inequality also holds. This proves part~(iii).
\end{proof}

\subsection{Lemma~\ref{lem:adjusted_ratio_reduction}: Exact reduction for adjusted ratios}

\begin{lemma}[Exact ratio identity]
\label{lem:adjusted_ratio_reduction}
Fix a realized adjustment vector \(G_i\) and suppose that \(\hat\tau_{G,R}\) is linear in the response \(R\). Let \(A_i=Y_i-\tau D_i\) and \(A_i^{(s)}=Y_i-sD_i\), where \(s\) is any realized scalar. Then
\[
\hat\tau_{G,A^{(s)}}=\hat\tau_{G,A}-(s-\tau)\hat\tau_{G,D}.
\]
Whenever \(\hat\tau_{G,D}\ne0\),
\[
s+\frac{\hat\tau_{G,A^{(s)}}}{\hat\tau_{G,D}}=\frac{\hat\tau_{G,Y}}{\hat\tau_{G,D}}=\tau+\frac{\hat\tau_{G,A}}{\hat\tau_{G,D}}.
\]
\end{lemma}

\begin{proof}
The response-linearity of ordinary least squares follows from its standard matrix representation; see \citet[Section~3.10]{hansen2022econometrics}. Since
\[
A_i^{(s)}=A_i-(s-\tau)D_i,\qquad Y_i=A_i+\tau D_i,
\]
linearity gives
\[
\hat\tau_{G,A^{(s)}}=\hat\tau_{G,A}-(s-\tau)\hat\tau_{G,D},\qquad \hat\tau_{G,Y}=\hat\tau_{G,A}+\tau\hat\tau_{G,D}.
\]
Dividing the second identity by \(\hat\tau_{G,D}\) and substituting the first proves the result.
\end{proof}

\subsection{Lemma~\ref{lem:sample_variance_stability}: Feasible residuals and sample variances}

\begin{lemma}[Feasible residuals and sample variances]
\label{lem:sample_variance_stability}
Let \(G_i\in\mathbb R^k\) be deterministic and fixed dimensional, let \(\hat G_{i,\mathrm{cf}}\) be a generated adjustment vector, and let \(\delta_z\in\mathbb R^k\). Suppose, for \(z\in\{0,1\}\),
\begin{align*}
\frac{1}{N_z}
\sum_{i:Z_i=z}
\|\hat G_{i,\mathrm{cf}}-G_i\|_2^2
&=
 o_p(1),
\\
\bar{\hat G}_{z,\mathrm{cf}}-\bar G
&=
 o_p(1),
\\
\hat b_z-\delta_z
&=
 o_p(1),
\\
\hat\theta-\tau
&=
 o_p(1).
\end{align*}
Suppose $N_z/N$ is bounded away from zero and $N_z^{-1}\sum_{i:Z_i=z}D_i(z)^2=O_p(1)$. Assume also that
\[
\|\delta_z\|_2=O(1),
\qquad
\frac{1}{N}\sum_{i=1}^N\|G_i-\bar G\|_2^2=O(1).
\]
For \(i\) with \(Z_i=z\), define
\[
\hat e_i
=
Y_i-\hat\theta D_i
-
(\hat G_{i,\mathrm{cf}}-\bar{\hat G}_{z,\mathrm{cf}})'\hat b_z,
\]
and define the oracle residual
\[
e_i(z)
=
A_i(z)-(G_i-\bar G)'\delta_z.
\]
Then
\[
\frac{1}{N_z}
\sum_{i:Z_i=z}
\{\hat e_i-e_i(z)\}^2
=
 o_p(1).
\]
If, in addition,
\[
S^2_{e(z)}=O(1),
\qquad
\max_{1\le i\le N}
\frac{\{e_i(z)-\bar e(z)\}^2}{N}
\to0,
\]
then
\[
\frac{1}{N_z-1}
\sum_{i:Z_i=z}
(\hat e_i-\bar{\hat e}_z)^2
-
S^2_{e(z)}
\xrightarrow{p}0.
\]
The last conclusion also holds for any feasible scalar sequence \(\hat e_i\) satisfying the first empirical \(L_2\) conclusion, without requiring the displayed regression representation.
\end{lemma}

\begin{proof}
\textbf{Step 1: Empirical \(L_2\) convergence of feasible residuals.}
\leavevmode\par
For \(i\) with \(Z_i=z\),
\begin{align*}
\hat e_i-e_i(z)
&=
-(\hat\theta-\tau)D_i(z)
\\
&\quad-
\left[
(\hat G_{i,\mathrm{cf}}-\bar{\hat G}_{z,\mathrm{cf}})'\hat b_z
-
(G_i-\bar G)'\delta_z
\right].
\end{align*}
The adjustment difference equals
\begin{align*}
&
(\hat G_{i,\mathrm{cf}}-G_i)'\hat b_z
-
(\bar{\hat G}_{z,\mathrm{cf}}-\bar G)'\hat b_z
+
(G_i-\bar G)'(\hat b_z-\delta_z).
\end{align*}
The empirical second-moment bound for $D_i(z)$, $\hat b_z=O_p(1)$, and the bounded oracle second moment of $G_i$, together with the convergence conditions in Lemma~\ref{lem:sample_variance_stability} and the Cauchy--Schwarz inequality, give
\[
\frac{1}{N_z}
\sum_{i:Z_i=z}
\{\hat e_i-e_i(z)\}^2
=
 o_p(1).
\]

\textbf{Step 2: Sample-variance convergence.}
\leavevmode\par
It remains to prove the sample-variance statement. Let
\[
\|u\|_{z,N}^2
=
\frac{1}{N_z}
\sum_{i:Z_i=z}u_i^2
\]
for a scalar sequence observed in arm \(z\), and let \(C_z\) denote sample centering within that arm. The centering map is an orthogonal projection, so
\[
\|C_z(\hat e-e(z))\|_{z,N}
\le
\|\hat e-e(z)\|_{z,N}
=
 o_p(1).
\]
Therefore,
\begin{align*}
&\left|
\frac{1}{N_z-1}\sum_{i:Z_i=z}(\hat e_i-\bar{\hat e}_z)^2
-
\frac{1}{N_z-1}\sum_{i:Z_i=z}\{e_i(z)-\bar e_z^{\mathrm{obs}}(z)\}^2
\right|
\\
&\qquad\le
\frac{N_z}{N_z-1}
\|C_z(\hat e-e(z))\|_{z,N}
\left\{
\|C_z\hat e\|_{z,N}
+
\|C_ze(z)\|_{z,N}
\right\}
\\
&\qquad=
 o_p(1),
\end{align*}
because the two centered norms are \(O_p(1)\).

For the oracle sample variance, set
\[
v_i=\{e_i(z)-\bar e(z)\}^2.
\]
The maximal condition and boundedness of $S^2_{e(z)}$ in Lemma~\ref{lem:sample_variance_stability} imply
\[
\frac{1}{N^2}\sum_{i=1}^Nv_i^2
\le
\left
\{
\max_i\frac{v_i}{N}
\right\}
\left
\{
\frac{1}{N}\sum_{i=1}^Nv_i
\right\}
\to0.
\]
The finite-population variance formula therefore gives
\[
\frac{1}{N_z}\sum_{i:Z_i=z}v_i
-
\frac{1}{N}\sum_{i=1}^Nv_i
\xrightarrow{p}0.
\]
Also,
\[
\bar e_z^{\mathrm{obs}}(z)-\bar e(z)=O_p(N^{-1/2}).
\]
Using
\[
\frac{1}{N_z-1}
\sum_{i:Z_i=z}
\{e_i(z)-\bar e_z^{\mathrm{obs}}(z)\}^2
=
\frac{N_z}{N_z-1}
\left[
\frac{1}{N_z}\sum_{i:Z_i=z}v_i
-
\{\bar e_z^{\mathrm{obs}}(z)-\bar e(z)\}^2
\right]
\]
proves convergence to \(S^2_{e(z)}\). Combining the two parts completes the proof.
\end{proof}

\subsection{Lemma~\ref{lem:cf_oracle_equivalence}: First-order equivalence under conditional cross-fitting}

\begin{lemma}[First-order equivalence under conditional cross-fitting]
\label{lem:cf_oracle_equivalence}
Suppose the conditions of Lemma~\ref{lem:cf_moment_stability} hold and \(S_G^2\to\Sigma_G\), where \(\Sigma_G\) is positive definite. Using the population projection coefficients \(\gamma_R(z)\), define
\[
R_{G,i}(z)=R_i(z)-\tilde G_i'\gamma_R(z),\qquad \hat\tau_{R_G}=\frac{1}{N_1}\sum_{i:Z_i=1}R_{G,i}(1)-\frac{1}{N_0}\sum_{i:Z_i=0}R_{G,i}(0).
\]
Then, for \(z\in\{0,1\}\),
\[
\hat\gamma_{\hat G_{\mathrm{cf}},R}(z)=\gamma_R(z)+o_p(1),\qquad \hat\tau_{\hat G_{\mathrm{cf}},R}=\hat\tau_{R_G}+o_p(N^{-1/2}).
\]
\end{lemma}

\begin{proof}
Lemma~\ref{lem:cf_moment_stability} gives \(\widehat S_{\hat G,z}^{\,2}=S_G^2+o_p(1)\) and \(\widehat S_{\hat G,R,z}=S_{G,R(z)}+o_p(1)\). The positive-definiteness condition in Lemma~\ref{lem:cf_oracle_equivalence} and continuity of inversion therefore imply
\[
\hat\gamma_{\hat G_{\mathrm{cf}},R}(z)=(\widehat S_{\hat G,z}^{\,2})^{-1}\widehat S_{\hat G,R,z}=\gamma_R(z)+o_p(1).
\]
The same lemma gives
\[
\bar{\hat G}_{z,\mathrm{cf}}-\bar{\hat G}_{\mathrm{cf}}=\bar G_z-\bar G+o_p(N^{-1/2}),\qquad \bar G_z-\bar G=O_p(N^{-1/2}).
\]
The bounded cross moments in Lemma~\ref{lem:cf_moment_stability} and the positive-definiteness condition in Lemma~\ref{lem:cf_oracle_equivalence} imply $\gamma_R(z)=O(1)$. It follows that
\begin{align*}
&(\bar{\hat G}_{z,\mathrm{cf}}-\bar{\hat G}_{\mathrm{cf}})'\hat\gamma_{\hat G_{\mathrm{cf}},R}(z)-(\bar G_z-\bar G)'\gamma_R(z)\\
&\quad=\left\{(\bar{\hat G}_{z,\mathrm{cf}}-\bar{\hat G}_{\mathrm{cf}})-(\bar G_z-\bar G)\right\}'\hat\gamma_{\hat G_{\mathrm{cf}},R}(z)+(\bar G_z-\bar G)'\left\{\hat\gamma_{\hat G_{\mathrm{cf}},R}(z)-\gamma_R(z)\right\}\\
&\quad=o_p(N^{-1/2}).
\end{align*}
Substituting into the two arm-specific adjusted means gives \(\hat\tau_{\hat G_{\mathrm{cf}},R}=\hat\tau_{R_G}+o_p(N^{-1/2})\).
\end{proof}

\subsection{Lemma~\ref{lem:split_sampling}: Marginal sampling induced by the split}

\begin{lemma}[Marginal sampling induced by the split]
\label{lem:split_sampling}
Under complete randomization, suppose Algorithm~\ref{alg:cf_late} partitions each realized arm uniformly over partitions with fixed fold sizes. For each $q\in\{1,\ldots,Q\}$ and $z\in\{0,1\}$, the following statements hold.
\begin{itemize}
\item[(i)] The training sample $\{i\in S^{[-q]}:Z_i=z\}$ is marginally a simple random sample without replacement of size $N_z^{[-q]}$ from $\{1,\ldots,N\}$.
\item[(ii)] Let $k\in\{1,\ldots,K_{\mathrm{cv}}\}$ index the inner cross-validation folds. Conditional on the training sample, suppose the validation fold $V_{z,k}^{[-q]}$ is sampled uniformly with fixed size $N_{z,k}^{[-q]}$, and let
\[
U_{z,k}^{[-q]}
=
\{i\in S^{[-q]}:Z_i=z\}
\setminus
V_{z,k}^{[-q]}.
\]
Then $V_{z,k}^{[-q]}$ and $U_{z,k}^{[-q]}$ are each marginally simple random samples without replacement from $\{1,\ldots,N\}$, of sizes $N_{z,k}^{[-q]}$ and $N_z^{[-q]}-N_{z,k}^{[-q]}$, respectively.
\end{itemize}
\end{lemma}

\begin{proof}
\textbf{Step 1: Outer training set.}
\leavevmode\par
Fix $q$ and $z$. Under complete randomization, the realized arm $\{i:Z_i=z\}$ is uniform over all subsets of size $N_z$. Under the uniform partition in Algorithm~\ref{alg:cf_late}, conditional on this arm, $\{i\in S^{[-q]}:Z_i=z\}$ is uniform over its subsets of size $N_z^{[-q]}$. In the sums below, $J$ ranges over subsets of $\{1,\ldots,N\}$ satisfying the indicated conditions. For any fixed $I\subseteq\{1,\ldots,N\}$ with $|I|=N_z^{[-q]}$,
\begin{align*}
\Pr(\{i\in S^{[-q]}:Z_i=z\}=I)
&=
\sum_{\substack{I\subseteq J\\|J|=N_z}}
\Pr(\{i:Z_i=z\}=J)
\Pr(\{i\in S^{[-q]}:Z_i=z\}=I\mid\{i:Z_i=z\}=J)
\\
&=
\sum_{\substack{I\subseteq J\\|J|=N_z}}
\frac{1}{\binom{N}{N_z}}
\frac{1}{\binom{N_z}{N_z^{[-q]}}}
\\
&=
\frac{1}{\binom{N}{N_z}}
\frac{1}{\binom{N_z}{N_z^{[-q]}}}
\sum_{\substack{I\subseteq J\\|J|=N_z}}1
\\
&=
\frac{\binom{N-N_z^{[-q]}}{N_z-N_z^{[-q]}}}
{\binom{N}{N_z}\binom{N_z}{N_z^{[-q]}}}
\\
&=
\frac{(N-N_z^{[-q]})!}{(N_z-N_z^{[-q]})!(N-N_z)!}
\frac{N_z!(N-N_z)!}{N!}
\frac{N_z^{[-q]}!(N_z-N_z^{[-q]})!}{N_z!}
\\
&=
\frac{N_z^{[-q]}!(N-N_z^{[-q]})!}{N!}
\\
&=
\binom{N}{N_z^{[-q]}}^{-1}.
\end{align*}
Thus $\{i\in S^{[-q]}:Z_i=z\}$ is a simple random sample without replacement.

\textbf{Step 2: Inner training and validation sets.}
\leavevmode\par
Fix $k\in\{1,\ldots,K_{\mathrm{cv}}\}$. For any fixed $I\subseteq\{1,\ldots,N\}$ with $|I|=N_{z,k}^{[-q]}$, Step~1 and the conditional uniform sampling of $V_{z,k}^{[-q]}$ give
\begin{align*}
\Pr(V_{z,k}^{[-q]}=I)
&=
\sum_{\substack{I\subseteq J\\|J|=N_z^{[-q]}}}
\Pr(\{i\in S^{[-q]}:Z_i=z\}=J)
\Pr(V_{z,k}^{[-q]}=I\mid\{i\in S^{[-q]}:Z_i=z\}=J)
\\
&=
\sum_{\substack{I\subseteq J\\|J|=N_z^{[-q]}}}
\frac{1}{\binom{N}{N_z^{[-q]}}}
\frac{1}{\binom{N_z^{[-q]}}{N_{z,k}^{[-q]}}}
\\
&=
\frac{1}{\binom{N}{N_z^{[-q]}}}
\frac{1}{\binom{N_z^{[-q]}}{N_{z,k}^{[-q]}}}
\sum_{\substack{I\subseteq J\\|J|=N_z^{[-q]}}}1
\\
&=
\frac{\binom{N-N_{z,k}^{[-q]}}{N_z^{[-q]}-N_{z,k}^{[-q]}}}
{\binom{N}{N_z^{[-q]}}\binom{N_z^{[-q]}}{N_{z,k}^{[-q]}}}
\\
&=
\frac{(N-N_{z,k}^{[-q]})!}
{(N_z^{[-q]}-N_{z,k}^{[-q]})!(N-N_z^{[-q]})!}
\frac{N_z^{[-q]}!(N-N_z^{[-q]})!}{N!}
\frac{N_{z,k}^{[-q]}!(N_z^{[-q]}-N_{z,k}^{[-q]})!}
{N_z^{[-q]}!}
\\
&=
\frac{N_{z,k}^{[-q]}!(N-N_{z,k}^{[-q]})!}{N!}
\\
&=
\binom{N}{N_{z,k}^{[-q]}}^{-1}.
\end{align*}
Hence $V_{z,k}^{[-q]}$ is a simple random sample without replacement. Conditional on the outer training sample, its complement $U_{z,k}^{[-q]}$ is also uniform over subsets of size $N_z^{[-q]}-N_{z,k}^{[-q]}$. The same calculation therefore proves the claim for $U_{z,k}^{[-q]}$.
\end{proof}

\subsection{Lemma~\ref{lem:training_ratio_rate}: Training-fold ratio rate}

\begin{lemma}[Training-fold ratio rate]
\label{lem:training_ratio_rate}
Suppose Assumptions~\ref{ass:id}, \ref{ass:asymp}(i)--(ii), and \ref{ass:conditional_crossfit}(ii) hold, and the split is generated by Algorithm~\ref{alg:cf_late}. Then, for $R\in\{Y,D\}$,
\[
\max_{1\le q\le Q}|\hat\tau_R^{[-q]}-\tau_R|=O_p(N^{-1/2}),
\]
and
\[
\max_{1\le q\le Q}|\hat\tau^{[-q]}-\tau|=O_p(N^{-1/2}).
\]
\end{lemma}

\begin{proof}
Lemma~\ref{lem:split_sampling} and Assumptions~\ref{ass:asymp}(i) and \ref{ass:conditional_crossfit}(ii) give $\min_{q,z}N_z^{[-q]}/N\ge c>0$. For $R\in\{Y,D\}$, marginal simple random sampling yields
\begin{align*}
\mathbb E\left[\frac{1}{N_z^{[-q]}}\sum_{i\in S^{[-q]}:Z_i=z}R_i(z)\right]
&=\frac{1}{N_z^{[-q]}}\sum_{i=1}^NR_i(z)
\Pr(i\in S^{[-q]},Z_i=z)\\
&=\frac{1}{N_z^{[-q]}}\sum_{i=1}^NR_i(z)\frac{N_z^{[-q]}}{N}\\
&=\bar R(z),\\
\operatorname{Var}\left(\frac{1}{N_z^{[-q]}}\sum_{i\in S^{[-q]}:Z_i=z}R_i(z)\right)
&=\frac{1}{(N_z^{[-q]})^2}\Biggl[
\frac{N_z^{[-q]}}{N}\sum_{i=1}^N\{R_i(z)-\bar R(z)\}^2\\
&\qquad+\frac{N_z^{[-q]}(N_z^{[-q]}-1)}{N(N-1)}
\sum_{i\ne j}\{R_i(z)-\bar R(z)\}\{R_j(z)-\bar R(z)\}
\Biggr]\\
&=\frac{N-N_z^{[-q]}}{N_z^{[-q]}N(N-1)}
\sum_{i=1}^N\{R_i(z)-\bar R(z)\}^2\\
&=\left(\frac{1}{N_z^{[-q]}}-\frac{1}{N}\right)S^2_{R(z)}\\
&=O(N^{-1}).
\end{align*}
The second equality uses $\sum_{i\ne j}\{R_i(z)-\bar R(z)\}\{R_j(z)-\bar R(z)\}=-\sum_i\{R_i(z)-\bar R(z)\}^2$. For every $M>0$, Chebyshev's inequality gives
\begin{align*}
&\Pr\left(\left|\frac{1}{N_z^{[-q]}}\sum_{i\in S^{[-q]}:Z_i=z}R_i(z)-\bar R(z)\right|>\frac{M}{\sqrt N}\right)\\
&\quad\le\frac{N}{M^2}
\left(\frac{1}{N_z^{[-q]}}-\frac{1}{N}\right)S^2_{R(z)}\\
&\quad\le\frac{S^2_{R(z)}}{cM^2}.
\end{align*}
Since $S^2_{R(z)}=O(1)$ under Assumption~\ref{ass:asymp}(ii) and $Q$ is fixed,
\begin{align*}
\max_{q,z}\left|\frac{1}{N_z^{[-q]}}\sum_{i\in S^{[-q]}:Z_i=z}R_i(z)-\bar R(z)\right|
&=O_p(N^{-1/2}),\\
\max_q|\hat\tau_R^{[-q]}-\tau_R|
&=O_p(N^{-1/2}).
\end{align*}
Assumptions~\ref{ass:id} and \ref{ass:asymp}(ii) give $\tau=O(1)$. Moreover,
\[
\Pr\left(\min_q|\hat\tau_D^{[-q]}|\ge\frac{C}{2}\right)\to1,
\qquad
\max_q|\hat\tau_D^{[-q]}|^{-1}\le\frac{2}{C}
\quad\text{on this event}.
\]
On this event,
\begin{align*}
\max_q|\hat\tau^{[-q]}-\tau|
&=\max_q\left|
\frac{\{\hat\tau_Y^{[-q]}-\tau_Y\}-\tau\{\hat\tau_D^{[-q]}-\tau_D\}}
{\hat\tau_D^{[-q]}}
\right|\\
&=O_p(N^{-1/2}).
\end{align*}
\end{proof}

\subsection{Lemma~\ref{lem:finite_population_concentration}: Maximal concentration under sampling without replacement}

\begin{lemma}[Maximal concentration under sampling without replacement]
\label{lem:finite_population_concentration}
Let $J_N\ge1$ be an integer and let $v_{i,j}$ be fixed finite-population arrays satisfying
\[
\max_{\substack{1\le i\le N\\1\le j\le J_N}}|v_{i,j}|\le H_N
\]
for a deterministic bound $H_N>0$. Let $\mathcal I_N$ be a collection of a fixed number of marginal simple random samples without replacement from $\{1,\ldots,N\}$. For each $I\in\mathcal I_N$, let $N_I$ denote its fixed sample size and suppose $N_I\ge cN$ for a constant $c>0$. Then
\[
\max_{I\in\mathcal I_N}\max_{1\le j\le J_N}
\left|
\frac{1}{N_I}\sum_{i\in I}v_{i,j}
-\frac{1}{N}\sum_{i=1}^Nv_{i,j}
\right|
=
O_p\left\{
H_N\sqrt{\frac{\log(J_N+1)}{N}}
\right\}.
\]
\end{lemma}

\begin{proof}
For each $I\in\mathcal I_N$ and $j\in\{1,\ldots,J_N\}$, applying Proposition~1.2 of \citet{bardenet2015concentration_hoeffding} to $v_{i,j}$ and $-v_{i,j}$ gives, for every $t>0$,
\begin{align*}
&\Pr\left(
\left|
\frac{1}{N_I}\sum_{i\in I}v_{i,j}
-\frac{1}{N}\sum_{i=1}^Nv_{i,j}
\right|\ge t
\right)\\
&\quad\le
\Pr\left(
\frac{1}{N_I}\sum_{i\in I}v_{i,j}
-\frac{1}{N}\sum_{i=1}^Nv_{i,j}\ge t
\right)
+
\Pr\left(
\frac{1}{N}\sum_{i=1}^Nv_{i,j}
-\frac{1}{N_I}\sum_{i\in I}v_{i,j}\ge t
\right)\\
&\quad\le
2\exp\left\{-\frac{2N_It^2}{(2H_N)^2}\right\}\\
&\quad=
2\exp\left(-\frac{N_It^2}{2H_N^2}\right).
\end{align*}
The first inequality separates the two tails. The second uses simple random sampling without replacement and the fact that the range of each coordinate is at most $2H_N$.

The union bound over samples and coordinates therefore gives
\begin{align*}
&\Pr\left(
\max_{I\in\mathcal I_N}\max_{1\le j\le J_N}
\left|
\frac{1}{N_I}\sum_{i\in I}v_{i,j}
-\frac{1}{N}\sum_{i=1}^Nv_{i,j}
\right|\ge t
\right)\\
&\quad\le
\sum_{I\in\mathcal I_N}\sum_{j=1}^{J_N}
\Pr\left(
\left|
\frac{1}{N_I}\sum_{i\in I}v_{i,j}
-\frac{1}{N}\sum_{i=1}^Nv_{i,j}
\right|\ge t
\right)\\
&\quad\le
2\sum_{I\in\mathcal I_N}\sum_{j=1}^{J_N}
\exp\left(-\frac{N_It^2}{2H_N^2}\right)\\
&\quad\le
2J_N\exp\left(-\frac{cNt^2}{2H_N^2}\right)
\sum_{I\in\mathcal I_N}1.
\end{align*}
The last inequality uses $N_I\ge cN$. 

Let $K_{\mathcal I}$ denote the fixed number of samples in $\mathcal I_N$. Setting $t=MH_N\sqrt{\log(J_N+1)/N}$ yields, for $M\ge\sqrt{2/c}$,
\begin{align*}
&\Pr\left(
\max_{I\in\mathcal I_N}\max_{1\le j\le J_N}
\left|
\frac{1}{N_I}\sum_{i\in I}v_{i,j}
-\frac{1}{N}\sum_{i=1}^Nv_{i,j}
\right|
\ge MH_N\sqrt{\frac{\log(J_N+1)}{N}}
\right)\\
&\quad\le
2K_{\mathcal I}J_N
\exp\left\{-\frac{cM^2}{2}\log(J_N+1)\right\}\\
&\quad=
2K_{\mathcal I}J_N(J_N+1)^{-cM^2/2}\\
&\quad\le
2K_{\mathcal I}(J_N+1)^{1-cM^2/2}\\
&\quad\le
K_{\mathcal I}2^{\,2-cM^2/2}.
\end{align*}
The penultimate inequality uses $J_N\le J_N+1$. The last uses $J_N+1\ge2$ and $1-cM^2/2\le0$. Since $K_{\mathcal I}$ is fixed, the final bound tends to zero as $M\to\infty$, uniformly in $N$. This proves the stated stochastic order.
\end{proof}

\subsection{Lemma~\ref{lem:lasso_path_geometry}: Population penalized path}

\begin{lemma}[Population penalized path]
\label{lem:lasso_path_geometry}
For $z\in\{0,1\}$ and $\lambda>0$, let
\[
L_{N,z}(\lambda)=\frac{1}{N}\sum_{i=1}^N\{A_i(z)-\mu_{A,i}(z;\lambda)\}^2.
\]
For $0<\lambda_1\le\lambda_2$,
\begin{align*}
\|\beta_A(z;\lambda_2)\|_1&\le\|\beta_A(z;\lambda_1)\|_1,\\
L_{N,z}(\lambda_1)&\le L_{N,z}(\lambda_2),\\
\|\mu_A(z;\lambda_2)-\mu_A(z;\lambda_1)\|_N^2
&\le L_{N,z}(\lambda_2)-L_{N,z}(\lambda_1).
\end{align*}
\end{lemma}

\begin{proof}
For $j=1,2$, write $\beta_j=\beta_A(z;\lambda_j)$ and $L_j=L_{N,z}(\lambda_j)$. Population optimality gives
\begin{align*}
L_1+\lambda_1\|\beta_1\|_1
&\le L_2+\lambda_1\|\beta_2\|_1,\\
L_2+\lambda_2\|\beta_2\|_1
&\le L_1+\lambda_2\|\beta_1\|_1.
\end{align*}
If $\lambda_1=\lambda_2$, uniqueness of the fitted-value vector and equality of the minimized objectives give the claim. Otherwise, the two inequalities imply
\begin{align*}
(\lambda_2-\lambda_1)(\|\beta_2\|_1-\|\beta_1\|_1)&\le0,\\
L_1-L_2&\le\lambda_1(\|\beta_2\|_1-\|\beta_1\|_1)\le0.
\end{align*}
Let $s\in\partial\|\beta_1\|_1$ satisfy
\[
2\Sigma_{X,N}\beta_1
-\frac{2}{N}\sum_{i=1}^N\tilde X_iA_i(z)
+\lambda_1s=0.
\]
The subgradient inequality implies
\begin{align*}
L_2-L_1
&=(\beta_2-\beta_1)'\Sigma_{X,N}(\beta_2-\beta_1)
-\lambda_1s'(\beta_2-\beta_1)\\
&\ge\|\mu_A(z;\lambda_2)-\mu_A(z;\lambda_1)\|_N^2
-\lambda_1(\|\beta_2\|_1-\|\beta_1\|_1)\\
&\ge\|\mu_A(z;\lambda_2)-\mu_A(z;\lambda_1)\|_N^2.
\end{align*}
\end{proof}

\subsection{Lemma~\ref{lem:lasso_validation_concentration}: Uniform validation concentration}

\begin{lemma}[Uniform validation concentration]
\label{lem:lasso_validation_concentration}
Suppose Assumptions~\ref{ass:id}, \ref{ass:asymp}(i)--(ii), \ref{ass:conditional_crossfit}(i)--(ii), \ref{ass:lasso_persistence}, and \ref{ass:lasso_cv_grid} hold, and construct the folds as specified in Algorithm~\ref{alg:cf_late} and Section~\ref{sec:lasso}. Then, for each $z\in\{0,1\}$,
\[
\max_{\substack{1\le q\le Q\\1\le k\le K_{\mathrm{cv}}}}
\sup_{\lambda\in\Lambda_N}
\left|
\frac{1}{N_{z,k}^{[-q]}}
\sum_{i\in V_{z,k}^{[-q]}}
\mu_{A,i}(z;\lambda)^2
-
\frac{1}{N}\sum_{i=1}^N
\mu_{A,i}(z;\lambda)^2
\right|
\xrightarrow{p}0.
\]
\end{lemma}

\begin{proof}
Fix $z$. Assumptions~\ref{ass:id} and \ref{ass:asymp}(ii) imply
\[
\frac{1}{N}\sum_{i=1}^N A_i(z)^2=O(1).
\]
Choose a finite constant $C_z$ such that
\[
\frac{1}{N}\sum_{i=1}^N A_i(z)^2\le C_z
\]
for all sufficiently large $N$. Population optimality relative to $b=0$ gives, uniformly over $\lambda>0$,
\begin{align*}
0\le L_{N,z}(\lambda)
&\le
\frac{1}{N}\sum_{i=1}^N
\{A_i(z)-\bar A(z)\}^2
\le C_z,
\\
\lambda\|\beta_A(z;\lambda)\|_1
&\le
\frac{1}{N}\sum_{i=1}^N
\{A_i(z)-\bar A(z)\}^2
\le C_z,
\\
\|\mu_A(z;\lambda)\|_N
&\le
\left\{
\frac{1}{N}\sum_{i=1}^N A_i(z)^2
\right\}^{1/2}
+
L_{N,z}(\lambda)^{1/2}
\le
2\sqrt{C_z}.
\end{align*}
Consequently,
\begin{align*}
\sup_{\lambda\ge\lambda_N}
\max_i|\mu_{A,i}(z;\lambda)|
&\le
|\bar A(z)|
+
K_{X,N}
\sup_{\lambda\ge\lambda_N}
\|\beta_A(z;\lambda)\|_1
\\
&\le
\sqrt{C_z}
+
\frac{C_zK_{X,N}}{\lambda_N},
\\
\frac{1}{N}
\left(
\sqrt{C_z}
+
\frac{C_zK_{X,N}}{\lambda_N}
\right)^2
&\to0
\end{align*}
by Assumption~\ref{ass:lasso_persistence}(ii).

Fix $\varepsilon>0$. Partition $[0,C_z]$ into at most
\[
1+\left\lceil\frac{C_z}{\varepsilon^2}\right\rceil
\]
subintervals of length at most $\varepsilon^2$. Choose one deterministic penalty from $\Lambda_N$ for each subinterval containing a value $L_{N,z}(\lambda)$. For every $\lambda\in\Lambda_N$, let $\lambda'$ be the chosen penalty from the same subinterval. The number of chosen penalties is bounded independently of $N$ and $|\Lambda_N|$. Lemma~\ref{lem:lasso_path_geometry} gives
\[
\|\mu_A(z;\lambda)-\mu_A(z;\lambda')\|_N^2
\le
|L_{N,z}(\lambda)-L_{N,z}(\lambda')|
\le
\varepsilon^2.
\]

Since $N_{z,k}^{[-q]}\ge c_{\mathrm{cv}}N$, Cauchy--Schwarz gives, uniformly over $q,k,\lambda$,
\begin{align*}
&\left|
\left(
\frac{1}{N_{z,k}^{[-q]}}
\sum_{i\in V_{z,k}^{[-q]}}
-
\frac{1}{N}\sum_{i=1}^N
\right)
\left\{
\mu_{A,i}(z;\lambda)^2
-
\mu_{A,i}(z;\lambda')^2
\right\}
\right|
\\
&\quad\le
(1+c_{\mathrm{cv}}^{-1})
\|\mu_A(z;\lambda)-\mu_A(z;\lambda')\|_N
\left\{
\|\mu_A(z;\lambda)\|_N
+
\|\mu_A(z;\lambda')\|_N
\right\}
\\
&\quad\le
4\sqrt{C_z}(1+c_{\mathrm{cv}}^{-1})\varepsilon.
\end{align*}

For each chosen penalty, Lemma~\ref{lem:split_sampling} makes
$V_{z,k}^{[-q]}$ marginally a simple random sample. The
finite-population sampling variance formula gives
\begin{align*}
&\operatorname{Var}\left(
\frac{1}{N_{z,k}^{[-q]}}
\sum_{i\in V_{z,k}^{[-q]}}
\mu_{A,i}(z;\lambda')^2
\right)
\\
&\quad\le
\left(
\frac{1}{N_{z,k}^{[-q]}}
-
\frac{1}{N}
\right)
\frac{1}{N-1}
\sum_{i=1}^N
\mu_{A,i}(z;\lambda')^4
\\
&\quad\le
O(N^{-1})
\frac{N}{N-1}
\max_i\mu_{A,i}(z;\lambda')^2
\|\mu_A(z;\lambda')\|_N^2
=
o(1).
\end{align*}
For every $t>0$, Chebyshev's inequality and the union bound give
\begin{align*}
&\Pr\left(
\max_{q,k,\lambda'}
\left|
\frac{1}{N_{z,k}^{[-q]}}
\sum_{i\in V_{z,k}^{[-q]}}
\mu_{A,i}(z;\lambda')^2
-
\frac{1}{N}
\sum_{i=1}^N
\mu_{A,i}(z;\lambda')^2
\right|
>t
\right)
\\
&\quad\le
\frac{1}{t^2}
\sum_{q=1}^Q
\sum_{k=1}^{K_{\mathrm{cv}}}
\sum_{\lambda'}
\operatorname{Var}\left(
\frac{1}{N_{z,k}^{[-q]}}
\sum_{i\in V_{z,k}^{[-q]}}
\mu_{A,i}(z;\lambda')^2
\right)
\\
&\quad\le
\frac{QK_{\mathrm{cv}}}{t^2}
\left(
1+\left\lceil\frac{C_z}{\varepsilon^2}\right\rceil
\right)
o(1)
=
o(1).
\end{align*}
The preceding deterministic bound is at most
\[
4\sqrt{C_z}(1+c_{\mathrm{cv}}^{-1})\varepsilon.
\]
Letting $N\to\infty$ for fixed $\varepsilon$ and then
$\varepsilon\downarrow0$ proves the result.
\end{proof}

\subsection{Lemma~\ref{lem:lasso_cv_uniform_stability}: Uniform stability of cross-validation fits}

\begin{lemma}[Uniform stability of cross-validation fits]
\label{lem:lasso_cv_uniform_stability}
Suppose Assumptions~\ref{ass:id}, \ref{ass:asymp}(i)--(ii),
\ref{ass:conditional_crossfit}(i)--(ii),
\ref{ass:lasso_persistence}, and \ref{ass:lasso_cv_grid}(i) hold, and
construct the outer and inner folds as specified in
Algorithm~\ref{alg:cf_late} and Section~\ref{sec:lasso}.
Then, for each $z\in\{0,1\}$,
\begin{align*}
&\max_{q,k}\sup_{\lambda\ge\lambda_N}
\|\hat\mu_A^{[-q,-k]}(z;\lambda)-\mu_A(z;\lambda)\|_N^2
=o_p(1),
\\
&\max_q\sup_{\lambda\ge\lambda_N}
\|\hat\mu_A^{[-q]}(z;\lambda)-\mu_A(z;\lambda)\|_N^2
=o_p(1).
\end{align*}
Moreover,
\begin{align*}
\sup_{\lambda\ge\lambda_N}
\|\beta_A(z;\lambda)\|_1
&=
O(\lambda_N^{-1}),
\\
\max_{q,k}\sup_{\lambda\ge\lambda_N}
\|\hat\beta_A^{[-q,-k]}(z;\lambda)\|_1
+
\max_q\sup_{\lambda\ge\lambda_N}
\|\hat\beta_A^{[-q]}(z;\lambda)\|_1
&=
O_p(\lambda_N^{-1}),
\end{align*}
and
\[
\sup_{\lambda\ge\lambda_N}
\|\mu_A(z;\lambda)\|_N
+
\max_{q,k}\sup_{\lambda\ge\lambda_N}
\|\hat\mu_A^{[-q,-k]}(z;\lambda)\|_N
=
O_p(1).
\]
\end{lemma}

\begin{proof}
Fix $z\in\{0,1\}$. For each $q$, let $I$ denote either an inner
training sample $U_{z,k}^{[-q]}$ or the outer training sample
$\{i\in S^{[-q]}:Z_i=z\}$, and write $n_I=|I|$.
Let $\hat\beta_I(z;\lambda)$ denote the corresponding
LASSO slope fitted on $I$, and define
\[
\bar{\hat A}_I^{[-q]}
=
\frac{1}{n_I}
\sum_{j\in I}\hat A_j^{[-q]},
\qquad
\hat\mu_{I,i}^{[-q]}(z;\lambda)
=
\bar{\hat A}_I^{[-q]}
+
\tilde X_i'\hat\beta_I(z;\lambda),
\quad i=1,\ldots,N.
\]
Write
\[
\hat\mu_I^{[-q]}(z;\lambda)
=
\left(
\hat\mu_{I,i}^{[-q]}(z;\lambda)
\right)_{i=1}^N.
\]
When $I=U_{z,k}^{[-q]}$, these coincide with
$\hat\beta_A^{[-q,-k]}(z;\lambda)$ and
$\hat\mu_A^{[-q,-k]}(z;\lambda)$; when
$I=\{i\in S^{[-q]}:Z_i=z\}$, they coincide with
$\hat\beta_A^{[-q]}(z;\lambda)$ and
$\hat\mu_A^{[-q]}(z;\lambda)$. All maxima over $q$ and $I$ below
range over this fixed collection of inner and outer training samples.

By Lemma~\ref{lem:split_sampling}, each such $I$ is marginally a
simple random sample without replacement from the finite population.
Assumption~\ref{ass:lasso_cv_grid}(i),
Assumptions~\ref{ass:asymp}(i) and
\ref{ass:conditional_crossfit}(ii), and fixed $Q$ and
$K_{\mathrm{cv}}$ imply
\[
\min_{q,I}\frac{n_I}{N}\ge c>0
\]
for some constant $c>0$ and all sufficiently large $N$.

Assumptions~\ref{ass:id} and \ref{ass:asymp}(ii) imply
$\tau=O(1)$ and
\[
\frac{1}{N}\sum_{i=1}^N
\{A_i(z)^2+D_i(z)^2\}
=
O(1).
\]
Hence, for $R\in\{A,D\}$,
\begin{align*}
\max_{q,I}
\frac{1}{n_I}\sum_{i\in I}R_i(z)^2
&\le
\frac{N}{\min_{q,I}n_I}
\frac{1}{N}\sum_{i=1}^N R_i(z)^2
\\
&=
O(1).
\end{align*}

For every such sample $I$,
\begin{align*}
\frac{1}{n_I}\sum_{i\in I}\hat A_i^{[-q]}-\bar A(z)
&=
\frac{1}{n_I}\sum_{i\in I}
\left\{
A_i(z)-(\hat\tau^{[-q]}-\tau)D_i(z)
\right\}
-\bar A(z)
\\
&=
\left\{
\frac{1}{n_I}\sum_{i\in I}A_i(z)-\bar A(z)
\right\}
-
(\hat\tau^{[-q]}-\tau)
\frac{1}{n_I}\sum_{i\in I}D_i(z).
\end{align*}
The first equality substitutes the definition of
$\hat A_i^{[-q]}$, and the second separates the two terms.

By Lemma~\ref{lem:split_sampling},
\begin{align*}
\mathbb E\left[
\frac{1}{n_I}\sum_{i\in I}A_i(z)
\right]
&=
\bar A(z),
\\
\operatorname{Var}\left(
\frac{1}{n_I}\sum_{i\in I}A_i(z)
\right)
&=
\left(
\frac{1}{n_I}-\frac{1}{N}
\right)
S_{A(z)}^2
=
O(N^{-1}),
\end{align*}
where $S_{A(z)}^2=O(1)$ follows from
Assumptions~\ref{ass:id} and \ref{ass:asymp}(ii).
Because the number of samples $I$ is fixed, Chebyshev's inequality
and the union bound give
\[
\max_{q,I}
\left|
\frac{1}{n_I}\sum_{i\in I}A_i(z)-\bar A(z)
\right|
=
O_p(N^{-1/2}).
\]
Moreover, Cauchy--Schwarz gives
\[
\max_{q,I}
\left|
\frac{1}{n_I}\sum_{i\in I}D_i(z)
\right|
\le
\max_{q,I}
\left\{
\frac{1}{n_I}\sum_{i\in I}D_i(z)^2
\right\}^{1/2}
=
O(1),
\]
and Lemma~\ref{lem:training_ratio_rate} gives
\[
\max_q|\hat\tau^{[-q]}-\tau|
=
O_p(N^{-1/2}).
\]
Therefore,
\[
\max_{q,I}
\left|
\frac{1}{n_I}\sum_{i\in I}\hat A_i^{[-q]}-\bar A(z)
\right|
=
O_p(N^{-1/2}).
\]

Next, define
\[
g_{N,z}
=
\frac{1}{N}\sum_{i=1}^N\tilde X_iA_i(z),
\qquad
\hat g_{I,z}^{[-q]}
=
\frac{1}{n_I}\sum_{i\in I}\tilde X_i\hat A_i^{[-q]}.
\]
Using the definition of $\hat A_i^{[-q]}$,
\begin{align*}
\hat g_{I,z}^{[-q]}-g_{N,z}
&=
\frac{1}{n_I}\sum_{i\in I}
\tilde X_i
\left\{
A_i(z)-(\hat\tau^{[-q]}-\tau)D_i(z)
\right\}
-g_{N,z}
\\
&=
\left\{
\frac{1}{n_I}\sum_{i\in I}\tilde X_iA_i(z)-g_{N,z}
\right\}
-
(\hat\tau^{[-q]}-\tau)
\frac{1}{n_I}\sum_{i\in I}\tilde X_iD_i(z).
\end{align*}
Hence,
\begin{align*}
\max_{q,I}
\|\hat g_{I,z}^{[-q]}-g_{N,z}\|_\infty
&\le
\max_{q,I}
\left\|
\frac{1}{n_I}\sum_{i\in I}\tilde X_iA_i(z)-g_{N,z}
\right\|_\infty
\\
&\quad+
K_{X,N}
\max_q|\hat\tau^{[-q]}-\tau|
\max_{q,I}
\left\{
\frac{1}{n_I}\sum_{i\in I}D_i(z)^2
\right\}^{1/2}.
\end{align*}
The inequality follows from the triangle inequality,
Cauchy--Schwarz, and
$|\tilde X_{ij}|\le K_{X,N}$ from
Assumption~\ref{ass:lasso_persistence}(i).

Lemma~\ref{lem:finite_population_concentration}, applied with
\[
v_{i,j}=\tilde X_{ij}A_i(z),
\qquad
H_N=K_{X,N}K_{A,N},
\qquad
J_N=p_N,
\]
gives
\[
\max_{q,I}
\left\|
\frac{1}{n_I}\sum_{i\in I}\tilde X_iA_i(z)-g_{N,z}
\right\|_\infty
=
O_p\left\{
K_{X,N}K_{A,N}
\sqrt{\frac{\log(p_N+1)}{N}}
\right\}.
\]
Together with Lemma~\ref{lem:training_ratio_rate} and the preceding
second-moment bound,
\begin{align*}
\max_{q,I}
\|\hat g_{I,z}^{[-q]}-g_{N,z}\|_\infty
&=
O_p\left[
K_{X,N}
\left\{
K_{A,N}\sqrt{\frac{\log(p_N+1)}{N}}
+
\frac{1}{\sqrt N}
\right\}
\right]
\\
&=
o_p(\lambda_N),
\end{align*}
where the last equality follows from
Assumption~\ref{ass:lasso_persistence}(ii).

For $\lambda\ge\lambda_N$, write
\[
\beta=\beta_A(z;\lambda),
\qquad
\hat\beta_I=\hat\beta_I(z;\lambda).
\]
For the population minimizer, comparison with the feasible choice
$b=0$ gives
\begin{align*}
\lambda\|\beta\|_1
&\le
\frac{1}{N}\sum_{i=1}^N
\{A_i(z)-\bar A(z)\}^2
\\
&=
O(1).
\end{align*}
Therefore,
\[
\sup_{\lambda\ge\lambda_N}
\|\beta_A(z;\lambda)\|_1
=
O(\lambda_N^{-1}).
\]

For the sample minimizer, comparison with $b=0$ gives
\begin{align*}
\lambda\|\hat\beta_I\|_1
&\le
\frac{2}{n_I}\sum_{i\in I}
\hat A_i^{[-q]}\tilde X_i'\hat\beta_I
-
\frac{1}{N}\sum_{i=1}^N
(\tilde X_i'\hat\beta_I)^2
\\
&\le
2
\left\{
\frac{N}{n_I^2}
\sum_{i\in I}(\hat A_i^{[-q]})^2
\right\}^{1/2}
\left\{
\frac{1}{N}\sum_{i=1}^N
(\tilde X_i'\hat\beta_I)^2
\right\}^{1/2}
\\
&\quad-
\frac{1}{N}\sum_{i=1}^N
(\tilde X_i'\hat\beta_I)^2
\\
&\le
\frac{N}{n_I^2}
\sum_{i\in I}(\hat A_i^{[-q]})^2
\\
&\le
\frac{2N}{n_I^2}
\sum_{i\in I}
\left\{
A_i(z)^2
+
(\hat\tau^{[-q]}-\tau)^2D_i(z)^2
\right\}
\\
&=
O_p(1).
\end{align*}
The second inequality uses Cauchy--Schwarz, the third uses
$2uv-v^2\le u^2$, and the fourth uses
$(a-b)^2\le2a^2+2b^2$.
The final bound follows from
Lemma~\ref{lem:training_ratio_rate}, the preceding second-moment
bounds, and $\min_{q,I}n_I/N\ge c>0$.
Since $\lambda\ge\lambda_N$,
\[
\max_{q,I}\sup_{\lambda\ge\lambda_N}
\|\hat\beta_I\|_1
=
O_p(\lambda_N^{-1}).
\]

The population and sample criteria have the same quadratic term
$\Sigma_{X,N}$ and differ only through
$g_{N,z}$ and $\hat g_{I,z}^{[-q]}$.
Their optimality conditions and the subgradient inequality give
\begin{align*}
\frac{1}{N}\sum_{i=1}^N
\{\tilde X_i'(\hat\beta_I-\beta)\}^2
&=
(\hat\beta_I-\beta)'
\Sigma_{X,N}
(\hat\beta_I-\beta)
\\
&\le
2
\|\hat g_{I,z}^{[-q]}-g_{N,z}\|_\infty
\left(
\|\hat\beta_I\|_1+\|\beta\|_1
\right).
\end{align*}
Therefore,
\begin{align*}
&\max_{q,I}\sup_{\lambda\ge\lambda_N}
\frac{1}{N}\sum_{i=1}^N
\{\tilde X_i'(\hat\beta_I-\beta)\}^2
\\
&\quad=
O_p\left[
\frac{K_{X,N}}{\lambda_N}
\left\{
K_{A,N}\sqrt{\frac{\log(p_N+1)}{N}}
+
\frac{1}{\sqrt N}
\right\}
\right]
=
o_p(1)
\end{align*}
by Assumption~\ref{ass:lasso_persistence}(ii).

Since $N^{-1}\sum_i\tilde X_i=0$,
\begin{align*}
\|\hat\mu_I^{[-q]}(z;\lambda)-\mu_A(z;\lambda)\|_N^2
&=
\left\{
\frac{1}{n_I}\sum_{j\in I}\hat A_j^{[-q]}
-
\bar A(z)
\right\}^2
\\
&\quad+
\frac{1}{N}\sum_{i=1}^N
\{\tilde X_i'(\hat\beta_I-\beta)\}^2
\\
&=
o_p(1)
\end{align*}
uniformly over the inner and outer training samples and
$\lambda\ge\lambda_N$. This proves the first two claims.

Finally, population optimality relative to $b=0$ gives
\[
L_{N,z}(\lambda)
\le
\frac{1}{N}\sum_{i=1}^N
\{A_i(z)-\bar A(z)\}^2
=
O(1),
\]
uniformly over $\lambda>0$. Hence,
\begin{align*}
\sup_{\lambda\ge\lambda_N}\|\mu_A(z;\lambda)\|_N
&\le
\|A(z)\|_N
+
\sup_{\lambda\ge\lambda_N}
\|A(z)-\mu_A(z;\lambda)\|_N
\\
&=
O(1).
\end{align*}
The triangle inequality and the uniform prediction bound then give
\[
\max_{q,k}\sup_{\lambda\ge\lambda_N}
\|\hat\mu_A^{[-q,-k]}(z;\lambda)\|_N
=
O_p(1).
\]
\end{proof}
\end{document}